\documentclass[%
 reprint,
 amsmath,amssymb,
pra,
]{revtex4-2}

\usepackage{graphicx}% Include figure files
\usepackage{dcolumn}% Align table columns on decimal point
\usepackage{bm}% bold math
\usepackage{caption}
\usepackage{subcaption}
\usepackage{float}
\usepackage{wrapfig}
\usepackage{blindtext}

\begin{document}

%\preprint{APS/123-QED}

\title{Beyond universality in repulsive SU(N) Fermi gases}
\author{Jordi Pera}
\author{Joaquim Casulleras}
\author{Jordi Boronat}%
\affiliation{%
 Departament de F\'\i sica, Campus Nord B4-B5, Universitat Polit\`ecnica de 
Catalunya, E-08034 Barcelona, Spain
}%

\date{\today}% It is always \today, today,

\begin{abstract}
Itinerant ferromagnetism in dilute Fermi gases is predicted to emerge at values 
of the gas parameter where
second-order perturbation theory is not accurate enough to properly 
describe the system. We have revisited
perturbation theory for SU(N) fermions and derived its generalization up to 
third order both in terms of the
gas parameter and the polarization. 
Our results agree satisfactorily with quantum Monte Carlo results for 
hard-sphere and soft-sphere potentials
for $S = 1/2$. Although the nature of the phase transition depends on the 
interaction potential, we find that for a
hard-sphere potential a phase transition is guaranteed to occur. While for 
$S= 1/2$ we observe a
quasi-continuous transition, for spins $3/2$ and $5/2$, a first-order phase 
transition is found.  For larger spins,
a double transition (combination of continuous and discontinuous) occurs. The 
critical density reduces
drastically when the spin increases, making the phase transition more accessible 
to experiments with ultracold
dilute Fermi gases. Estimations for Fermi gases of Yb and Sr with spin $5/2$ 
and $9/2$, respectively, are reported.

\end{abstract}

\maketitle

%\tableofcontents

\section{Introduction}

The well-known Stoner model of itinerant ferromagnetism~\cite{stoner}
predicts a transition to a ferromagnetic phase for an
electron gas when density is increased, due to the interplay
between potential and kinetic energies.
Trapped cold Fermi gases offer an ideal platform to study itinerant 
ferromagnetism that, in real materials,
has proven to be extremely elusive~\cite{pfleiderer,leduc}. This transition must
happen in the repulsive branch, which is metastable with respect to the  
formation of spin-up spin-down dimers~\cite{ji}. A first 
pioneering observation of the ferromagnetic transition in cold Fermi 
gases \cite{jo} was revised and concluded that pair formation hinders the 
achievement of the gas parameters required for observing this 
transition \cite{sanner}. Recently, it has been reported that the ferromagnetic 
state is effectively observed in a Fermi $^7$Li gas around a gas parameter 
$x=k_Fa_0\simeq1$, with $k_F$ the Fermi momentum and $a_0$ the $s$-wave
scattering length~\cite{valtolina}. This transition has
been extensively studied from a theoretical point of view, mainly using quantum 
Monte Carlo 
methods~\cite{conduit,pilati,chang,conduit2,massignan,cui,arias,pilati2}. These 
numerical estimations agree to localize the ferromagnetic transition around 
$x\simeq1$.

Dilute Fermi gases can also be studied using perturbation theory. At first 
order, the famous Stoner model~\cite{stoner} predicts a continuous phase 
transition for $S = 1/2$ and a first-order one for $S > 1/2$. As the Stoner 
model (Hartree-Fock approximation) in not accurate enough
when the gas parameter grows, a second-order theory was developed long time ago~\cite{huangyang,leeyang,galitskii,abrikosov}, but not applicable when the number of particles in each spin is different. Recently, this perturbative correction for SU(N) gases has been derived in a fully analytical
form in terms of both polarization and gas parameter~\cite{Pera}. Previous 
results by Kanno~\cite{kanno} were limited to the hard-sphere potential and $S = 
1/2$. At second order one observes that the ferromagnetic transition for $S = 
1/2$ turns to be discontinuous, breaking the asymmetry produced by the Stoner 
model for different spin values.

Second-order perturbation energies improve substantially the Stoner model but
still are not accurate enough to study Fermi gases close to the expected 
critical densities. Old results, at third-order of perturbation theory, are 
reported in Ref. ~\cite{Bishop}. However, to fully characterize the magnetic 
behavior of the Fermi gases it is fundamental to know the dependency of the 
energy on the gas parameter but also on the spin 
polarization~\cite{piotr2,lianyi}. In the present work, we derive the energy as 
a function of both parameters. Unlike the second order, we have not managed to 
derive a fully analytical expression. It is worth noticing that, going beyond
second order breaks universality, meaning that the energy
dependence is no longer solely determined by the $s$-wave
scattering length. As we will demonstrate, two additional
scattering parameters come into play in the description: the
$s$-wave effective range and the $p$-wave scattering length. It is worth mentioning that, for a spin balanced gas, the fourth order term has been fully derived in Ref.~\cite{4t-ordre}.

In recent years, the experimental production of SU(N) fermions has renewed the 
theoretical interest in their study. In particular, Ytterbium~\cite{pagano}, 
with spin 5/2, and Strontium~\cite{goban}, with spin
9/2, are now available for studying Fermi gases with spin degeneracy never 
achieved before. Importantly, Cazalilla \textit{et 
al.}~\cite{cazalilla_1,cazalilla_2}  showed  that Fermi gases made of alkaline 
atoms with two electrons in the external shell, such as $^{173}$Yb, present an 
SU(N) emergent symmetry. They also argued that the ferromagnetic
transition must be of first order when $S > 1/2$, based on the
significant dissimilarities in the mathematical structure
between SU(N$>$2) and SU(2).
The role of interaction effects in SU(N) Fermi gases  as a function of $N$ were 
also  studied in Ref.~\cite{effect_vary_N}. Collective excitations in SU(N) 
Fermi gases with tunable spin were investigated in 
Ref.~\cite{collective_excitations}. On the other hand, 
the prethermalization of these systems was analyzed~\cite{prethermalization},
finding that, under some conditions, the imbalanced initial state could be 
stabilized for a certain time. Recently, the thermodynamics of  $^{87}$Sr for 
which $N$ can be tuned up to 10 was thoroughly analyzed in 
Ref.~\cite{estronci_N_10}. For temperatures above the 
super-exchange energy, the behavior of the thermodynamic quantities was found to 
be universal with respect to N~\cite{universal_behavior}. The   
temperature dependence of itinerant ferromagnetism in SU(N)-symmetric Fermi 
gases at 
second order of perturbation theory has been studied recently by Huang and 
Cazalilla~\cite{cazalilla-t}.

In the present work, we compare our predictions with diffusion Monte Carlo 
(DMC) results~\cite{pilati} for $S = 1/2$. 
We achieve a satisfactory agreement when considering hard-sphere
fermions up to  $k_Fa_0 \simeq 1$, where the ferromagnetic transition 
occurs. Interestingly, in contrast to the observations in second-order
perturbation theory, fermions with $S = 1/2$ exhibit a
quasi-continuous phase transition into the ferromagnetic state.
An important result of our study is that  the critical density for itinerant 
ferromagnetism decreases with $N$, with values that are systematically smaller 
than the ones obtained in second order~\cite{Pera}. Our 
results are derived for a generic spin and thus, can be applied to SU(N) 
fermions. This generalization allows, for instance, the study of dilute Fermi 
gases of Ytterbium~\cite{pagano}, with spin 5/2, and Strontium~\cite{goban}, 
with spin 9/2.

\section{Methodology}
We study a repulsive Fermi gas at zero temperature with spin $S$ and spin  
degeneracy $\nu=2 S +1$. We have used perturbation theory in our analysis, 
resulting in a combination of analytic results and numerical estimations as 
the final outcome.  In the dilute gas regime, only particles 
with different $z$-spin component interact via a central potential 
$V(r)$ ($s$-wave scattering). 
However, since our objective is to obtain the energy up to
the third order in the gas parameter, we will have to
consider interactions between particles with varying $z$-spin
components, involving not only the $s$-wave scattering length
but also the $s$-wave effective range and the $p$-wave scattering
length. We discuss the respective terms and carry out their
calculations, which require the evaluation of several
challenging integrals.

The first and second order terms have 
already been calculated in Ref.~\cite{Pera}. The terms coming from the 
$s$-wave effective range and the $p$-wave scattering length can be obtained 
analytically. However, the other contributions to the third-order 
expansion require a  
combination of analytical derivation and numerical integration.
The number of particles in each spin channel is 
$N_\lambda=C_\lambda N/\nu$, with $N$ the total number of particles and   
$C_\lambda$ being the fraction of $\lambda$ particles (normalized to be one if the 
system is unpolarized, $N_\lambda=N/\nu$, $\forall \, \lambda$).  The Fermi  
momentum of each species is $k_{F,\lambda}=k_F C_\lambda^{1/3}$, $k_F$ being 
$(6\pi^2n/\nu)^{1/3}$. The kinetic energy is readily obtained as it corresponds 
to the one of the free Fermi gas,
\begin{equation}
\frac{T}{N}=\frac{3}{5}\epsilon_F\frac{1}{\nu}\sum_{\lambda}C_{\lambda}^{5/3} \ 
,
\label{ekin}
\end{equation}
$\epsilon_F=\hbar^2 k_F^2/ 2m$ being the Fermi energy, and 
where the summation is extended to include all the
$z$-spin degrees of freedom. However, obtaining the potential
energy is more challenging, requiring the application of
perturbation theory.
The formalism that we use is based on previous 
works~\cite{Bishop,efimov}. In essence, what is done is to  
calculate the Feynman diagrams that contribute to each order of the 
expansion, and then to substitute the interaction by the $\mathcal{K}$-matrix, which 
depends on the low-energy scattering parameters of the potential $V$. We have 
generalized 
this procedure considering that all the Fermi species can have any 
spin-channel occupation and thus including the polarization as a new variable. 
The potential energy can be written in terms of the $\mathcal{K}$-matrix,
\begin{equation}
\begin{aligned}
V=\frac{\hbar^2 
\Omega}{2m}\sum_{\lambda_1,\lambda_2}\int\frac{d\textbf{p}}{(2\pi)^3} 
n_p\int\frac{d\textbf{p}'}{(2\pi)^3}n_{p'}
     \hspace{2cm}
     \\ 
\times \{ \mathcal{K}(\textbf{p},\textbf{p}';\textbf{p},\textbf{p}')-\delta_{\lambda_1,
\lambda_2 } \mathcal{K}(\textbf{p} , \textbf{p}';\textbf{p}' , \textbf{p}) \} \ ,
\end{aligned}
\label{kmatrix}
\end{equation}
with $\Omega$ the volume, and $n_p$ and $n_{p'}$ the momentum distributions of the 
free Fermi gas. Therefore,  we have to integrate over two particles 
($\lambda_1,\lambda_2$) 
that are interacting between themselves through the $\mathcal{K}$-matrix, which brings 
information on the potential.
The $\mathcal{K}$-matrix up to third order is written as~\cite{Bishop,Baker,efimov}
\begin{widetext}
\begin{equation}
\label{matriuk}
\begin{aligned}
    \mathcal{K}(\textbf{Q},\textbf{P};\textbf{Q'},\textbf{P'})=4\pi a_0+(4\pi a_0)^2I(Q,P)+4\pi \textbf{Q}^2\frac{r_0}{2}a_0^2+4\pi a_1^3\textbf{Q}\cdot \textbf{Q'}+(4\pi a_0)^3I^2(Q,P)\\
    +4\pi a_0\int\frac{2 \, 
d\textbf{m}d\textbf{m}'}{(2\pi)^3}\frac{(1-n_m)(1-n_{m'})}{p^2+p'^2-m^2-m'^2}
\delta(\textbf{p}+\textbf{p}'-\textbf{m}-\textbf{m}')\bigg\{(4\pi a_0)^2\int 
\frac{2 \, 
d\textbf{p}_1d\textbf{p}_1'}{(2\pi)^3}\frac{n_{p_1}n_{p_1'}\delta(\textbf{p}
_1+\textbf{p}_1'-\textbf{m}-\textbf{m}')}{p_1^2+p_1'^2-m^2-m'^2}\\
    +(4\pi a_0^{(13)})(4\pi 
a_0^{(23)})\sum_{\lambda_3}(2-3\delta_{\lambda_1,\lambda_3}-3\delta_{\lambda_2,
\lambda_3})\int 
\frac{d\textbf{p}_1d\textbf{m}_1}{(2\pi)^3}n_{p_1}(1-n_{m_1})\bigg[\frac{
\delta(\textbf{p}+\textbf{p}_1-\textbf{m}-\textbf{m}_1)}{p^2+p_1^2-m^2-m_1^2}
+\frac{\delta(\textbf{p}'+\textbf{p}_1-\textbf{m}'-\textbf{m}_1)}{
p'^2+p_1^2-m'^2-m_1^2}\bigg]\bigg\}\\+O(a_0^4)\ .
\end{aligned}
\end{equation}
\end{widetext}

In Eq. (\ref{matriuk}), $\textbf{Q} = (\textbf{p}-\textbf{p}')/2  $ and 
$\textbf{P}=\textbf{P'}=\textbf{p}+\textbf{p}' $ are the relative momentum and 
center of mass momentum, respectively. The parameters $r_0$ and $a_1$ in that 
equation are the $s$-wave effective range and $p$-wave scattering length, 
respectively. For 
$\mathcal{K}(\textbf{p},\textbf{p}';\textbf{p},\textbf{p}')$, $\textbf{Q}'= \textbf{Q}$, 
whereas for  $\mathcal{K}(\textbf{p} , \textbf{p}';\textbf{p}' , 
\textbf{p})$,  $\textbf{Q}'= -\textbf{Q}$. The scattering length $a_0^{(13)}$ 
corresponds to the interaction between the first and third particles and 
$a_0^{(23)}$ to the second and third ones. When the interactions between 
different channels are the same,  $a_0^{(13)}=a_0^{(23)}=a_0$. The term 
proportional to $a_0^{(13)}a_0^{(23)}$, in the expression of the $\mathcal{K}$-matrix, is 
a three-body interaction term.  On the other hand, the function $I(Q,P)$ in Eq. 
(\ref{matriuk}) is defined as
\begin{eqnarray}
\lefteqn{ I(Q,P)  = \frac{1}{(2\pi)^3}}   \\
       & &     \times \int 2 \, d\textbf{q}d\textbf{q}'\frac{1-(1-n_q)(1-n_{q'})}{q^2+q'^2-p^2-p'^2}\delta(\textbf{q}+\textbf{q}'-\textbf{p}-\textbf{p}')  \ . \nonumber
\label{Iint}          
\end{eqnarray}
Considering only the first term in the expansion of the $\mathcal{K}$-matrix (\ref{matriuk}), 
one gets for the potential energy (\ref{kmatrix}) the well-know Hartree-Fock 
energy~\cite{stoner},
\begin{equation}
 \left( \frac{V}{N} \right)_1=   \frac{2 \epsilon_F}{3\pi}  \left[ \frac{1}{\nu}\sum_{\lambda_1,\lambda_2}
 C_{\lambda_1}C_{\lambda_2} (1-\delta_{\lambda_1,\lambda_2}) \right] \, x \ ,
 \label{hfenergy}
\end{equation}
with $x \equiv k_F a_0$ the gas parameter of the Fermi gas.

The second-order term in the gas parameter $x$ derives from  the second term of 
the $\mathcal{K}$-matrix. This second order term reads
\begin{equation}
\left( \frac{V}{N} \right)_2= \frac{\epsilon_F}{k_F^7} \left[ \frac{1}{\nu}\sum_{\lambda_1,\lambda_2}I_2(k_{F,\lambda_1},k_{F,\lambda_2})(1-\delta_{\lambda_1,\lambda_2}) \right] x^2 \ ,
\label{2ndorder}
\end{equation}
with
\begin{equation}
\begin{aligned}
I_2(k_{F,\lambda_1},k_{F,\lambda_2})=\frac{3}{16\pi^5}\int d\textbf{p} \, 
n_p\int 
d\textbf{p}' \, n_{p'} \int 2 \, d\textbf{q}d\textbf{q}'\\
     \times\frac{1-(1-n_q)(1-n_{q'})}{q^2+q'^2-p^2-p'^2}\delta(\textbf{q}+\textbf{q}'-\textbf{p}-\textbf{p}') \ .
\end{aligned}
\label{i2tot}
\end{equation}

This integral $I_2$ was already obtained in Ref.~\cite{Pera}. In terms of 
$C_{\lambda_1}$ and $C_{\lambda_2}$, 
\begin{equation}
I_2(C_{\lambda_1},C_{\lambda_2})  
=\frac{4k_F^7}{35\pi^2}C_{\lambda_1}C_{\lambda_2}\frac{C_{\lambda_1}^{1/3} +C_ 
{\lambda_2}^{1/3}}{2} \, F(y) \ ,
\label{i2final}
\end{equation}
with
\begin{widetext}
\begin{equation}
F(y)= \frac{1}{4}\big(15y^2-19y+52-19y^{-1}+15y^{-2}\big)   
+\frac{7}{8}y^{-2}\big(y-1\big)^4\big(y+3+y^{-1}\big)\ln{\bigg\vert\frac{1-y}{
1+y}\bigg\vert}
    -\frac{2y^4}{1+y}\ln{\bigg\vert 
1+\frac{1}{y}\bigg\vert}-\frac{2y^{-4}}{1+y^{-1}}\ln{\bigg\vert 1+y\bigg\vert}
\label{funy}
\end{equation}
\end{widetext}
and $y\equiv (C_{\lambda_1}/C_{\lambda_2})^{1/3}$. It is worth noticing that 
for $S=1/2$  this term agrees with the Kanno result~\cite{kanno}.

In the following subsections, we calculate the new terms, that are the ones 
contributing to the third order in the gas parameter.

\subsection{S-wave effective range term}
Beyond second-order, one needs to introduce additional scattering parameters 
other than the $s$-wave scattering length and thus  
the expression is no longer universal. The effective range term in the 
$\mathcal{K}$-matrix is $4\pi \textbf{Q}^2\frac{1}{2}r_0a_0^2$. First of all, we 
express  $\textbf{Q}^2$ in terms of $\textbf{p}$ and $\textbf{p}'$,
\begin{equation}
    \textbf{Q}^2=\frac{1}{4}(\textbf{p}-\textbf{p}')^2=\frac{1}{4}(p^2+p'^2-2pp'\cos{\theta})
\end{equation}

Then, we substitute the value of $\mathcal{K}$ in Eq. (\ref{kmatrix}) and integrate it, 
\begin{equation}
\begin{aligned}
     V_{r_0}
=\frac{\hbar^2 
\Omega}{2m}\sum_{\lambda_1,\lambda_2}\int\frac{d\textbf{p}}{(2\pi)^3} 
n_p\int\frac{d\textbf{p}'}{(2\pi)^3}n_{p'}\\
     \times\frac{4\pi r_0a_0^2}{8}(p^2+p'^2-2pp'\cos{\theta})(1-\delta_{\lambda_1,\lambda_2})\\
  =\frac{\hbar^2 
\Omega }{16m \pi^3}r_0a_0^2  
\sum_{\lambda_1,\lambda_2} \bigg[ \bigg(\frac{k_{F, 
\lambda_1}^3}{3}\frac{k_{F,\lambda_2}^5}{5}+\frac{k_{F,\lambda_1}^5}{5}\frac{k_{ 
F,\lambda_2}^3}{3}\bigg)\\
     \times (1-\delta_{\lambda_1,\lambda_2}) \bigg] \ ,
\end{aligned}
\end{equation}
where the  term containing $\cos{\theta}$ gives zero after doing the angular 
integration.  The potential energy per particle as a function of the parameters 
$C_{\lambda}$ is finally
\begin{equation}
\begin{aligned}
 \left(\frac{V}{N} \right)_{r_0} =\epsilon_F 
\, \frac{1}{\nu}  \frac{1}{ 10\pi } \sum_{\lambda_1,\lambda_2} \bigg[ 
C_{\lambda_1}C_{\lambda_2}\bigg(\frac{C_{\lambda_1}^{2/3}+C_{\lambda_2}^{2/3}}{2
}\bigg)
    \\
     \times 
(k_Fr_0)(k_Fa_0)^2(1-\delta_{\lambda_1,\lambda_2}) \bigg] \ .
\end{aligned}
\end{equation}

\subsection{P-wave term}
The inclusion of the $p$-wave scattering length in the $\mathcal{K}$-matrix
 (\ref{kmatrix}) introduces additional complexity as it also depends on 
$\textbf{Q'}$. The two $\mathcal{K}$-matrix terms that we require differ by a 
minus sign: 
$\mathcal{K}(\textbf{p},\textbf{p}';\textbf{p},\textbf{p}')=4\pi 
a_1^3\textbf{Q}^2$,  
$\mathcal{K}(\textbf{p},\textbf{p}';\textbf{p}',\textbf{p})=-4\pi 
a_1^3\textbf{Q}^2$.
This negative sign causes the term proportional to
the Kronecker delta to change its sign, thereby introducing
interaction between particles of the same spin. Although this
fact may not be immediately evident, it becomes apparent once
we integrate and rearrange the terms. Apart from the negative
sign, the integral that needs to be performed is formally
similar to the one calculated for the effective range.
Hence, the potential energy per 
particle is
\begin{equation}
\begin{aligned}
 \left( \frac{V}{N} 
\right)_{a_1} =\epsilon_F\frac{1}{\nu} \frac{1}{5 
\pi }\sum_{\lambda_1,\lambda_2} \bigg[   
C_{\lambda_1}C_{\lambda_2}\bigg(\frac{C_{\lambda_1}^{2/3}+C_{\lambda_2}^{2/3}}{2
}\bigg)        \\
    \times (k_Fa_1)^3 (1+\delta_{\lambda_1,\lambda_2}) \bigg]  \ .
\end{aligned}
    \label{eq.eneonaP}
\end{equation}

 The term $(1+\delta_{\lambda_1,\lambda_2})$ in Eq.(\ref{eq.eneonaP}) 
is 
equivalent to $2\delta_{\lambda_1,\lambda_2}+(1-\delta_{\lambda_1,\lambda_2})$. 
Then, there are interaction between pairs of different spin, as in 
all previous 
terms, which are all proportional to 
$(1-\delta_{\lambda_1,\lambda_2})$. But now we also have a term 
$2\delta_{\lambda_1,\lambda_2}$, which will give rise to an extra interaction 
term between particles of same spin. Particularizing Eq. (\ref{eq.eneonaP}) 
for the latter contribution, between particles of same spin, 
we obtain
\begin{equation}
\begin{aligned}
 \left(\frac{V}{N}\right) 
=\epsilon_F\frac{1}{\nu} \frac{ 1 } { 5\pi } \sum_{\lambda_1,\lambda_2} \bigg[ 
C_ { 
\lambda_1}C_{\lambda_2}\bigg(\frac{C_{\lambda_1}^{2/3}+C_{\lambda_2}^{2/3}}{2} 
\bigg)
\\
\times (k_Fa_1)^3(2\delta_{\lambda_1,\lambda_2}) \bigg]\\
   = \epsilon_F\frac{1}{\nu} 
\frac{2}{5 \pi} \sum_{\lambda} C_{\lambda}^{8/3} (k_Fa_1)^3 \ .
\end{aligned}
\label{eq.onap-samev}
\end{equation}

If we split Eq. (\ref{eq.eneonaP}) in two parts, one part containing the
interaction between particles of different spin and another one with this new 
contribution, one gets
\begin{equation}
\begin{aligned}
    \left(\frac{V}{N}\right)_{a_1} =\frac{3}{5}\epsilon_F \frac{1}{\nu} 
    \left\{
\frac { 2 } { 3\pi } \sum_{ \lambda } C_{\lambda}^{8/3}(k_Fa_1)^3 \right.
     \hspace{3.5cm}\\
 \hspace{-0.3cm} +  \left. \frac{1}{3 \pi}
\sum_{\lambda_1,\lambda_2}  \bigg[ C_{\lambda_1}C_{\lambda_2}
\bigg(\frac{C_{\lambda_1}^{2/3}+C_{\lambda_2}^{2/3}}{2}
\bigg)(k_Fa_1)^3(1-\delta_{\lambda_1,\lambda_2})\bigg] \right\}  \ .
\end{aligned}
\end{equation}

The primary focus of our work is the investigation of highly
degenerate and extremely dilute Fermi gases, assuming that
the $p$-wave interaction among particles of the same spin can
be neglected. Consequently, this term  (Eq. \ref{eq.onap-samev})   will not be
taken into account in the Results section.

\subsection{3rd order terms depending on $\bm{a_0}$}
Due to their intricate mathematical nature, we have not been able to 
calculate the 
remaining  third order terms in a fully analytical form. These terms are 
exclusively dependent  on the $s$-wave scattering length. We used a Monte Carlo 
integration tool to calculate these terms. 

The first term that requires consideration arises from  
the $(4\pi a_0)^3I^2(Q,P)$ term in the $\mathcal{K}$-matrix expansion. Inserted 
in Eq. (\ref{kmatrix}), one obtains
\begin{equation}
\begin{aligned}
V_3=\frac{\hbar^2
\Omega}{2m}\sum_{\lambda_1,\lambda_2}\int\frac{d\textbf{p}}{(2\pi)^3} 
n_p\int\frac{d\textbf{p}'}{(2\pi)^3}n_{p'}
     \hspace{2cm}
     \\ 
\times (4\pi a_0)^3I^2(Q,P)(1-\delta_{\lambda_1,
\lambda_2 }) \ .
\end{aligned}
\end{equation}
Rearranging the integrals, one can write a more manageable expression,
\begin{equation}
%\begin{aligned}
\left(\frac{V}{N}\right)_3 =\epsilon_F\frac{3x^3}{32\pi^7}\frac{1}{\nu}\sum_{\lambda_1,\lambda_2} \left[ (1-\delta_{\lambda_1,\lambda_2})E_3(C_{\lambda_1},C_{\lambda_2}) \right] \ ,
\end{equation}
with
\begin{equation}
\begin{aligned}
E_3(C_{\lambda_1},C_{\lambda_2})=\frac{1}{k_F^8}\int d\textbf{p}n_p\int d\textbf{p}'n_{p'}\\
\times \bigg[\int 2 \, d\textbf{q}d\textbf{q}' (1-(1-n_q)(1-n_{q'})) \frac{\delta(\textbf{q}+\textbf{q}'-\textbf{p}-\textbf{p}'}  {q^2+q'^2-p^2-p'^2}  )\bigg]^2 \ .
\end{aligned}
\end{equation}

Going back to the expression of the $\mathcal{K}$-matrix (\ref{kmatrix}) one can see that 
there is another pair-like term,  given by
\begin{equation}
\begin{aligned}
V_4=\frac{\hbar^2
\Omega}{2m}\sum_{\lambda_1,\lambda_2}\int\frac{d\textbf{p}}{(2\pi)^3}
n_p\int\frac{d\textbf{p}'}{(2\pi)^3}n_{p'}(1-\delta_{\lambda_1,\lambda_2 })(4\pi a_0)\\
    \times\int\frac{2 \, d\textbf{m}d\textbf{m}'}{(2\pi)^3}\frac{(1-n_m)(1-n_{m'})}{p^2+p'^2-m^2-m^2}\delta(\textbf{p}+\textbf{p}'-\textbf{m}-\textbf{m}')\\
    \times(4\pi a_0)^2\int \frac{2 \, d\textbf{p}_1d\textbf{p}_1'}{(2\pi)^3}n_{p_1}n_{p_1'} \frac{\delta(\textbf{p}_1+\textbf{p}_1'-\textbf{m}-\textbf{m}')}{p_1^2+p_1'^2-m^2-m'^2}\ .
\end{aligned}
\end{equation}
Observing  that $\textbf{p}$ and $\textbf{p}_1$ run over the same values, and the same happens for $\textbf{p}'$ and $\textbf{p}_1'$, we can interchange the integrals and rewrite the whole expression as
\begin{equation}
%\begin{aligned}
\left( \frac{V}{N} \right)_4 =\epsilon_F\frac{3x^3}{32\pi^7}\frac{1}{\nu}\sum_{\lambda_1,\lambda_2} \left[ (1-\delta_{\lambda_1,\lambda_2})E_4(C_{\lambda_1},C_{\lambda_2}) \right] \ ,
\end{equation}
with
\begin{equation}
\begin{aligned}
E_4(C_{\lambda_1},C_{\lambda_2})=\frac{1}{k_F^8}\int d\textbf{m}(1-n_m)\int d\textbf{m}'(1-n_{m'})\\
    \times\bigg[\int 2 \, d\textbf{p}d\textbf{p}' n_p n_{p'}  \frac{\delta(\textbf{p}+\textbf{p}'-\textbf{m}-\textbf{m}')}{p^2+p'^2-m^2-m^2}\bigg]^2
\end{aligned}
\end{equation}
Finally, the last term in Eq. (\ref{kmatrix}) contains the interaction between three
particles. At third order, this is the only three-body interacting term. Its expression is more involved than the previous ones,
\begin{widetext}
\begin{equation}
\begin{aligned}
V_5=\frac{\hbar^2
\Omega}{2m}\sum_{\lambda_1,\lambda_2}\int\frac{d\textbf{p}}{(2\pi)^3}
n_p\int\frac{d\textbf{p}'}{(2\pi)^3}n_{p'}(1-\delta_{\lambda_1,\lambda_2 })(4\pi a_0)^3\int\frac{2 \, d\textbf{m}d\textbf{m}'}{(2\pi)^3} (1-n_m)(1-n_{m'})  \frac{\delta(\textbf{p}+\textbf{p}'-\textbf{m}-\textbf{m}')} {p^2+p'^2-m^2-m^2}  \\
    \times\sum_{\lambda_3}(2-3\delta_{\lambda_1,\lambda_3}-3\delta_{\lambda_2,\lambda_3})\int \frac{d\textbf{p}_1d\textbf{m}_1}{(2\pi)^3}n_{p_1} (1-n_{m_1})\bigg[\frac{\delta(\textbf{p}+\textbf{p}_1-\textbf{m}-\textbf{m}_1)}{p^2+p_1^2-m^2-m_1^2}+\frac{\delta(\textbf{p}'+\textbf{p}_1-\textbf{m}'-\textbf{m}_1)}{p'^2+p_1^2-m'^2-m_1^2}\bigg] \ .
\end{aligned}
\end{equation}
After rearranging, we can write it in a more compact form
\begin{equation}
%\begin{aligned}
\left( \frac{V}{N} \right)_5=\epsilon_F\frac{3x^3}{32\pi^7}\frac{1}{\nu}\sum_{\lambda_1,\lambda_2,\lambda_3} \left[ (1-\delta_{\lambda_1,\lambda_2 })(2-3\delta_{\lambda_1,\lambda_3}-3\delta_{\lambda_2,\lambda_3})E_5(C_{\lambda_1},C_{\lambda_2},C_{\lambda_3}) \right] \ ,
\end{equation}
with
\begin{equation}
\begin{aligned}
 E_5(C_{\lambda_1},C_{\lambda_2},C_{\lambda_3})=
\frac{1}{2k_F^8}\bigg\{\int d\textbf{p}n_p\int d\textbf{m}(1-n_m)\\
\times\bigg[\int2d\textbf{m}'d\textbf{p}'(1-n_{m'})n_{p'} \frac{\delta(\textbf{p}+\textbf{p}'-\textbf{m}-\textbf{m}'} {p^2+p'^2-m^2-m'^2}  )\bigg]\bigg[\int2d\textbf{m}_1d\textbf{p}_1 (1-n_{m_1})n_{p_1} \frac{ \delta(\textbf{p}+\textbf{p}_1-\textbf{m}-\textbf{m}_1}  {p^2+p_1^2-m^2-m_1^2} )\bigg]\\
     +\int d\textbf{p}'n_{p'}\int d\textbf{m}'(1-n_{m'})\\
     \times\bigg[\int2d\textbf{m}d\textbf{p} (1-n_m) n_p  \frac{ \delta(\textbf{p}+\textbf{p}'-\textbf{m}-\textbf{m}'} {p^2+p'^2-m^2-m'^2} )\bigg]\bigg[\int2d\textbf{m}_1d\textbf{p}_1 (1-n_{m_1})n_{p_1} \frac{\delta(\textbf{p}'+\textbf{p}_1-\textbf{m}'-\textbf{m}_1}  {p'^2+p_1^2-m'^2-m_1^2} )\bigg]\bigg\}
\end{aligned}
\end{equation}
\end{widetext}

The  integrals $E_3$, $E_4$, and $E_5$ were already calculated  previously for a
non-polarized gas and $S=1/2$ 
\cite{leeyang,huangyang,galitskii,abrikosov,efimov,efimovsamu}. In order to 
calculate numerically the integrals for spins greater
than 1/2, we have made the assumption that as the concentration of one 
species 
increases, all the other species diminish in the same manner. This particular 
configuration minimizes the energy when the total number of particles remains 
constant~\cite{Pera}. Under these conditions, the
concentrations $C_{\lambda}$ for a given polarization $P$ are 
\begin{eqnarray}
C_+ & = & 1+ |P| \, (\nu-1) \\
C_{\lambda \ne +} & = & 1-|P| \ ,
\label{cesp3}
\end{eqnarray}
with subindex $+$ standing for the spin state with the largest population. The explicit way in which these integrals are calculated can be found in the Appendices. We have computed these integrals for different values of the polarization and a range of spin values. More precisely, we have calculated 201 points between $P=0$ and $P=1$, with $10^8$ sampling points using an accurate adaptative Monte Carlo integration~\cite{vegas,vegas2}.

\subsection{Energy up to third order}
Collecting the different terms discussed in previous subsections, we can write a final expression for the energy of a Fermi gas up to third order in the gas parameter $k_F a_0$, and for any spin degeneracy and polarization,
\begin{widetext}
\begin{equation}
\label{eq.forfinal}
\begin{aligned}
     \frac{E}{N}=\frac{3}{5}\epsilon_F\bigg\{ \frac{1}{\nu}\sum_{\lambda}C_{\lambda}^{5/3}+\frac{1}{\nu}\sum_{\lambda}\frac{2}{3\pi}C_{\lambda}^{8/3}(k_Fa_1)^3
   +\frac{5}{3\nu}\sum_{\lambda_1,\lambda_2} \bigg[ \bigg(  \frac{2}{3\pi}(k_Fa_0)C_{\lambda_1}C_{\lambda_2}+\frac{4}{35\pi^2}C_{\lambda_1}C_{\lambda_2}\frac{C_{\lambda_1}^{1/3} +C_
{\lambda_2}^{1/3}}{2} \, F(y)(k_Fa_0)^2\\
   +\frac{1}{10\pi}C_{\lambda_1}C_{\lambda_2}\bigg(\frac{C_{\lambda_1}^{2/3}+C_{\lambda_2}^{2/3}}{2}\bigg)\bigg[\frac{r_0}{a_0}+2\frac{a_1^3}{a_0^3}\bigg](k_Fa_0)^3
   +\frac{3}{32\pi^7}\big[E_3+E_4+\sum_{\lambda_3}( 
(2-3\delta_{\lambda_1,\lambda_3} 
-3\delta_{\lambda_2,\lambda_3})E_5) \big](k_Fa_0)^3\bigg)(1-\delta_{\lambda_1, 
\lambda_2})\bigg] \bigg\} \ .
\end{aligned}
\end{equation}
\end{widetext}

\section{Results}

In this section, we discuss the main results of SU(N)
Fermi gases using the framework established in the previous
section. Unlike the second order analysis, we now introduce
two additional scattering parameters that characterize the
interaction: the $s$-wave effective range ($r_0$) and the $p$-wave
scattering length ($a_1$). This implies that the specific
potential model has an influence on the derived results
beyond the simple dependence on the $s$-wave scattering length.
Considering our focus on a repulsive gas, unless otherwise
stated, we will assume a hard-sphere potential ($r_0=2\,a_0/3$, $a_1=a_0$), 
as it provides a somewhat more universal model.

\begin{figure}[h]
    \centering
    \includegraphics[width=0.48\textwidth]{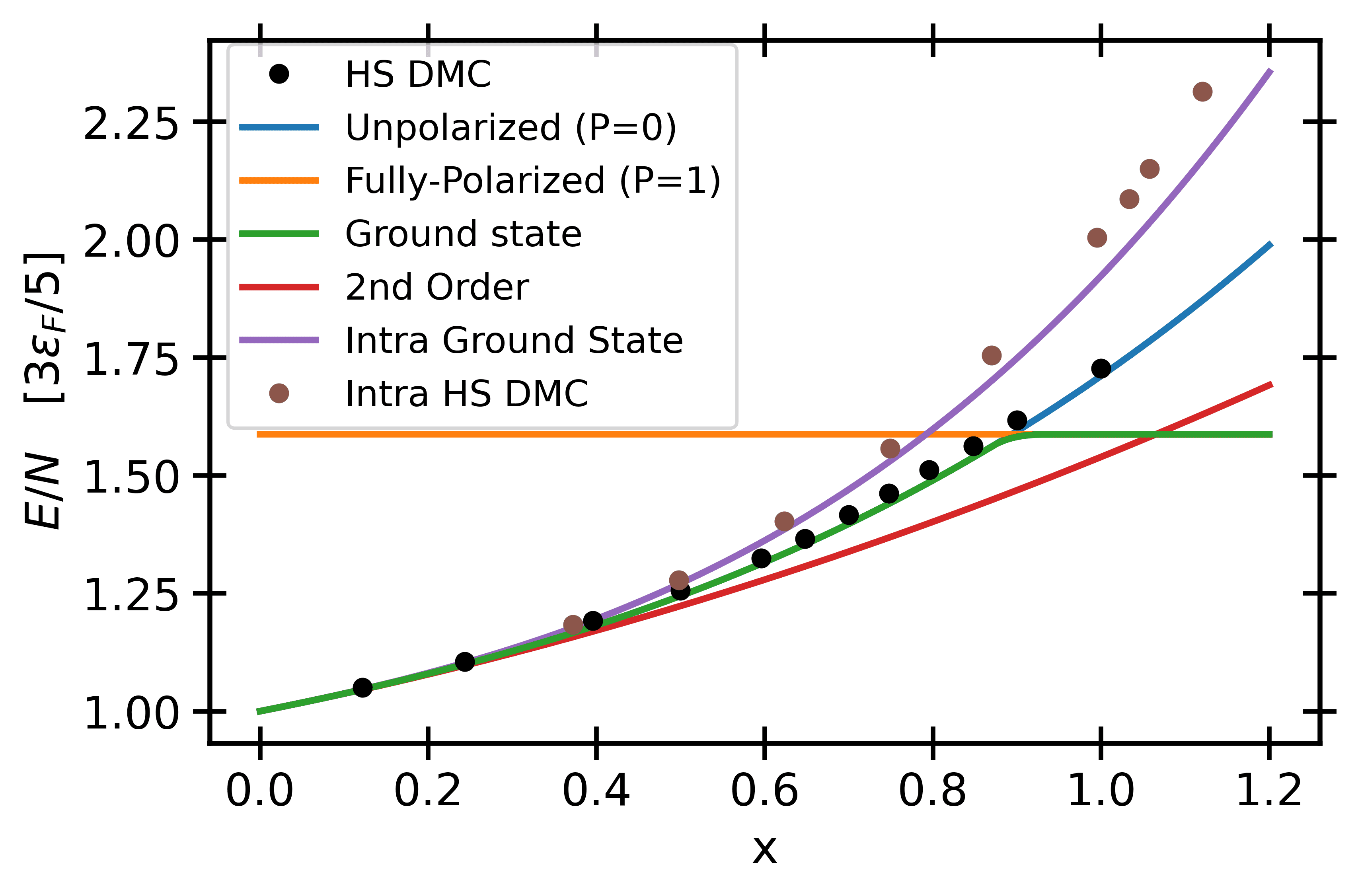}
    \caption{Energy for $S=1/2$ as a function of the gas parameter $x=k_F a_0$. 
We plot three models: in red, the second-order 
model; in green (and blue), the third-order one; and in purple, the third-order one with $P$-wave intra-species interaction. The black and brown points are DMC results from Ref. \cite{pilati} and \cite{bertaina} respectively. The potential used 
is hard-spheres ($r_0=2\,a_0/3$, $a_1=a_0$).}
    \label{fig.HS}
\end{figure}

Having a perturbative prediction at our disposal, we can
make a comparison between our results and the existing
quantum Monte Carlo data. We compare two sets of values in Fig. \ref{fig.HS}. The black 
points are diffusion Monte Carlo (DMC) data for spin 1/2~\cite{pilati}. This set does 
not include $P$-wave scattering terms between particles with the same $z$-spin
component. The brown points in the same figure are DMC data by Bertaina et 
al.~\cite{bertaina} which include intra-species interaction.
The blue and purple lines are our theoretical predictions for these two cases, 
respectively. Both results correspond to non-polarized gases. As we can see, 
the energy is higher when the intra-species interaction is considered. Moreover, 
although it is not shown here, intra-species interaction in the  
fully-polarized gas make the energy increase with the density~\cite{bertaina,pol-gas}. 
In the rest of our results, we do not include this contribution. 
The orange line corresponds to the fully-polarized gas, the green one stands 
for the configuration of minimum 
energy, that is, at each 
value of the density, we select the polarization that minimizes the energy. And 
finally, the red line corresponds to the second-order energy 
(universal expansion) to show the differences  with the third-order expansion. 
We can see how the second-order  and the third-order expansions 
reproduce the same energy for values of $k_Fa_0 \lesssim 0.4$. Beyond that, the 
difference intensifies with increasing density~\cite{arias}. Concerning the DMC 
points, although they are  upper-bounds to the exact energy due to 
the sign problem, they fit pretty well the third-order curves.

\begin{figure}[h]
    \centering
    \includegraphics[width=0.48\textwidth]{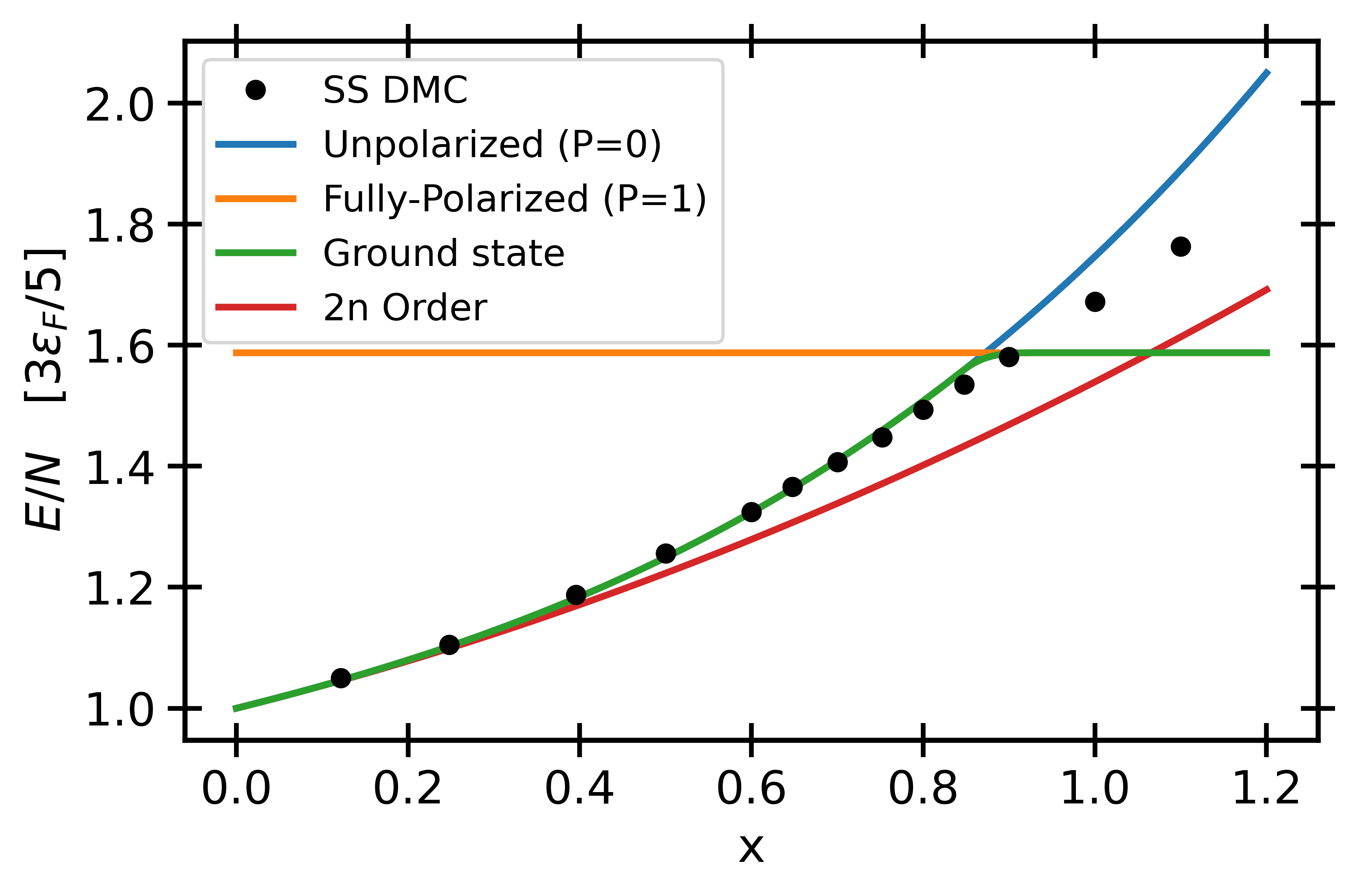}
    \caption{Energy for $S=1/2$ as a function of the gas parameter. We 
plot two models: in green, the third-order model; in red, the second-order one. 
The points are  DMC results from Ref. \cite{pilati}. The potential used 
is a soft-sphere model with  $r_0=0.424\,a_0$, and $a_1=1.1333\,a_0$.}
    \label{fig.SS}
\end{figure}

In Fig. \ref{fig.SS}, we analyze the dependence of the energy with the gas 
parameter for a soft-sphere potential. The points are DMC data from 
Ref.~\cite{pilati}. Here, the scattering parameters are $r_0=0.424\,a_0$ 
and $a_1=1.1333\,a_0$. We see that the third-order energies reproduce accurately 
the DMC data up to a value of $k_Fa_0 \simeq 0.8$ and, after 
that, Eq. \ref{eq.forfinal}  starts to depart from the DMC energies. 
The reason is that our perturbative expansion works for $k_Fa_0<1$, but also for 
$k_Fa_1<1$. As now $a_1$ is larger than $a_0$, the range of convergence has been 
reduced.

\begin{figure}
    \centering
    \includegraphics[width=0.48\textwidth]{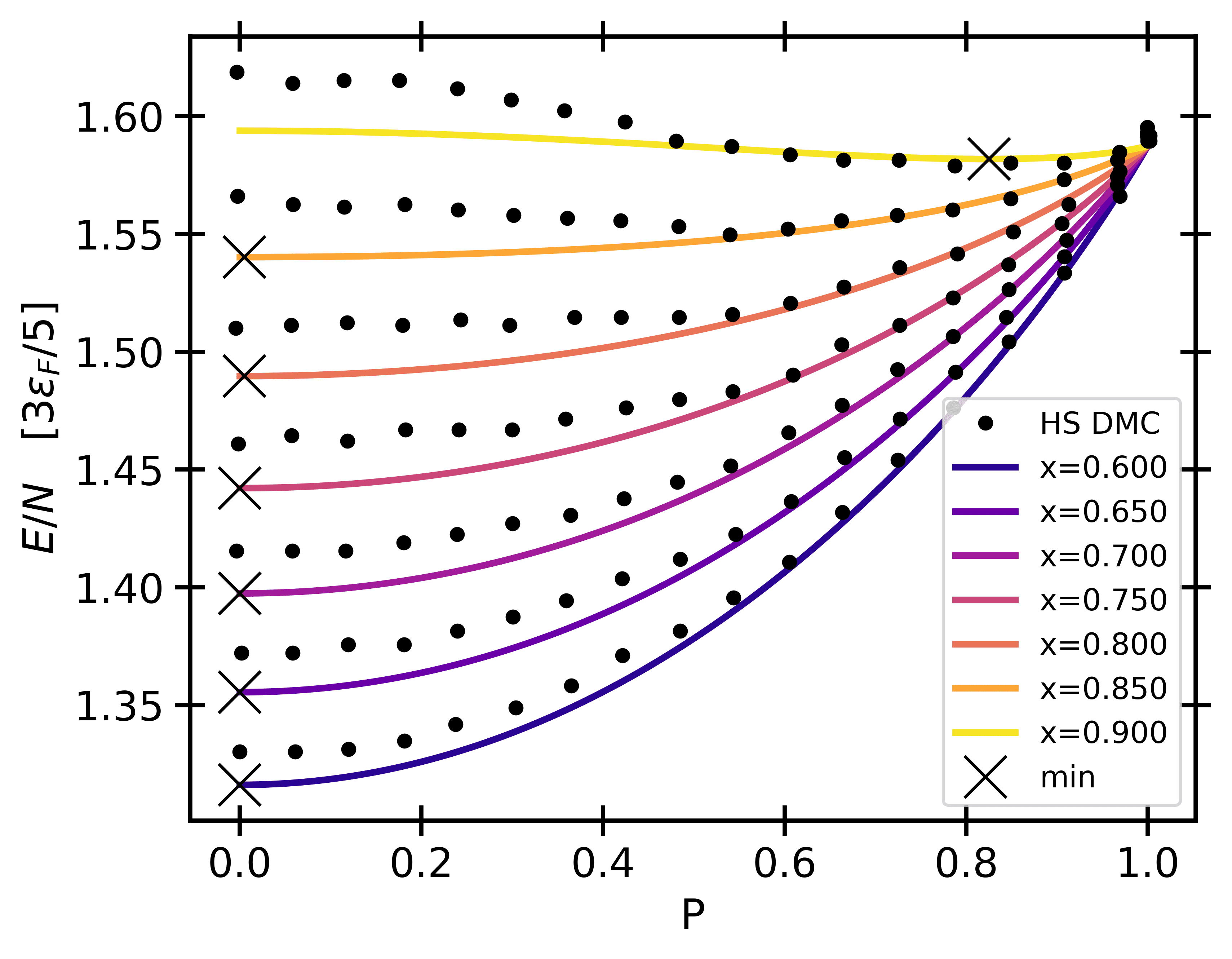}
    \caption{Energy per particle of a $S=1/2$ Fermi gas as a function of the 
polarization and for different values of the gas parameter. The solid points 
stand for DMC results~\cite{pilati} and the crosses to our prediction for the 
polarization giving the minimum energy. The potential is a hard-sphere one.}
    \label{fig:HSpol}
\end{figure}

Equation \ref{eq.forfinal} gives the energy as a function of the gas 
parameter but importantly also as a function of the polarization. 
In Fig. \ref{fig:HSpol}, we show our results for different 
gas parameters. One can see that, until a gas parameter of $k_Fa_0=0.85$, the 
polarization that minimizes the energy (shown as crosses in the figure) is 
zero. However, at $k_Fa_0=0.9$, the 
minimum of the energy moves to a larger polarization. As it is clear in the 
figure, the DMC 
points~\cite{pilati}  fit progressively better to our results when the 
polarization increases. A possible explanation for this behavior could be the 
different quality of the model nodal surface used in DMC calculations: it is 
well known that the plane-waves Slater determinant used in those DMC 
calculations is better for a polarized than for an unpolarized Fermi gas~\cite{backflow,slater_det}.

\begin{figure}
    \centering
    \includegraphics[width=0.48\textwidth]{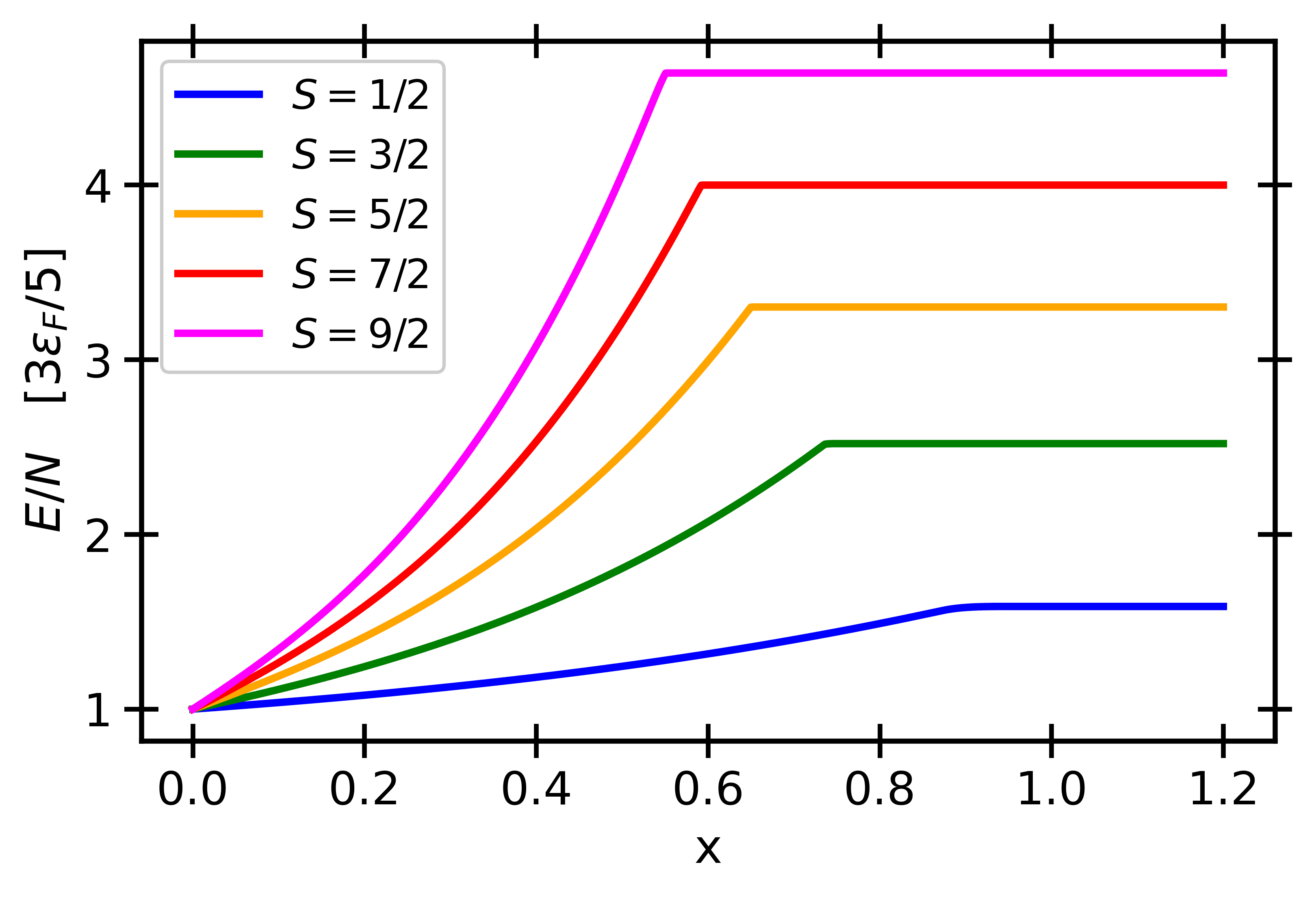}
    \caption{Energy per particle as a function of the gas parameter for spins 
$S=1/2, 3/2, 5/2, 7/2$, and $9/2$.}
    \label{fig:ed}
\end{figure}

In Fig. (\ref{fig:ed}), we show the energy as a function of the gas parameter 
for five spin values $S=1/2, 3/2, 5/2, 7/2$, and 
$9/2$. It is worth noticing that for $S>1/2$ there are not available DMC to 
compare with. The results reported in the figure correspond to a 
hard-sphere interaction. As one can see,  the third-order model predicts a 
ferromagnetic phase transition for the five spin values, since the curve 
becomes flat after a certain characteristic $x$. As observed also in second 
order ~\cite{Pera}, the ferromagnetic transition occurs 
at lower values of the gas parameter when $S$ increases. For a same spin value, 
the critical density decreases when third-order terms are introduced in the 
expansion.

\begin{figure}[h]
    \centering
    \includegraphics[width=0.48\textwidth]{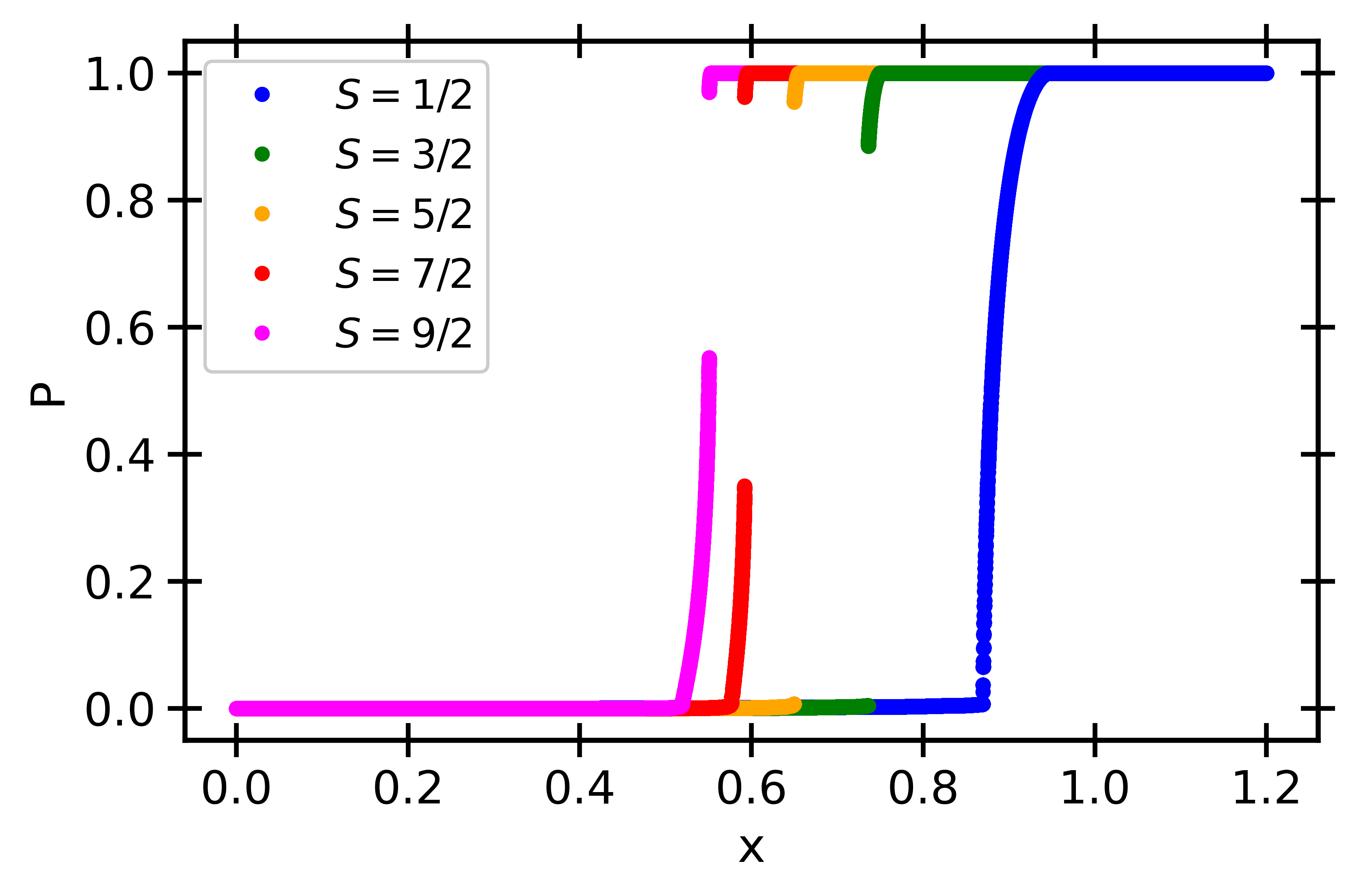}
    \caption{Polarization as a function of the gas 
parameter for spins $S=1/2, 3/2, 5/2, 7/2$, and $9/2$.}
    \label{fig:pd}
\end{figure}

In order to gain a deeper understanding of the phase
transition, we plot in Fig. \ref{fig:pd} the order parameter (in
this instance, the polarization) as a function of the gas
parameter $x$. The polarization we plot is the one that minimizes the energy at a given $x$. Indeed, as previously commented, all gases evolve
from a non-polarized to a fully-polarized state. However,
there are notable distinctions among them.
For spin $1/2$, the transition is quasi-continuous, this fact 
contrasts 
with the second-order  result reported in Ref.~\cite{Pera} where the 
transition for spin $1/2$ was first order and had a polarization jump of 
0.545. By quasi-continuous, we mean that the transition could be discontinuous, but 
with a tiny jump. Apparently our results show a continuous transition, but at 
third order we no longer have a fully analytical expression and, hence, 
our prediction has the limits of our numerical accuracy. For spin $3/2$ and $5/2$, we have partial discontinuous transitions, as 
there is a polarization jump, but it does not reach $P=1$, hence, the label 'partial'. If it reached $P=1$ directly, it would be a total discontinuous transition. Astonishingly, the 
gases with the last two largest spins, $S=7/2$ and $9/2$, experience a 
continuous transition when increasing the density, but the continuous transition 
is truncated because a discontinuous transition occurs before $P$ reaches 1.

\begin{figure}[h]
    \centering
    \includegraphics[width=0.48\textwidth]{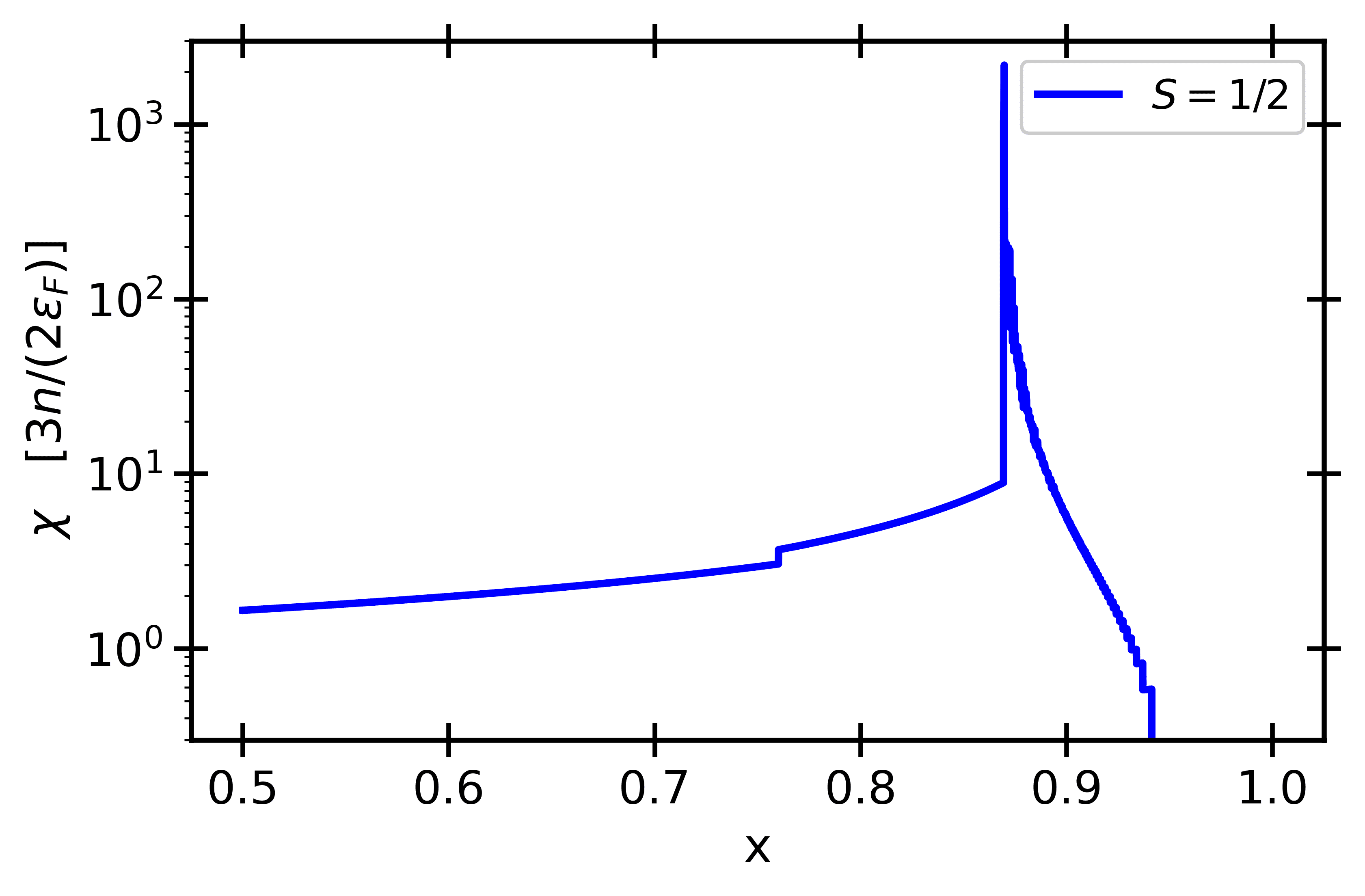}
    \caption{Magnetic susceptibility as a function of the 
gas parameter for spin $S=1/2$. The y-scale is semi-log.}
    \label{fig:my_cdv2}
\end{figure}

\begin{figure}[h]
    \centering
    \includegraphics[width=0.48\textwidth]{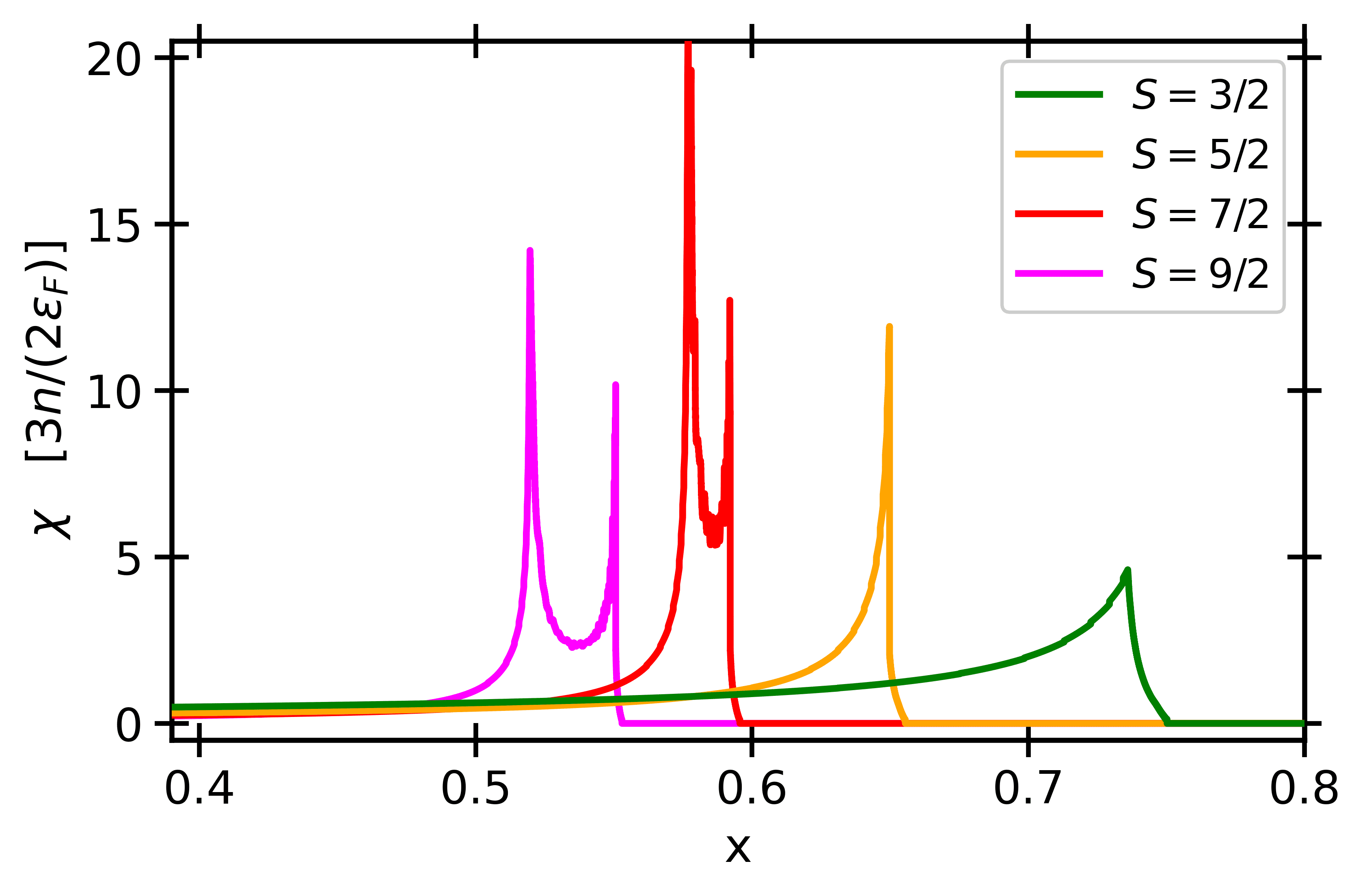}
    \caption{Magnetic susceptibility as a function of the 
gas parameter for spins $S=3/2, 5/2, 7/2$, and $9/2$.}
    \label{fig:my_cdv6-10}
\end{figure}

Fundamental information on the magnetic properties of the Fermi gases 
is contained in  the magnetic susceptibility $\chi$, 
\begin{equation}
   \frac{1}{\chi}=\frac{1}{n}\left(\frac{\partial^2 (E/N)}{\partial 
P^2} \right)_x \ .
\label{suscept}
\end{equation}
Third-order results of $\chi$ for  spin $1/2$ are shown 
in Fig. \ref{fig:my_cdv2}, and the ones for larger spins in Fig. 
\ref{fig:my_cdv6-10}. 
We split the results in two figures because, as spin 1/2 
suffers a quasi-continuous transition, the magnetic susceptibility diverges. We recall that, for second-order phase transitions, the susceptibility must diverge.
Notice, however, 
that we obtain a very large value~($\sim2e3$) at $k_Fa_0=0,85$ and not a real divergence 
due to our finite numerical precision. If the accuracy is improved, the critical value increases. The behavior of the magnetic 
susceptibility for spin $1/2$ confirms the behavior of the 
polarization shown in Fig.~\ref{fig:pd}. 

The other $\chi$ results for larger spins (Fig.~\ref{fig:my_cdv6-10})  
exhibit the behavior we have predicted above. For spins 3/2 and 5/2, we only 
have a finite peak, telling us that there is a discontinuous transition. For 
spins 7/2 and 9/2, we see the singular double transition (two peaks) that  we 
have mentioned before. With increasing density, the first peak corresponds to 
the truncated continuous transition, and the second peak to the latter 
first-order phase transition. The two peaks point to the existence of an intermediate phase between the non-polarized phase and the fully-polarized one. This phase would be located around the bottom that lies between peaks in Fig.~\ref{fig:my_cdv6-10}. And, according to Fig.~\ref{fig:pd}, this intermediate phase would have a partial polarization, hence, the Fermi gas would exhibit some kind of magnetic ordering. The rich phase diagram that appears in SU(N) Fermi systems have also been pointed out in Ref.~\cite{intermed_phase}.

\begin{figure}[h]
    \centering
    \includegraphics[width=0.48\textwidth]{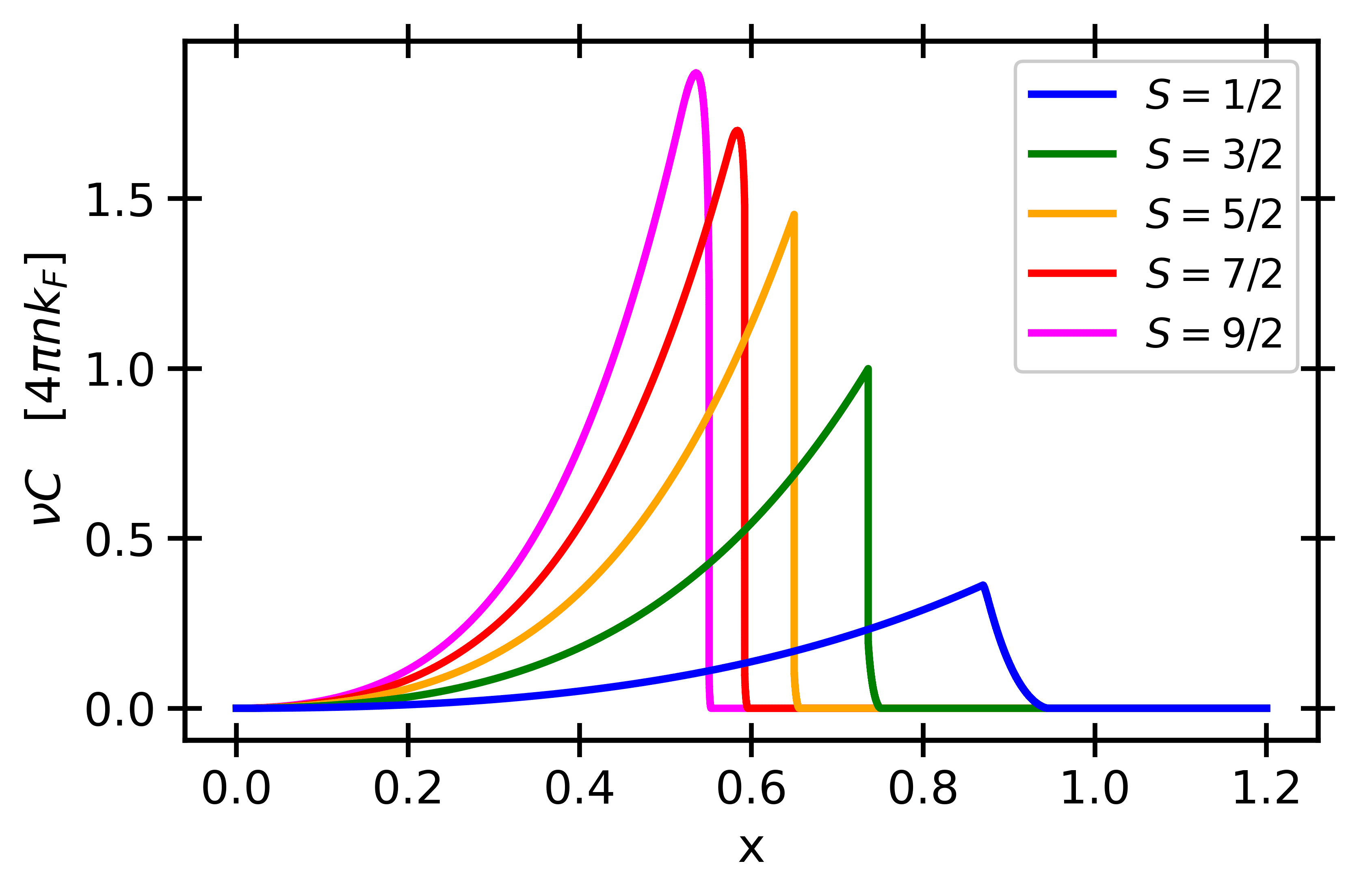}
    \caption{Tan's constant as a function of the gas 
parameter for spins $S=1/2, 3/2, 5/2, 7/2$, and $9/2$).}
    \label{fig:my_td}
\end{figure}

The Tan's constant,
\begin{equation}
   C=\frac{8\pi ma_0^2}{\nu\hbar^2}\frac{N}{V}\frac{\partial 
(E/N)}{\partial a_0} \ ,
\end{equation}
is a very good tool to locate  the density at which the itinerant 
ferromagnetism transition occurs. It is 
so, because right before 
the transition, the Tan's constant value is maximum (see Fig. \ref{fig:my_td}). 
In terms of the gas parameter, it is
\begin{equation}
C=\frac{4\pi nk_F}{\nu}\frac{x^2}{\epsilon_F}\frac{\partial (E/N)}{\partial x} \ .
\end{equation} 
The Tan's constant results for five values of the spin are shown in Fig. 
\ref{fig:my_td}.  One can see again that the transition happens at lower 
densities when the degeneracy increases.

\begin{figure}[h]
    \centering
    \includegraphics[width=0.48\textwidth]{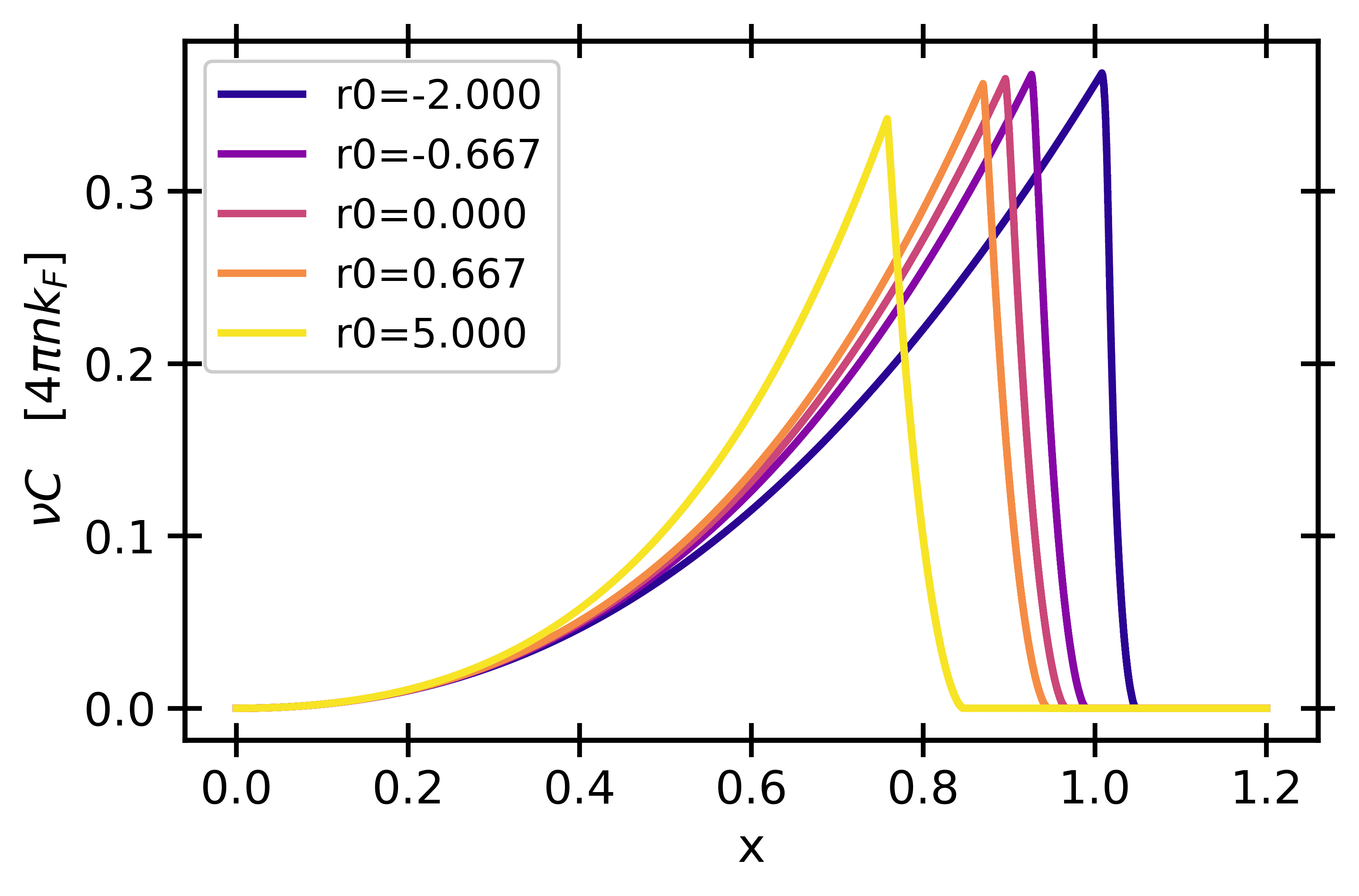}
    \caption{Tan's constant as a function of the gas 
parameter for spin 1/2. The $P$-wave scattering length is 
$a_1=a_0$ as for the hard-sphere interaction. The effective range varies from 
$-2\,a_0$ to $5\,a_0$.}
    \label{fig.tanr0}
\end{figure}

As we are now beyond universality, it is interesting to explore the role of the 
$s$-wave effective range and $p$-wave scattering length in the location of the 
critical density. To reduce the dimensionality of the problem, we have analyzed 
the particular case of $a_1=a_0$ and spin $1/2$. In Fig. \ref{fig.tanr0}, 
we plot the Tan's constant by changing only the effective range. 
As one can see, the critical density changes significantly with $r_0$. 
Interestingly, increasing $r_0$ the ferromagnetic transition moves to lower 
densities making it somehow more accessible in experiments. 

\begin{figure}
    \centering
    \includegraphics[width=0.48\textwidth]{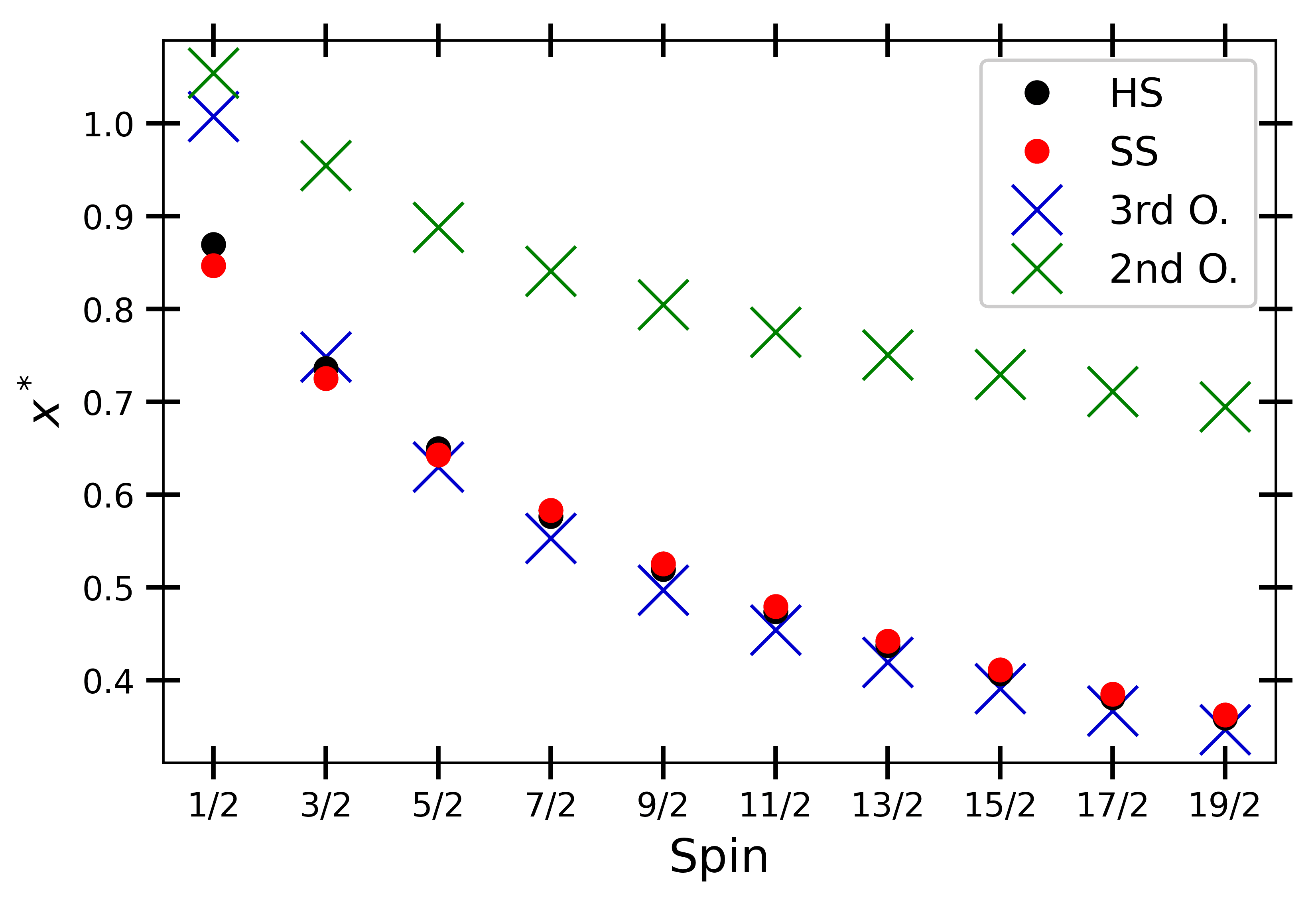}
    \caption{Critical gas parameter as a function of the 
spin value.  The black and red points stand for  the hard-sphere ($r_0=2a_0/3$, $a_1=a_0$) and 
soft-sphere potentials ($r_0=0.424\,a_0$, $a_1=1.1333\,a_0$), respectively. The green crosses are 
second-order results. The blue crosses are third-order results including only 
the terms that depend on $a_0$.}
    \label{fig.tall}
\end{figure}

As commented before, the ferromagnetic phase transition 
happens at lower densities if we increase the spin of the gas. In Fig. 
\ref{fig.tall}, we show the critical gas parameter as a function of 
the spin value, up to spin $19/2$. The inclusion of third-order terms reduces 
the critical $x^*$, with an effect that increases with the spin due to the 
enhanced role of interactions between different spin channels. We report the 
critical gas parameters obtained using both the hard-sphere and soft-sphere 
potentials. The soft-sphere potential is the same we used in Fig. \ref{fig.SS}, this means $r_0=0.424\,a_0$ and $a_1=1.1333\,a_0$. The difference between these two potentials is almost imperceptible 
except somehow for $S=1/2$ where the critical value for soft-spheres is 
slightly below the hard-sphere one. It is interesting to explore what is the 
relevance of the effective range and $p$-wave contributions with respect to the 
one coming from the third-order terms that depend only on $a_0$ on the critical 
value. To this end, we show in Fig.~\ref{fig.tall} third-order results 
depending only on $a_0$ (blue crosses). As one can see, the role of $r_0$ and $a_1$ is small 
for $S>1/2$ but substantial for the particular case $S=1/2$.
Neglecting the effect of scattering parameters others than $a_0$, we estimate 
that the transition gas parameter is 0.65 for Ytterbium ($S=5/2$), and 0.53 for 
Strontium ($S=9/2$).

\section{Conclusions}

To summarize, we have derived the expression for the energy
of a repulsive SU(N) Fermi gas, incorporating terms up to
third order in the gas parameter and in terms of the
spin-channel occupations. We have extended up to third order
our second-order expansion~\cite{Pera}, written also in terms of the
spin polarization. We have included the analytic terms 
dependent on the $s$-wave effective range and the $p$-wave
scattering length, while the remaining third-order terms have
been computed numerically. Although these numerical terms
introduce some level of uncertainty, we have managed to
minimize it to the best of our abilities. The uncertainty 
in the values of $P$ are typically 0.005, and for the energy are about 
$10^{-6}$.

In order to study the Fermi gas, and to reduce the number
of variables to explore, we have selected the occupational
configuration that minimizes the energy. At $P = 0$, all species
are equally occupied, and as $P$ increases, one species
increases while all others decrease in the same manner.
However, we point out that our formalism 
can be applied to any occupational configuration. One just needs to find the 
new expressions for the fractions of $\lambda$ particles $C_{\lambda}$. In fact, 
in several experiments, the species' occupation can be tuned or controlled 
almost at will~\cite{collective_excitations,estronci_N_10} and there are works 
that have dealt with imbalanced systems~\cite{prethermalization}. 
However, we
point out that all these other configurations represent
excited states with higher energy compared to the
configuration we have chosen. Consequently, if a Fermi system
is capable of thermalization, it will converge to the
behavior predicted in our work. Notice that adopting a
different occupational configuration would yield distinct
critical values for the gas parameter  $x^*$.

The third-order expression allows to reproduce and test numerical values 
obtained with diffusion Monte Carlo. 
In the case of itinerant ferromagnetism, where three
scattering parameters come into play, the existence of the
phase transition could even not exist. This is not the case
for the interactions analyzed in the present work:
for a hard-sphere potential and using the third-order energies, 
the  spin $1/2$ Fermi gas exhibits a quasi-continuous magnetic phase transition, 
in contrast with 
the first-order phase  transition predicted by the second-order approximation 
(universal expansion)~\cite{Pera}. According to Ref.~\cite{patrick}, the order 
of the $S=1/2$ ferromagnetic transition is always first-order because 
the sign of the term $P^4 \ln P$ in the Landau expansion must be positive. 
However, and within our numerical precision, this term is in fact negative at 
third-order and thus, the transition becomes quasi-continuous. The same change of sign is observed in \cite{canvi-signe} where the authors apply a resummation method. 
For higher 
spins values, we have different situations. Spins $3/2$ and $5/2$ exhibit a 
first-order  transition. However, gases with larger spins show a double 
transition. First, a continuous transition happens, but, before reaching $P=1$, 
a discontinuous transition truncates it.

Importantly, the inclusion of the third-order terms
significantly reduces the critical value of the gas parameter
at which the ferromagnetic transition is expected. For spin
$5/2$, it occurs at $k_Fa_0=0.65$, and for spin $9/2$, at $k_Fa_0=0.53$. 
This finding reinforces the notion that the
observation of itinerant ferromagnetism may be more favorable
when working with highly degenerate gases like Yb~\cite{pagano}  and 
Sr~\cite{goban}. Beyond the second-order approximation, the energy
ceases to be universal in terms of the gas parameters because
the $s$-wave effective range and $p$-wave scattering length come
into play \cite{Bishop}. However, for larger spins, the region of
interest (where the transition occurs) tends to be at lower
densities, as mentioned earlier, leading these systems toward
a new form of universality.

\begin{acknowledgments}
Stimulating discussions with Miguel Cazalilla and Chen-How Huang  are 
aknowledged. 
We acknowledge financial support from   MCIN/AEI/10.13039/501100011033 
(Spain) Grant No.  PID2020-113565GB-C21 and from AGAUR- Generalitat de 
Catalunya Grant N0. 2021-SGR-01411. 
\end{acknowledgments}

\appendix
\clearpage
\begin{widetext}
\section{Numerical integration of third-order terms}
\label{appendix:a}

We have three terms contributing at third order of the perturbative series 
which are numerically integrated. The first two terms involve interaction 
between two particles, however, the third one is a three-body interaction term.
In the following subsections, each one of these terms is analyzed. 

\subsection{Term E3}
The expression of the $E_3$ term is
\begin{equation}
\begin{aligned}
\frac{E}{N}=\epsilon_F\frac{3x^3}{32\pi^7}\frac{1}{\nu}\sum_{\lambda_1,\lambda_2
}(1-\delta_{\lambda_1,\lambda_2})E_3(C_{\lambda_1},C_{\lambda_2})\ ,\
\end{aligned}
\end{equation}
with
\begin{equation}
\begin{aligned}
E_3(C_{\lambda_1},C_{\lambda_2})=\frac{1}{k_F^8}\int 
d\textbf{p}n_p\int d\textbf{p}'n_{p'} \bigg[\int 2 \, 
d\textbf{q}d\textbf{q}'\frac{1-(1-n_q)(1-n_{q'})}{q^2+q'^2-p^2-p'^2}
\delta(\textbf{q}+\textbf{q}'-\textbf{p}-\textbf{p}')\bigg]^2\ .
\end{aligned}
\end{equation}

We recall that $\textbf{p}$ and $\textbf{q}$ run over $k_{F,\lambda_1}$, while $\textbf{p}'$ and $\textbf{q}'$ run over $k_{F,\lambda_2}$.
If we expand the numerator, where the occupation functions are, we see that we 
can 
split the terms: two containing $n_q$ and the other one with $n_q n_{q'}$. This 
is telling us the volume where we have to integrate. For the first two 
terms, $n_q$ gives an sphere, and, for the third one, $n_q n_{q'}$ gives a more 
complicated volume, as it is the intersection of two spheres. Therefore, we have 
to solve two integrals, the one having the volume of a sphere (Sphere-Integral 
-SI-), and the other one with the intersection (Sphere-Intersection-Integral 
-SII-). Then,
\begin{equation}
   E_3(C_{\lambda_1},C_{\lambda_2}) = \frac{1}{k_F^8}\int d\textbf{p}n_p\int 
d\textbf{p}'n_{p'}\bigg[SI(p,p',k_{F,\lambda_1})+SI(p,p',k_{F,\lambda_2})-SII(p,p',k_{F
,\lambda_1},k_{F,\lambda_2})\bigg]^2
\end{equation}
The SI terms can be integrated in the following way,
\begin{equation*}
    SI(p,p',k_{F,\lambda})=\int2d\textbf{q}d\textbf{q}'\frac{n_q}{q^2+q'^2-p^2-p'^2}\delta(\textbf{q}+\textbf{q}'-\textbf{p}-\textbf{p}')=\int d\textbf{q}\frac{n_q}{q^2-\textbf{q}\cdot(\textbf{p}+\textbf{p}')+\textbf{p}\cdot\textbf{p}'}
\end{equation*}
\begin{equation*}
    =\int_0^{k_{F,\lambda}} q^2dq\int_0^{\pi}\sin{\theta}d\theta\frac{2\pi}{q^2-q\vert\textbf{p}+\textbf{p}'\vert+\textbf{p}\cdot\textbf{p}'}=2\pi\int_0^{k_{F,\lambda}} q^2dq\frac{1}{q\vert\textbf{p}+\textbf{p}'\vert}\ln{\bigg[\frac{q^2+q\vert\textbf{p}+\textbf{p}'\vert+\textbf{p}\cdot\textbf{p}'}{q^2-q\vert\textbf{p}+\textbf{p}'\vert+\textbf{p}\cdot\textbf{p}'}\bigg]}
\end{equation*}
\begin{equation*}
    =\frac{2\pi}{\vert\textbf{p}+\textbf{p}'\vert}\bigg\{\bigg(\frac{k_{F,\lambda}^2}{2}-\frac{p^2+p'^2}{4}\bigg)\ln{\bigg[\frac{k_{F,\lambda}^2+k_{F,\lambda}\vert\textbf{p}+\textbf{p}'\vert+\textbf{p}\cdot\textbf{p}'}{k_{F,\lambda}^2-k_{F,\lambda}\vert\textbf{p}+\textbf{p}'\vert+\textbf{p}\cdot\textbf{p}'}\bigg]}
\end{equation*}
\begin{equation}
    -\frac{\vert\textbf{p}+\textbf{p}'\vert\vert\textbf{p}-\textbf{p}'\vert}{4}\ln{\bigg[\frac{k_{F,\lambda}^2+k_{F,\lambda}\vert\textbf{p}-\textbf{p}'\vert-\textbf{p}\cdot\textbf{p}'}{k_{F,\lambda}^2-k_{F,\lambda}\vert\textbf{p}-\textbf{p'}\vert-\textbf{p}\cdot\textbf{p}'}\bigg]}+k_{F,\lambda}\vert\textbf{p}+\textbf{p'}\vert\bigg\}
\end{equation}
Now, we perform the following change of variable that will be useful later on,
\begin{equation}
    \begin{matrix}
        \textbf P=\textbf p+\textbf p'\\
        \\
        \textbf Q=\cfrac{\textbf p-\textbf p'}{2}
    \end{matrix}
    \quad\Rightarrow\quad
    \begin{matrix}
        \textbf p=\cfrac{\textbf P}{2}+\textbf Q\\
        \\
        \textbf p'=\cfrac{\textbf P}{2}-\textbf Q
    \end{matrix}
\end{equation}
With these new variables the sphere integral can be rewritten as follows,
\begin{equation}
\boxed{
\begin{aligned}
    SI(P,Q,k_{F,\lambda})=\frac{2\pi}{P}\bigg\{\bigg(\frac{k_{F,\lambda}^2}{2}-\frac{1}{2}\bigg(\frac{P^2}{4}+Q^2\bigg)\bigg)\ln{\bigg\vert\frac{\big(k_{F,\lambda}+\frac{P}{2}\big)^2-Q^2}{\big(k_{F,\lambda}-\frac{P}{2}\big)^2-Q^2}\bigg\vert}
     -\frac{P Q}{2}\ln{\bigg\vert\frac{\big(k_{F,\lambda}+Q\big)^2-\frac{P^2}{4}}{\big(k_{F,\lambda}-Q\big)^2-\frac{P^2}{4}}\bigg\vert}+k_{F,\lambda} P\bigg\}
\end{aligned}
    }
\end{equation}

When $P$ is zero, $SI$ could diverge, 
however, taking the limit of $P$ going to zero, one can see that the result is 
finite. This function when  $P \to 0$  is
\begin{equation}
    SI(0,Q,k_{F,\lambda})=2\pi\bigg[2k_{F,\lambda}-Q\ln{\bigg\vert\frac{k_{F,\lambda}+Q}{k_{F,\lambda}-Q}\bigg\vert}\bigg]
\end{equation}

Now, we analyze the $SII$ term. 
We first apply the change of variables introduced above, and then we integrate 
over $P$. Then, 
\begin{equation*}
    SII(p,p',k_{F,\lambda_1},k_{F,\lambda_2})=\int2d\textbf{q}d\textbf{q}'\frac{n_q n_{q'}}{q^2+q'^2-p^2-p'^2}\delta(\textbf{q}+\textbf{q}'-\textbf{p}-\textbf{p}')
\end{equation*}
\begin{equation}
    =\int d\textbf{Q}d\textbf{P}\frac{n_{P/2+Q} n_{P/2-Q}}{Q^2+P^2/4-2(p^2+p'^2)/4}\delta(\textbf{P}-\textbf{p}-\textbf{p}')=\int d\textbf{Q}\cfrac{n_{P/2+Q} n_{P/2-Q}}{Q^2-\frac{(p-p')^2}{4}}
\end{equation}
The term $n_{P/2+Q} n_{P/2-Q}$ is telling us the volume that we have to 
integrate. This comes from the intersection of spheres
\begin{align}
    I: \quad \bigg(Q_x+\frac{P_x}{2}\bigg)^2+\bigg(Q_y+\frac{P_y}{2}\bigg)^2+\bigg(Q_z+\frac{P_z}{2}\bigg)^2\leq k_{F,\lambda_1}^2
    \label{sph1}\\
    II: \quad \bigg(Q_x-\frac{P_x}{2}\bigg)^2+\bigg(Q_y-\frac{P_y}{2}\bigg)^2+\bigg(Q_z-\frac{P_z}{2}\bigg)^2\leq k_{F,\lambda_2}^2
    \label{sph2}
\end{align}
As we see, we have two spheres with different radii. The first one is centered 
at $\big(\frac{-P_x}{2},\frac{-P_y}{2},\frac{-P_z}{2}\big)$ with radius 
$k_{F,\lambda_1}$; and the second one is centered at 
$\big(\frac{P_x}{2},\frac{P_y}{2},\frac{P_z}{2}\big)$ with radius 
$k_{F,\lambda_2}$. And  $\textbf{Q}$ must satisfy both constraints, that is, we 
have to integrate the intersection of the two spheres. Depending on the 
distance between the spheres we will have different kind of intersections. The 
distance between the two centers is just $P$. If $P$ is larger than the sum of 
both radius, then there is no intersection and the integral will be zero. If 
$P$ is lower than the difference of radii, then the small sphere will be 
completely inside the big sphere, therefore the intersection will be the small 
sphere and the integral will be exactly $SI$ because the volume will 
be just an sphere. And finally, when $P$ is smaller than the sum of both radii 
and larger than their difference, the intersection corresponds to the sum of 
two spherical caps, one coming from each sphere. This explanation is summarized 
in Table \ref{tau1} and plotted  in Fig. \ref{fig1}.

\begin{table}[h]
\centering
\begin{tabular}{ |c|c| } 
 \hline
 \textbf{Condition} & \textbf{Intersection} \\ 
 \hline
 \hline
 $0\leq P<\vert k_{F,\lambda_1}-k_{F,\lambda_2}\vert$ & The small sphere \\ 
 \hline
 $P=\vert k_{F,\lambda_1}-k_{F,\lambda_2}\vert$ & The small sphere (Inner tangency) \\ 
 \hline
 $\vert k_{F,\lambda_1}-k_{F,\lambda_2}\vert< P< (k_{F,\lambda_1}+k_{F,\lambda_2})$ & Two spherical caps \\ 
 \hline
 $P= (k_{F,\lambda_1}+k_{F,\lambda_2})$ & A point (Outer tangency) \\ 
 \hline
 $P>(k_{F,\lambda_1}+k_{F,\lambda_2})$ & Null intersection \\ 
 \hline
\end{tabular}
\caption{Different types of intersection between two spheres 
depending on the distance between centers.}
\label{tau1}
\end{table}

Therefore, one only needs to analyze the case of 
spherical caps, $\vert k_{F,\lambda_1}-k_{F,\lambda_2}\vert< P< 
(k_{F,\lambda_1}+k_{F,\lambda_2})$.
As the coordinates of $P$ can take a large set of values and 
thus we can have many distributions of the spheres in the space, we will 
perform a 3D-rotation in order to set our system in a vertical position 
regardless of the initial configuration. More specifically, the sphere with 
radius $k_{F,\lambda_2}$ will be at the top, and the one with radius 
$k_{F,\lambda_1}$  at the bottom.

\begin{figure}[H]
    \centering
    \includegraphics[width=0.945\textwidth]{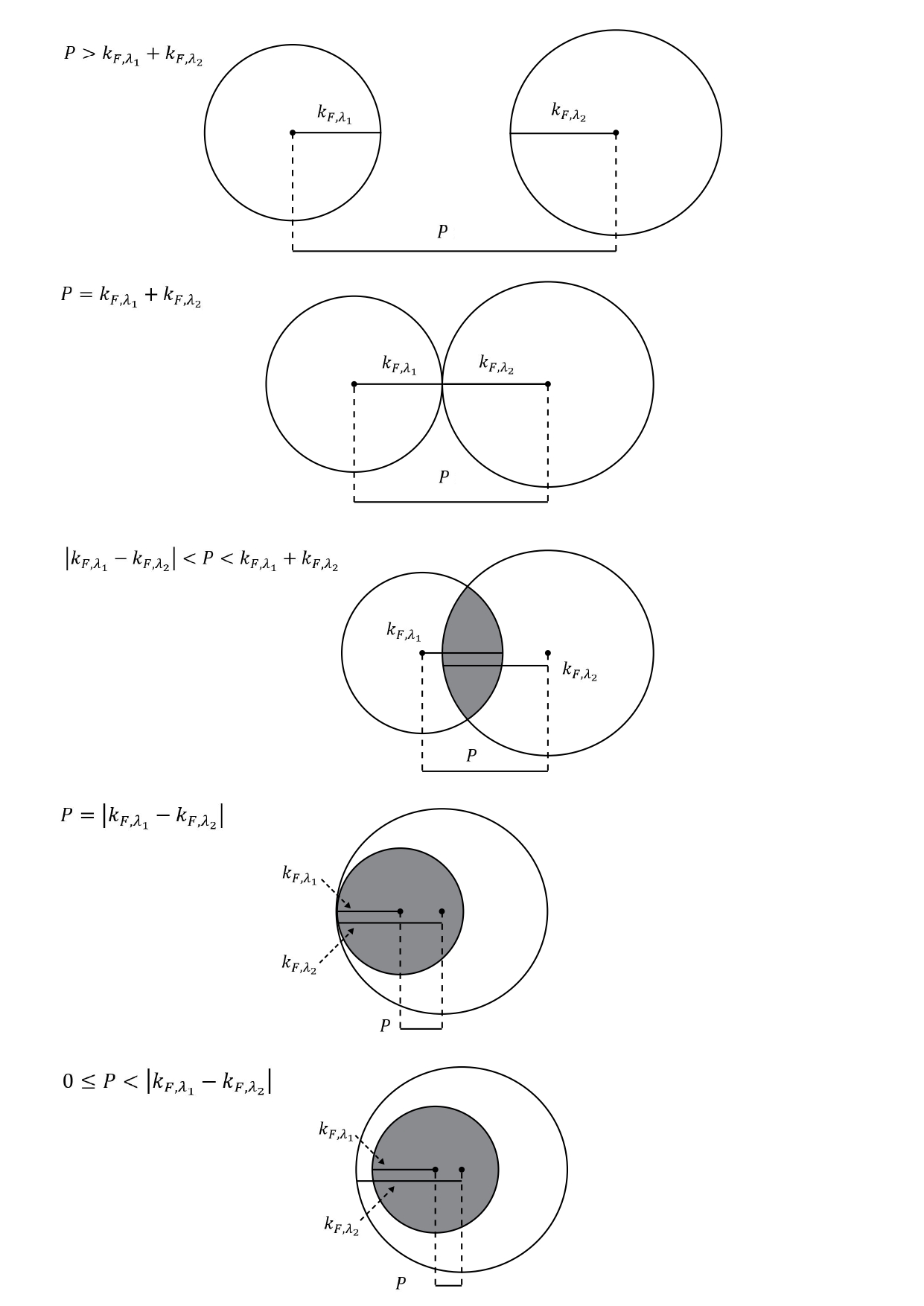}
    \caption{Sphere-sphere 
intersections. The grey area is the intersection we are looking for. $P$ is the 
distance between centers. $k_{F,\lambda_1}$ and $k_{F,\lambda_2}$ are the 
respective radii.}
    \label{fig1}
\end{figure}

We want the following transformations for the
rotation: $\big(\frac{-P_x}{2},\frac{-P_y}{2},\frac{-P_z}{2}\big) \Rightarrow 
\big(0,0,\frac{-P}{2}\big)$ and 
$\big(\frac{P_x}{2},\frac{P_y}{2},\frac{P_z}{2}\big)\Rightarrow\big(0,0,\frac{P}
{2}\big)$, but in fact they are the same transformation because the minus sign 
makes no difference, and also the one half multiplying can be taken out. 
Under this rotation,  the change of variables is
\begin{gather}
    Q_x=\cfrac{P_x}{\sqrt{P_x^2+P_y^2}}\cfrac{P_z}{P}Q_{\alpha x}-\cfrac{P_y}{\sqrt{P_x^2+P_y^2}}Q_{\alpha y}+\cfrac{P_x}{P}Q_{\alpha z}\\
    Q_y=\cfrac{P_y}{\sqrt{P_x^2+P_y^2}}\cfrac{P_z}{P}Q_{\alpha x}+\cfrac{P_x}{\sqrt{P_x^2+P_y^2}}Q_{\alpha y}+\cfrac{P_y}{P}Q_{\alpha z}\\
    Q_z=-\cfrac{\sqrt{P_x^2+P_y^2}}{P}Q_{\alpha x}+\cfrac{P_z}{P}Q_{\alpha z}
\end{gather}
Where $Q_{x,y,z}$ are the coordinates in the initial configuration, and 
$Q_{\alpha x,\alpha y,\alpha z}$ are the new coordinates after the rotation. 

Equations (\ref{sph1}) and (\ref{sph2}), which were the ones defining the two 
spheres, become in the new system of coordinates: (the sub-index $\alpha$ of 
$Q$ is omitted from now on)
\begin{align}
    I: \quad Q_x^2+Q_y^2+\bigg(Q_z+\frac{P}{2}\bigg)^2\leq k_{F,\lambda_1}^2\\
    II: \quad Q_x^2+Q_y^2+\bigg(Q_z-\frac{P}{2}\bigg)^2\leq k_{F,\lambda_2}^2
\end{align}
Changing to cylindrical coordinates,
\begin{align}
    I: \quad r^2+\bigg(z+\frac{P}{2}\bigg)^2\leq k_{F,\lambda_1}^2\quad\Rightarrow\quad
    r_{max}[z]=\sqrt{k_{F,\lambda_1}^2-\bigg(z+\frac{P}{2}\bigg)^2}\\
    II: \quad r^2+\bigg(z-\frac{P}{2}\bigg)^2\leq k_{F,\lambda_2}^2\quad\Rightarrow\quad
    r_{max}[z]=\sqrt{k_{F,\lambda_2}^2-\bigg(z-\frac{P}{2}\bigg)^2}
\end{align}
From the expressions above we can find the value of $z$ at which both spheres 
coincide, that is the value of $z$ that will separate the two spherical caps we 
need to integrate, 
\begin{equation}
    z_{\text{lim}}=\frac{k_{F,\lambda_1}^2-k_{F,\lambda_2}^2}{2P}
\end{equation}
We point out that $P$ can only be zero if both $k_{F,\lambda}$'s are equal, and 
in that case $z_{\text{lim}}$ would exactly be 0.

After all these manipulations, now we are ready to write the final expression of
the integral and the two regions of integration  ( 
Spherical Caps Integral -SCI-)
\begin{equation}
    SCI=\int r dr d\theta dz\frac{n_I n_{II}}{r^2+z^2-\frac{(p-p')^2}{4}}
\end{equation}
\begin{equation}
\begin{matrix}
I: & r\in\bigg[0,\sqrt{k_{F,\lambda_1}^2-\bigg(z+\cfrac{P}{2}\bigg)^2}\bigg] &  &  &  & II: & r\in\bigg[\sqrt{k_{F,\lambda_2}^2-\bigg(z-\cfrac{P}{2}\bigg)^2}\bigg]
\\
\\
 & \theta\in[0,2\pi] &  &  &  &  & \theta\in[0,2\pi]
\\
\\
 & z\in\bigg[\cfrac{k_{F,\lambda_1}^2-k_{F,\lambda_2}^2}{2P},k_{F,\lambda_1}-\cfrac{P}{2}\bigg] &  &  &  &  & z\in\bigg[\cfrac{P}{2}-k_{F,\lambda_2},\cfrac{k_{F,\lambda_1}^2-k_{F,\lambda_2}^2}{2P}\bigg]
\end{matrix}
\end{equation}
The final step is to integrate and sum. After some algebra, the final 
integrated expression is
\begin{equation}
\boxed{
\begin{aligned}
    SCI(P,Q,k_{F,\lambda_1},k_{F,\lambda_2})=2\pi\frac{1}{2}\bigg\{k_{F,\lambda_1}+k_{F,\lambda_2}-P+Q\ln{\bigg\vert\frac{Q-k_{F,\lambda_1}+P/2}{Q+k_{F,\lambda_1}-P/2}\bigg\vert}+Q\ln{\bigg\vert\frac{Q-k_{F,\lambda_2}+P/2}{Q+k_{F,\lambda_2}-P/2}\bigg\vert}\\
    \\
    +\frac{\big(Q^2+P^2/4-k_{F,\lambda_1}^2\big)}{P}\ln{\Bigg\vert\frac{\big(k_{F,\lambda_1}-P/2\big)^2-Q^2}{\cfrac{k_{F,\lambda_1}^2+k_{F,\lambda_2}^2}{2}-\cfrac{P^2}{4}-Q^2}\Bigg\vert}
    +\frac{\big(Q^2+P^2/4-k_{F,\lambda_2}^2\big)}{P}\ln{\Bigg\vert\frac{\big(k_{F,\lambda_2}-P/2\big)^2-Q^2}{\cfrac{k_{F,\lambda_1}^2+k_{F,\lambda_2}^2}{2}-\cfrac{P^2}{4}-Q^2}\Bigg\vert}
    \bigg\}
\end{aligned}
}
\label{eq.sci}
\end{equation}

Following the same reasoning we did in SI, at $P=0$ the 
expression we have obtained can diverge. If $k_{F,\lambda_1}$ is equal to 
$k_{F,\lambda_2}$, then $P$ can be 0. In that limit, SCI becomes
\begin{equation}
    SCI(0,Q,k_{F,\lambda_1}=k_{F,\lambda},k_{F,\lambda_1}=k_{F,\lambda})=2\pi\bigg\{2k_{F,\lambda}+Q\ln{\bigg\vert\frac{Q-k_{F,\lambda}}{Q+k_{F,\lambda}}\bigg\vert}\bigg\}
\end{equation}
A summary of the different integration domains is shown in Table 
\ref{tau.sii} 

\begin{table}[H]
\centering
\begin{tabular}{ |c|c| } 
 \hline
 \textbf{Condition} & \textbf{SII Integral} \\ 
 \hline
 \hline
 $0\leq P\leq\vert k_{F,\lambda_1}-k_{F,\lambda_2}\vert$ & $SI(P,Q,min(k_{F,\lambda_1},k_{F,\lambda_2}))$ \\ 
 \hline
 $\vert k_{F,\lambda_1}-k_{F,\lambda_2}\vert< P< (k_{F,\lambda_1}+k_{F,\lambda_2})$ & $SCI(P,Q,k_{F,\lambda_1},k_{F,\lambda_2})$ \\ 
 \hline
 $P\geq (k_{F,\lambda_1}+k_{F,\lambda_2})$ & 0 \\ 
 \hline
\end{tabular}
\caption{Summary of the SII integral depending on the value of P.}
\label{tau.sii}
\end{table}

We have been able to solve analytically SI and SII, however, 
now we should square it and integrate it again. This second part has not been 
analytically achieved and numerical integration has led to the solution. This 
is written as
\begin{equation}
   E_3(C_{\lambda_1},C_{\lambda_2})=\frac{1}{k_F^8}\int d\textbf{p}n_p\int d\textbf{p}'n_{p'}\bigg[SI(P,Q,k_{F,\lambda_1})+SI(P,Q,k_{F,\lambda_2})-SII(P,Q,k_{F,\lambda_1},k_{F,\lambda_2})\bigg]^2
\end{equation}
The final integrals can be simplified because the functions SI and SII only 
depend on the modules of $\textbf{p}$ and $\textbf{p}'$ and the angle $\theta$ 
between them. Moreover, if we perform the following change of variables 
$k_{F,\lambda_1}=k_FC_{\lambda}^{1/3}$, we can get rid of the Fermi momenta and 
express everything in terms of $C_{\lambda}$,
\begin{equation}
   E_3(C_{\lambda_1},C_{\lambda_2})=8\pi^2\int_0^{C_{\lambda_1}^{1/3}} 
p^2dp\int_0^{C_{\lambda_2}^{1/3}}p'^2dp'\int_{-1}^{1}dx 
\bigg[SI(P,Q,C_{\lambda_1}^{1/3})+SI(P,Q,C_{\lambda_2}^{1/3})-SII(P,Q,C_{
\lambda_1}^{1/3},C_{\lambda_2}^{1/3})\bigg]^2 \ ,
\end{equation}
with $x=-\cos{\theta}$, $P=\sqrt{p^2+p'^2-2pp'x}$, and 
$2Q=\sqrt{p^2+p'^2+2pp'x}$.

\subsection{Term E4}
 The E4 contribution to the energy is
\begin{equation}
\begin{aligned}
\frac{E}{N}=\epsilon_F\frac{3x^3}{32\pi^7}\frac{1}{\nu}\sum_{\lambda_1,\lambda_2
}(1-\delta_{\lambda_1,\lambda_2})E_4(C_{\lambda_1},C_{\lambda_2}) \ ,\
\end{aligned}
\end{equation}
with
\begin{equation}
\begin{aligned}
E_4(C_{\lambda_1},C_{\lambda_2})=\frac{1}{k_F^8}\int 
d\textbf{m}(1-n_m)\int d\textbf{m}'(1-n_{m'})\bigg[\int 2 \, 
d\textbf{p}d\textbf{p}'\frac{n_pn_{p'}\delta(\textbf{p}+\textbf{p}'-\textbf{m}
-\textbf{m}')}{p^2+p'^2-m^2-m^2}\bigg]^2 \ .
\end{aligned}
\end{equation}
The vectors $\textbf{m}$ and $\textbf{p}$ run over $k_{F,\lambda_1}$, and 
$\textbf{m}'$ and $\textbf{p}'$ run over $k_{F,\lambda_2}$. This term is similar 
to the previous term E3, but 
what is now different is the fact that the external integrals go to infinity 
and that the inner part is only the Sphere-Intersection-Integral (SII). Hence, 
we can rewrite it in terms of this known function  
(See Table \ref{tau.sii}),
\begin{equation}
    E_4(C_{\lambda_1},C_{\lambda_2})=\frac{1}{k_F^8}\int d\textbf{m}(1-n_m)\int d\textbf{m}'(1-n_{m'})\bigg[SII(P,Q,k_{F,\lambda_1},k_{F,\lambda_2})\bigg]^2
\end{equation}
As we have done for E3, we can introduce $C_{\lambda}$ and 
simplify the external integrals,
\begin{equation}
E_4(C_{\lambda_1},C_{\lambda_2})=8\pi^2\int_0^{C_{\lambda_1}^{1/3}+C_{\lambda_2}
^{1/3}}P^2dP\int_{-\infty}^{\infty}dz\int_{r_{min}}^{\infty}rdr\bigg[SII(P,Q,C_{
 \lambda_1}^{1/3},C_{\lambda_2}^{1/3})\bigg]^2 \ , 
\end{equation}
with  $\textbf{P}=\textbf{m}+\textbf{m}'$ and 
$\textbf{Q}=(\textbf{m}-\textbf{m}')/2$. Afterwards, the integrals in 
$\textbf{Q}$ are transformed into cylindrical coordinates. The three angular 
integrals can be done trivially. The relation between $Q$, $r$ and $z$ is 
$Q=\sqrt{r^2+z^2}$. The volume of integration of $\textbf{Q}$ comes from 
$(1-n_m)(1-n_{m'})=(1-n_{P/2+Q})(1-n_{P/2-Q})$, which is the space out of the 
intersection of two spheres. From here, we can know the integration limits of 
$r$ and $z$. The minimum value that $r$ can take depends on $P$ and $z$, as  
shown in Table \ref{tau.rmin}.

\begin{table}[H]
\centering
\begin{tabular}{ |c|c| } 
 \hline
 \textbf{Condition} & \textbf{$r_{min}$} \\ 
 \hline
 \hline
  & \\
 $P\leq\vert k_{F,\lambda_1}-k_{F,\lambda_2}\vert$ \& $z\in\bigg[\cfrac{P}{2}-k_M,\cfrac{P}{2}+k_M\bigg]$ & $\sqrt{k_M^2-\bigg(z-\cfrac{P}{2}\bigg)^2}$ \\
  & \\
 \hline
  & \\
 $\vert k_{F,\lambda_1}-k_{F,\lambda_2}\vert< P< (k_{F,\lambda_1}+k_{F,\lambda_2})$ \& $z\in\bigg[-\cfrac{P}{2}-k_{F,\lambda_1},\cfrac{k_{F,\lambda_1}^2-k_{F,\lambda_2}^2}{2P}\bigg]$ & $\sqrt{k_{F,\lambda_1}^2-\bigg(z+\cfrac{P}{2}\bigg)^2}$ \\
  & \\
 \hline
  & \\
 $\vert k_{F,\lambda_1}-k_{F,\lambda_2}\vert< P< (k_{F,\lambda_1}+k_{F,\lambda_2})$ \& $z\in\bigg[\cfrac{k_{F,\lambda_1}^2-k_{F,\lambda_2}^2}{2P},\cfrac{P}{2}+k_{F,\lambda_2}\bigg]$ & $\sqrt{k_{F,\lambda_2}^2-\bigg(z-\cfrac{P}{2}\bigg)^2}$ \\
  & \\
 \hline
 Else & 0 \\
 \hline
\end{tabular}
\caption{Values of $r_{min}$ depending on $P$ and $z$. The object 
$k_M$ is defined as $k_M=max(k_{F,\lambda_1},k_{F,\lambda_2})$.}
\label{tau.rmin}
\end{table}

\subsection{Term E5}
The last term that contributes to the third-order expansion is a 
three-body interaction term, 
%This can be seen in the fact that the vectors 
%$\textbf{p}_1$ and $\textbf{m}_1$ run over $C_{\lambda_3}$. The vectors 
%$\textbf{m}$ and $\textbf{p}$ run over $k_{F,\lambda_1}$, and $\textbf{m}'$ 
%and $\textbf{p}'$ run over $k_{F,\lambda_2}$, just as the previous integrals.
\begin{equation}
\begin{aligned}
\frac{E}{N}=\epsilon_F\frac{3x^3}{32\pi^7}\frac{1}{\nu}\sum_{\lambda_1,\lambda_2
,\lambda_3}(1-\delta_{\lambda_1,\lambda_2 
})(2-3\delta_{\lambda_1,\lambda_3}-3\delta_{\lambda_2,\lambda_3})E_5(C_{
\lambda_1},
C_{\lambda_2},C_{\lambda_3}) \ ,
\end{aligned}
\end{equation}
with
\begin{equation}
\begin{aligned}
E_5(C_{\lambda_1},C_{\lambda_2},C_{\lambda_3})=\frac{1}{2k_F^8}\bigg\{\int 
d\textbf{p}n_p\int d\textbf{m}(1-n_m)\\
\times\bigg[\int2d\textbf{m}'d\textbf{p}'\frac{(1-n_{m'})n_{p'}}{
p^2+p'^2-m^2-m'^2}\delta(\textbf{p}+\textbf{p}'-\textbf{m}-\textbf{m}')\bigg]    
 \bigg[\int2d\textbf{m}_1d\textbf{p}_1\frac{(1-n_{m_1})n_{p_1}}{ 
p^2+p_1^2-m^2-m_1^2}
\delta(\textbf{p}+\textbf{p}_1-\textbf{m}-\textbf{m}_1)\bigg]\\
     +\int d\textbf{p}'n_{p'}\int d\textbf{m}'(1-n_{m'})\\
\times\bigg[\int2d\textbf{m}d\textbf{p}\frac{(1-n_m)n_p}{p^2+p'^2-m^2-m'^2}\delta(\textbf{p}+\textbf{p}'-\textbf{m}-\textbf{m}')\bigg]
     \bigg[\int2d\textbf{m}_1d\textbf{p}_1\frac{(1-n_{m_1})n_{p_1}}{p'^2+p_1^2-m'^2-m_1^2}\delta(\textbf{p}'+\textbf{p}_1-\textbf{m}'-\textbf{m}_1)\bigg]\bigg\}
\end{aligned}
\end{equation}

One identifies four inner integrals that have the same 
formal shape. Each one is of the form 
\begin{equation}
  I_{5,\text{inner}}=\bigg[\int2d\textbf{m}'d\textbf{p}'\frac{(1-n_{m'})n_{p'}}{
p^2+p'^2-m^2-m'^2}\delta(\textbf{p}+\textbf{p}'-\textbf{m}-\textbf{m}')\bigg] \ 
.
\end{equation}
In this integral, we 
decompose the term $(1-n_{m'})n_{p'}= n_{p'}-n_{m'}n_{p'}$, so we 
have again a sphere and the intersection of two spheres.  One can 
proceed then in a similar way to previous integrals. 
The spherical part ($SI_5$) is
\begin{equation*}
   SI_5=\int2d\textbf{m}'d\textbf{p}'\frac{n_{p'}}{p^2+p'^2-m^2-m'^2}\delta(\textbf{p}+\textbf{p}'-\textbf{m}-\textbf{m}')=-\int d\textbf{p}'\frac{n_{p'}}{m^2+\textbf{p}'\cdot(\textbf{p}-\textbf{m})-\textbf{p}\cdot\textbf{m}}
\end{equation*}
\begin{equation*}
   =2\pi\int_0^{k_{F,\lambda}} p'^2dp'\frac{1}{p'\vert \textbf{m}-\textbf{p}\vert}\ln{\bigg\vert\frac{m^2-\textbf{p}\cdot\textbf{m}-p'\vert \textbf{m}-\textbf{p}\vert}{m^2-\textbf{p}\cdot\textbf{m}+p'\vert \textbf{m}-\textbf{p}\vert}\bigg\vert}
\end{equation*}
\begin{equation}
\hspace*{-0.3 cm}
    =2\pi\bigg\{-k_{F,\lambda}\frac{(m^2-\textbf{p}\cdot\textbf{m})}{\vert \textbf{m}-\textbf{p}\vert^2}+\bigg(\frac{k_{F,\lambda}^2}{2\vert \textbf{m}-\textbf{p}\vert}-\frac{(m^2-\textbf{p}\cdot\textbf{m})^2}{2\vert \textbf{m}-\textbf{p}\vert^3}\bigg)\ln{\bigg\vert\frac{m^2-\textbf{p}\cdot\textbf{m}-k_{F,\lambda}\vert \textbf{m}-\textbf{p}\vert}{m^2-\textbf{p}\cdot\textbf{m}+k_{F,\lambda}\vert \textbf{m}-\textbf{p}\vert}\bigg\vert}\bigg\}
\end{equation}
It can be rewritten as 
\begin{equation}
\boxed{
    SI_5=2\pi\bigg\{-k_{F,\lambda}\frac{(2Q^2-a)}{4Q^2}+\bigg(\frac{k_{F,\lambda}^2}{4Q}-\frac{(2Q^2-a)^2}{16Q^3}\bigg)\ln{\bigg\vert\frac{2Q^2-a-2k_{F,\lambda}Q}{2Q^2-a+2k_{F,\lambda}Q}\bigg\vert}\bigg\} \ ,}
\end{equation}
with $a=(p^2-m^2)/2$ and $Q=\vert \textbf{m}-\textbf{p}\vert/2$.

Before doing the sphere intersection part, we define a change of variables 
that are useful for this particular integral,
\begin{equation}
    \begin{matrix}
        \textbf s=\cfrac{\textbf p'+\textbf m'}{2}\\
        \\
        \textbf d=\textbf p'-\textbf m'
    \end{matrix}
    \quad\Rightarrow\quad
    \begin{matrix}
        \textbf p'=\textbf s+\cfrac{\textbf d}{2}\\
        \\
        \textbf m'=\textbf s-\cfrac{\textbf d}{2}
    \end{matrix}
\end{equation}
Then, 
\begin{equation*}
    SII_5=\int 2d\textbf{m}'d\textbf{p}'\frac{n_{m'}n_{p'}}{p^2+p'^2-m^2-m'^2}\delta(\textbf{p}+\textbf{p}'-\textbf{m}-\textbf{m}')
\end{equation*}
\begin{equation}
    =\int d\textbf{s}d\textbf{d}\frac{n_{s+d/2}n_{s-d/2}}{\cfrac{p^2-m^2}{2}+\textbf{s}\cdot\textbf{d}}\delta(\textbf{p}-\textbf{m}+\textbf{d})=\int d\textbf{s}\frac{n_{s+d/2}n_{s-d/2}}{\cfrac{p^2-m^2}{2}+\textbf{s}\cdot\textbf{d}}\quad where\quad \textbf{d}=\textbf{m}-\textbf{p}
\end{equation}
As commented previously, there are three types of intersection: null, two 
caps and a small sphere. The null is just zero and the small-sphere case does 
not happen here because the two radius are equal. 
Therefore just one case is missing, the one concerning the two spherical caps. 
So $SII_5$ will be just $SCI_5$. After some algebra, one arrives to 
\begin{equation*}
    SCI_5=2\pi\bigg\{\frac{8ak_{F,\lambda}d-4ad^2}{8d^3}-\frac{(4a^2-4ad^2-4k_{F,\lambda}^2d^2+d^4)}{8d^3}\ln{\bigg\vert\frac{a+\big(k_{F,\lambda}-d/2\big)d}{a}\bigg\vert}
\end{equation*}
\begin{equation}
    +\frac{(4a^2+4ad^2-4k_{F,\lambda}^2d^2+d^4)}{8d^3}\ln{\bigg\vert\frac{a+\big(d/2-k_{F,\lambda}\big)d}{a}\bigg\vert}\bigg\} \ .
\end{equation}
Changing variable to $Q=\big\vert\textbf{m}-\textbf{p}\big\vert/2=d/2$, we get
\begin{equation}
\boxed{
\begin{aligned}
    SCI_5=2\pi\bigg\{\frac{ak_{F,\lambda}Q-aQ^2}{4Q^3}-\frac{(a^2-4aQ^2-4k_{F,\lambda}^2Q^2+4Q^4)}{16Q^3}\ln{\bigg\vert\frac{a+2\big(k_{F,\lambda}-Q\big)Q}{a}\bigg\vert}\\
    \\
    +\frac{(a^2+4aQ^2-4k_{F,\lambda}^2Q^2+4Q^4)}{16Q^3}\ln{\bigg\vert\frac{a+2\big(Q-k_{F,\lambda}\big)Q}{a}\bigg\vert}\bigg\} \ ,
\end{aligned}
}
\end{equation}
with $a=(p^2-m^2)/2$.

Therefore, $I_{5,\text{inner}}$ is $SI_5-SCI_5$, but in both functions 
there is one situation that can cause problems, and that is the limit when $Q 
\to 0$. However, if one calculates 
separately the limits for $SI_5$ and $SCI_5$, both give $4\pi 
k_{F,\lambda}^3/(3a)$, so the total limit  is 
zero. The possible contributions to $I_{5,\text{inner}}$ are summarized in 
Table \ref{int5}.

\begin{table}[H]
\centering
\begin{tabular}{ |c|c| } 
 \hline
 \textbf{Condition} & \textbf{$I_{5,\text{inner}}$ Integral} \\ 
 \hline
 \hline
 $Q=0$ & 0 \\ 
 \hline
 $0<Q<k_{F,\lambda}$ & $SI_5-SCI_5$ \\ 
 \hline
 $Q\geq k_{F,\lambda}$ & $SI_5$ \\ 
 \hline
\end{tabular}
\caption{Summary of the $SI_{5,\text{inner}}$ integral depending on the 
value of $Q$.}
\label{int5}
\end{table}

If we look to $SI_5$ and $SCI_5$, we can see that the integrals 
depend on $k_{F,\lambda}$, $a$ and $Q$. Hence, they depend on the modulus of 
both $\textbf{m}$ and $\textbf{p}$, and also on the relative angle between both 
vectors. We recall that $a=(p^2-m^2)/2$ and $Q=\sqrt{p^2+m^2+2pmx}/2$ with 
$x = -\cos{\theta}$. In Eq. (\ref{eq.final.e5}), we show the entire expression of $E_5$, the inner integrals are expressed through the compact form $I_{5,inner}(R,a,k_{F,\lambda})$.
\begin{equation}
\begin{aligned}
E_5(C_{\lambda_1},C_{\lambda_2},C_{\lambda_3})=\frac{1}{2k_F^8}\bigg\{\int d\textbf{p}n_p\int d\textbf{m}(1-n_m)I_{5,inner}(R,a,k_{F,\lambda_2})I_{5,inner}(R,a,k_{F,\lambda_3})\\
+\int d\textbf{p}'n_{p'}\int d\textbf{m}'(1-n_{m'})I_{5,inner}(R',a',k_{F,\lambda_1})I_{5,inner}(R',a',k_{F,\lambda_3})\bigg\}
\end{aligned}
\label{eq.final.e5}
\end{equation}
We point out that $R'$ means $\sqrt{p'^2+m'^2+2p'm'x'}/2$ and $a'=(p'^2-m'^2)/2$. In terms of the concentrations  $C_{\lambda}$ one arrives 
to the final expression
\begin{equation}
\begin{aligned}
E_5(C_{\lambda_1},C_{\lambda_2},C_{\lambda_3})=4\pi^2\bigg\{\int_0^{C_{\lambda_1
}^{1/3}}dp \int_{C_{\lambda_1}^{1/3}}^{\infty} dm\int_{-1}^1dx 
I_{5,\text{inner}}
(Q,a,k_{F,\lambda_2})I_{5,\text{inner}}(Q,a,k_{F,\lambda_3})\\
+\int_0^{C_{\lambda_2}^{1/3}}dp' \int_{C_{\lambda_2}^{1/3}}^{\infty} 
dm'\int_{-1}^1dx' 
I_{5,\text{inner}}(Q',a',k_{F,\lambda_1})I_{5,\text{inner}}
(Q', a',k_{F,\lambda_3 } )\bigg\} \ .
\end{aligned}
\end{equation}

We would like to point out that for the three integrals ($E3$, $E4$, and $E5$) 
we have been able to reduce high dimensional integrals to just 3D integrals, 
which makes numerical integration less costly.

\section{Terms E3, E4, and E5 for the  ground state}
In Ref.~\cite{Pera}, we proved that a Fermi gas in the ground state chooses a 
certain spin occupational distribution. This configuration, that minimizes the 
energy of the system, is the following: one species increases, and the rest 
ones diminish equally. Under these conditions, the concentrations 
$C_{\lambda}$ for a given polarization $P$ are 
\begin{eqnarray}
C_+ & = & 1+ |P| \, (\nu-1) 
\label{cesp2}\\
C_{-} & = & 1-|P| \ ,
\label{cesp}
\end{eqnarray}
with subindex $+$ standing for the state with the larger population and $-$ 
for the rest. We point
out that, in this configuration, all the species that decrease have the same 
Fermi momentum or, equivalently, the same concentration $C_{\lambda}$. This is 
important because it reduces the number of integrals to be carried out. 
As commented before, the energy corresponding to these terms is
\begin{equation}
     \frac{E}{N}=\frac{3}{5}\epsilon_F\bigg[\frac{1}{\nu}\sum_{\lambda_1,\lambda_2}\bigg\{\frac{5}{32\pi^7}\Big(E_3+E_4+\sum_{\lambda_3}\big(2-3\delta_{\lambda_1,\lambda_3}-3\delta_{\lambda_2,\lambda_3}\big)E_5\Big)(k_Fa_0)^3\bigg\}(1-\delta_{\lambda_1,\lambda_2})\bigg] \ .
\end{equation}
For $E3$ and $E4$, we only have to do a sum over pairs. Therefore, we only have 
two terms, one taking into account the combination $+-$, and the other one 
accounting for $--$,
\begin{equation}
\frac{1}{2}\sum_{\lambda_1,\lambda_2}E_3(C_{\lambda_1},C_{\lambda_2})(1-\delta_{\lambda_1,\lambda_2})=(\nu-1)E_3(C_+,C_-)+\frac{(\nu-1)(\nu-2)}{2}E_3(C_-,C_-)
\end{equation}
Moreover, the second term can be simplified because both concentrations are the 
same. The function $E_3(C_-,C_-)$ is the same as its value at zero polarization 
times $C_-^{8/3}$,
\begin{equation}
\frac{1}{2}\sum_{\lambda_1,\lambda_2}E_3(C_{\lambda_1},C_{\lambda_2})(1-\delta_{\lambda_1,\lambda_2})=(\nu-1)E_3(C_+,C_-)+\frac{(\nu-1)(\nu-2)}{2}E_3(1,1)C_-^{8/3}
\end{equation}
The same reasoning done for $E3$ can be applied to $E4$,
\begin{equation}
\frac{1}{2}\sum_{\lambda_1,\lambda_2}E_4(C_{\lambda_1},C_{\lambda_2})(1-\delta_{\lambda_1,\lambda_2})=(\nu-1)E_4(C_+,C_-)+\frac{(\nu-1)(\nu-2)}{2}E_4(1,1)C_-^{8/3}
\end{equation}
For $E5$, as we have a three-body interaction, there are more possible 
combinations, but they can be grouped up to four terms,
\begin{equation}
\begin{aligned}
    \frac{1}{2}\sum_{\lambda_1,\lambda_2,\lambda_3}E_5(C_{\lambda_1},C_{\lambda_2},C_{\lambda_3})(1-\delta_{\lambda_1,\lambda_2})(2-3\delta_{\lambda_1,\lambda_3}-3\delta_{\lambda_2,\lambda_3})=-(\nu-1)E_5(C_+,C_-,C_+)\\
    +(\nu-1)(2\nu-5)E_5(C_+,C_-,C_-)+(\nu-1)(\nu-2)E_5(C_-,C_-,C_+)+(\nu-1)(\nu-2)(\nu-4)E_5(C_-,C_-,C_-)
\end{aligned}
\end{equation}
In the same manner as E3 and E4, the term $E_5(C_-,C_-,C_-)$ can be simplified 
into $E_5(1,1,1)C_-^{8/3}$, which is the value at 
zero polarization times $C_-^{8/3}$,
\begin{equation}
\begin{aligned}
    \frac{1}{2}\sum_{\lambda_1,\lambda_2,\lambda_3}E_5(C_{\lambda_1},C_{\lambda_2},C_{\lambda_3})(1-\delta_{\lambda_1,\lambda_2})(2-3\delta_{\lambda_1,\lambda_3}-3\delta_{\lambda_2,\lambda_3})=-(\nu-1)E_5(C_+,C_-,C_+)\\
    +(\nu-1)(2\nu-5)E_5(C_+,C_-,C_-)+(\nu-1)(\nu-2)E_5(C_-,C_-,C_+)+(\nu-1)(\nu-2)(\nu-4)E_5(1,1,1)C_-^{8/3}
\end{aligned}
\end{equation}
Therefore, in the ground state, of all the integrals we had to calculate, 
we only care about five, which are: $E_3(C_+,C_-)$, $E_4(C_+,C_-)$, 
$E_5(C_+,C_-,C_+)$, $E_5(C_+,C_-,C_-)$, and $E_5(C_-,C_-,C_+)$.

\section{Functions E3, E4, and E5 for different spins}

In this section, we plot the integrals obtained numerically for different spin 
values. To simplify the notation, we use E3 
for $E_3(C_+,C_-)$, E4 for $E_4(C_+,C_-)$, 
E5\_1 for $E_5(C_+,C_-,C_+)$, E5\_2 for $E_5(C_+,C_-,C_-)$, and E5\_3 for 
$E_5(C_-,C_-,C_+)$.
%\clearpage

\subsection{Spin $1/2$}
\begin{figure}[H]
    \centering
    \begin{subfigure}[b]{0.49\textwidth}
        \centering
        \includegraphics[width=\textwidth]{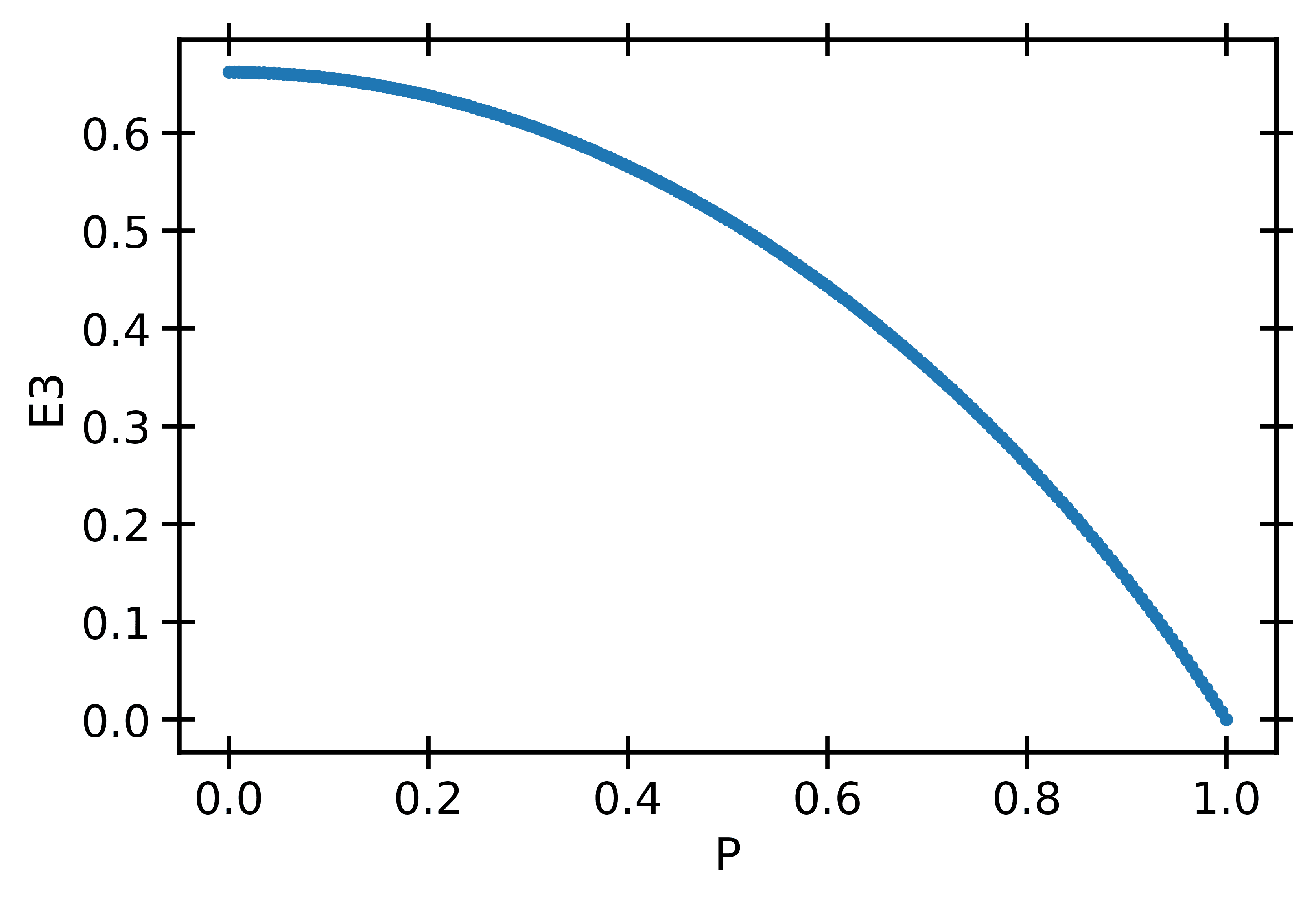}
    \end{subfigure}
    \begin{subfigure}[b]{0.49\textwidth}
        \centering
        \includegraphics[width=\textwidth]{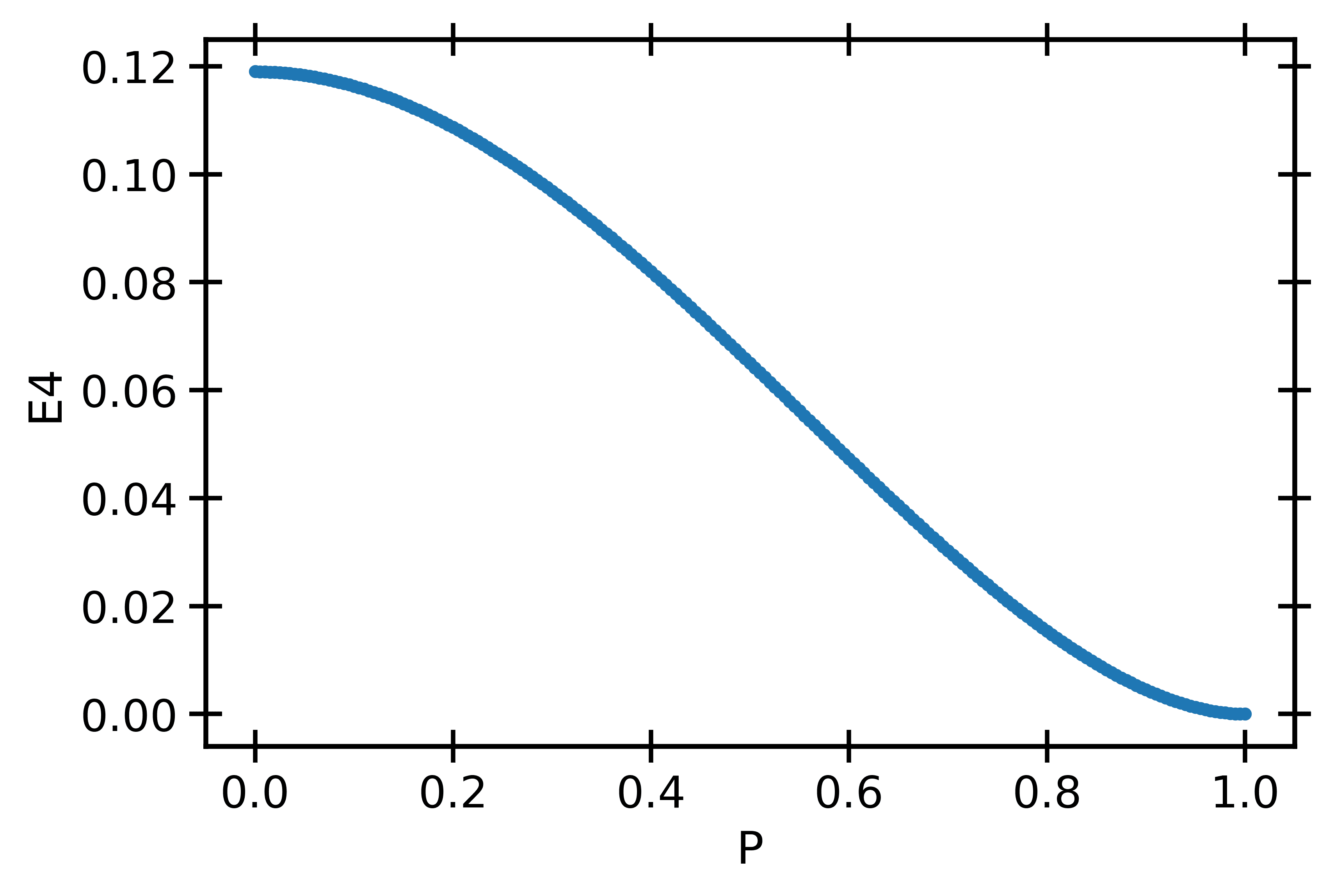}
    \end{subfigure}
    \\
    \begin{subfigure}[b]{0.49\textwidth}
        \centering
        \includegraphics[width=\textwidth]{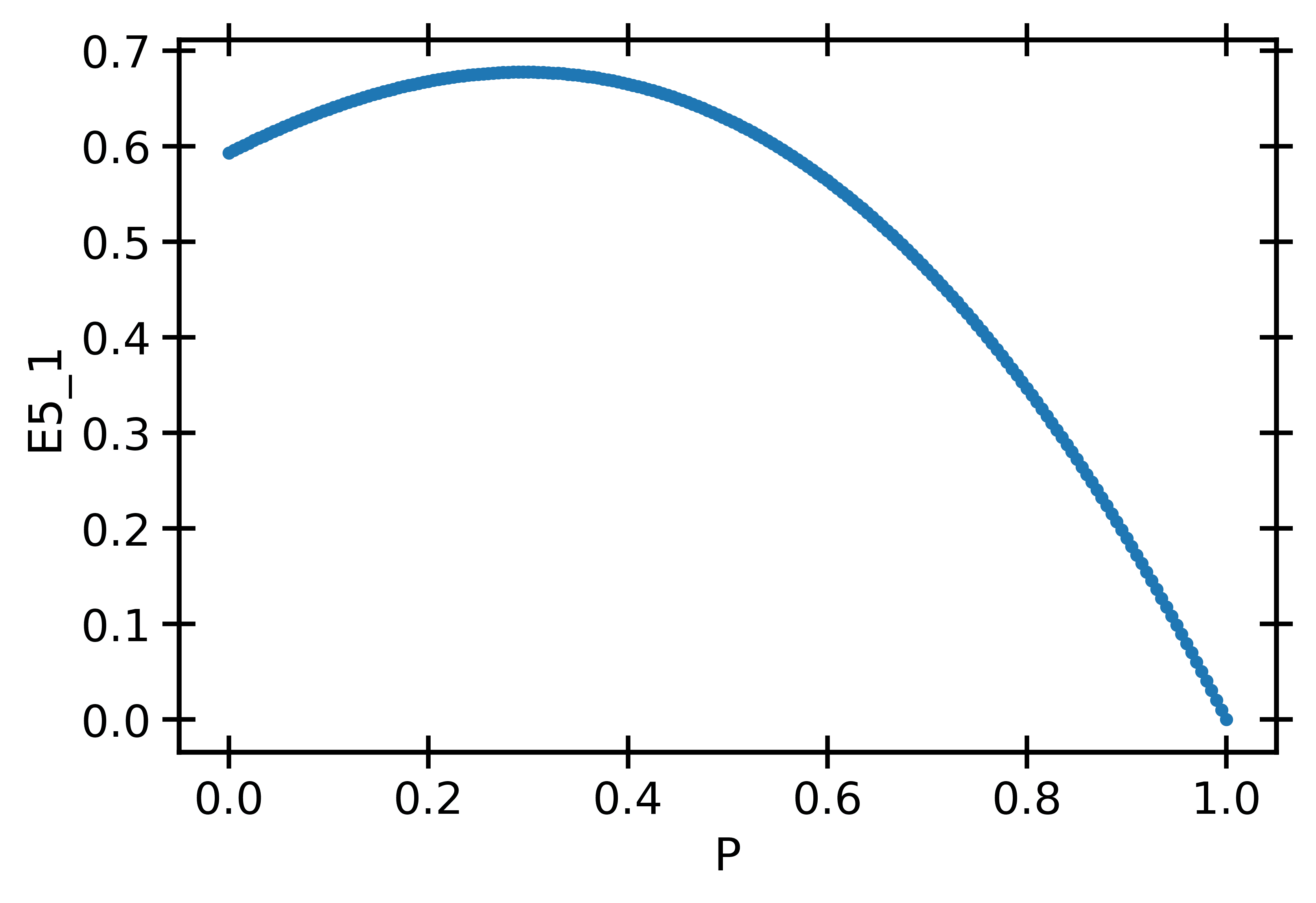}
    \end{subfigure}
    \begin{subfigure}[b]{0.49\textwidth}
        \centering
        \includegraphics[width=\textwidth]{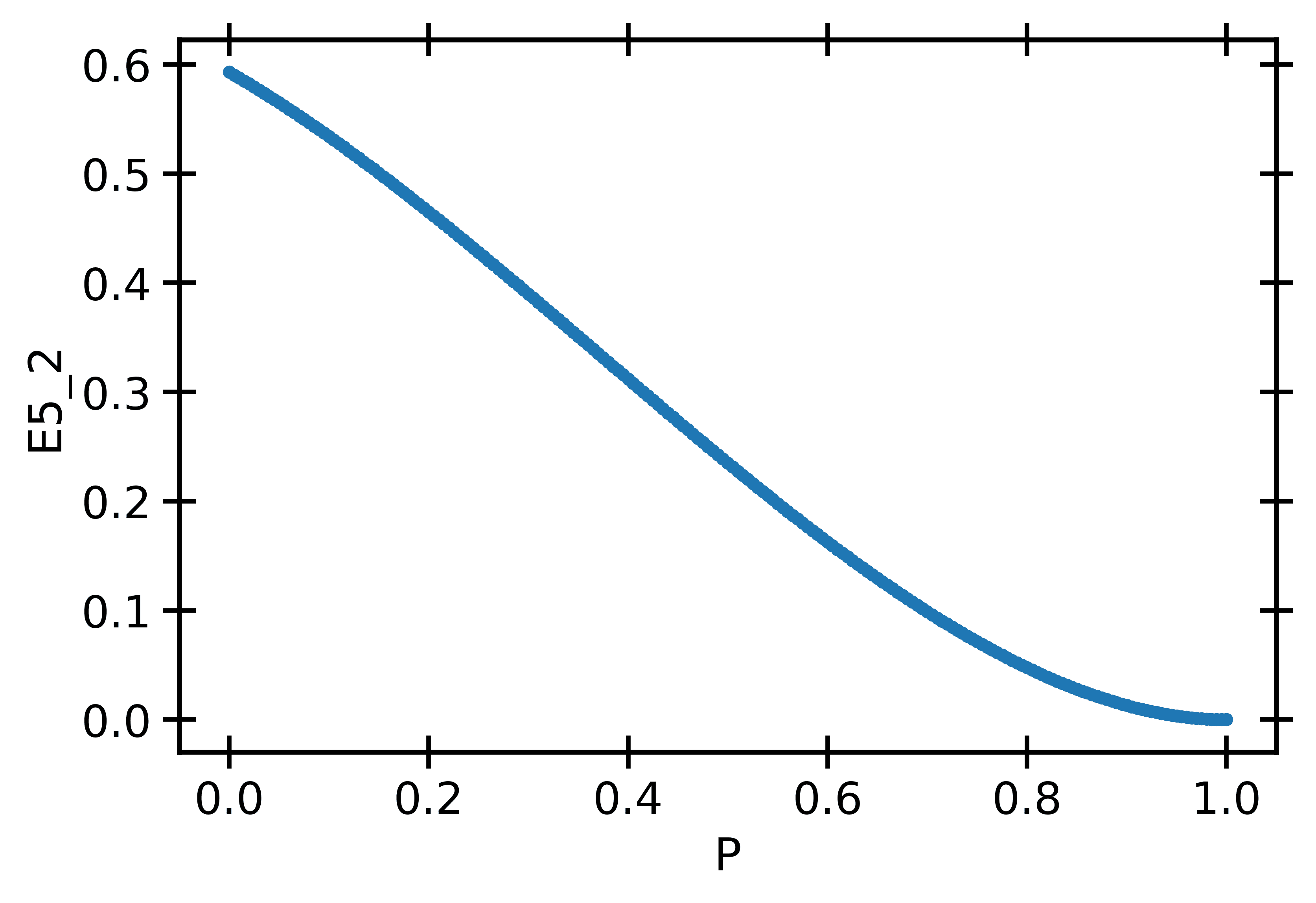}
    \end{subfigure}
    \\
    \begin{subfigure}[b]{0.49\textwidth}
        \centering
        \includegraphics[width=\textwidth]{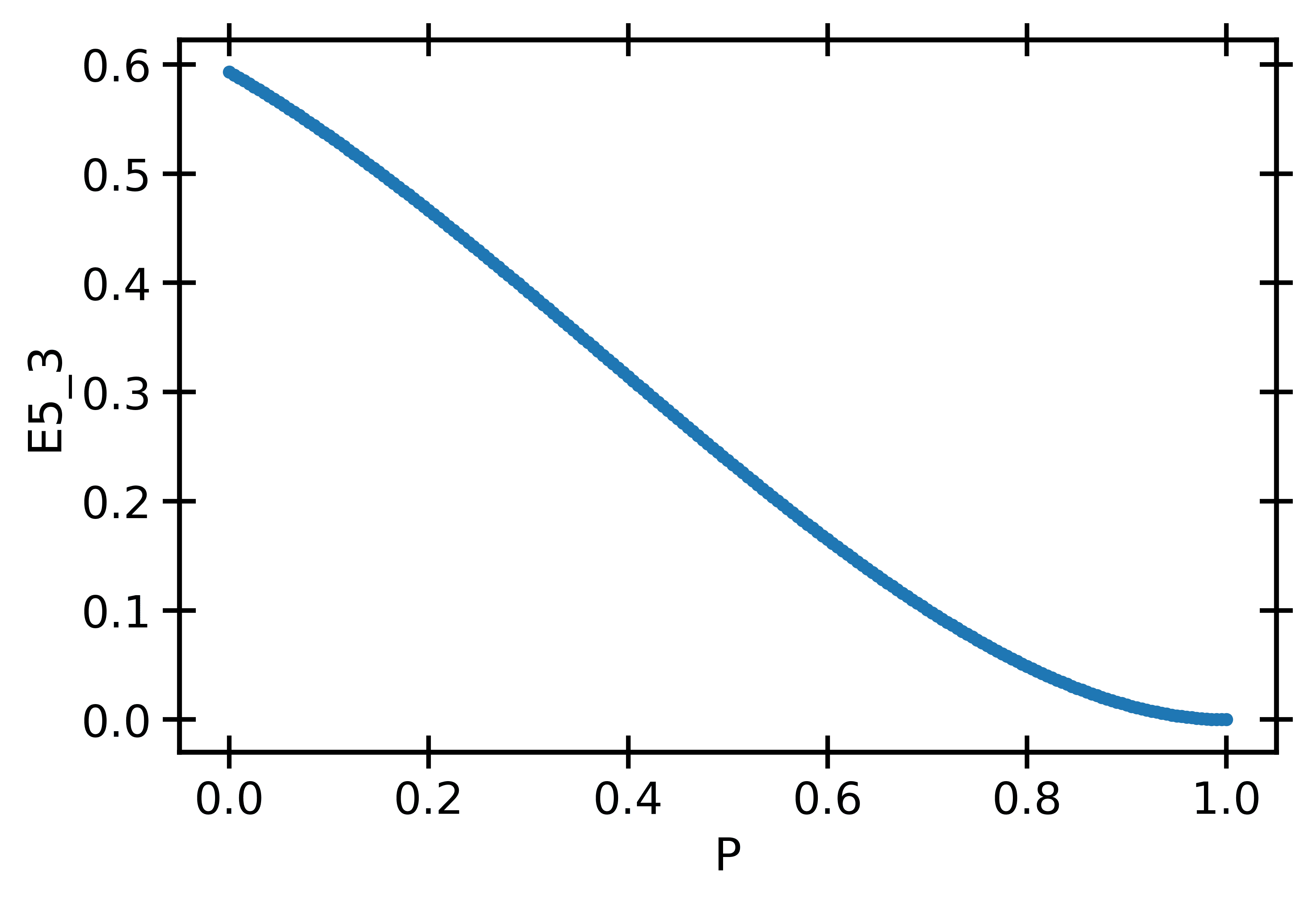}
    \end{subfigure}
    \caption{E3, E4, E5\_1, E5\_2 and E5\_3 in 
terms of the polarization $P$ for $S=1/2$. The error bars are smaller than 
the size of the symbols. }   
\label{fig.int_v2}
\end{figure}

\subsection{Spin $3/2$}
\begin{figure}[H]
    \centering
    \begin{subfigure}[b]{0.49\textwidth}
        \centering
        \includegraphics[width=\textwidth]{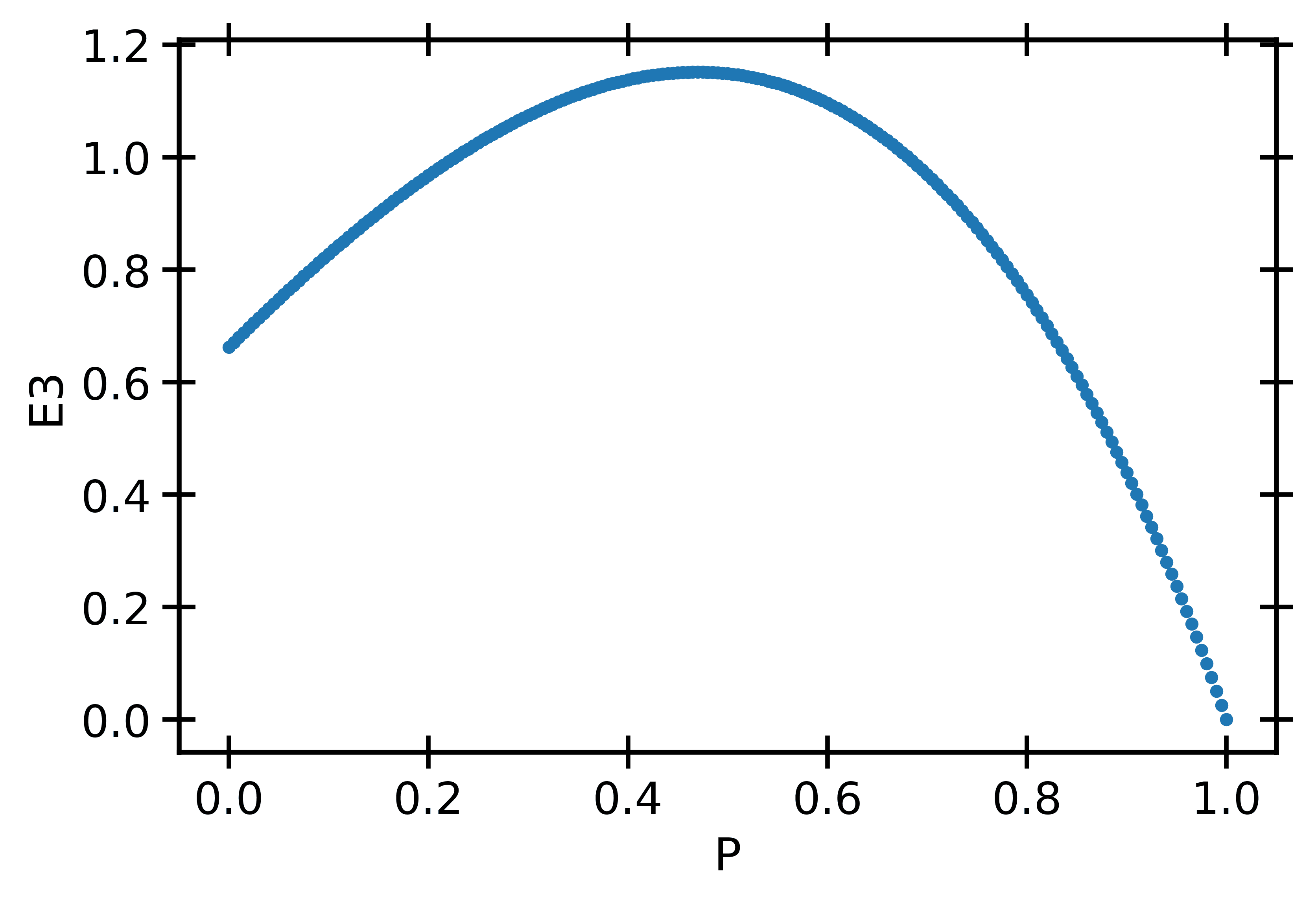}
    \end{subfigure}
    \begin{subfigure}[b]{0.49\textwidth}
        \centering
        \includegraphics[width=\textwidth]{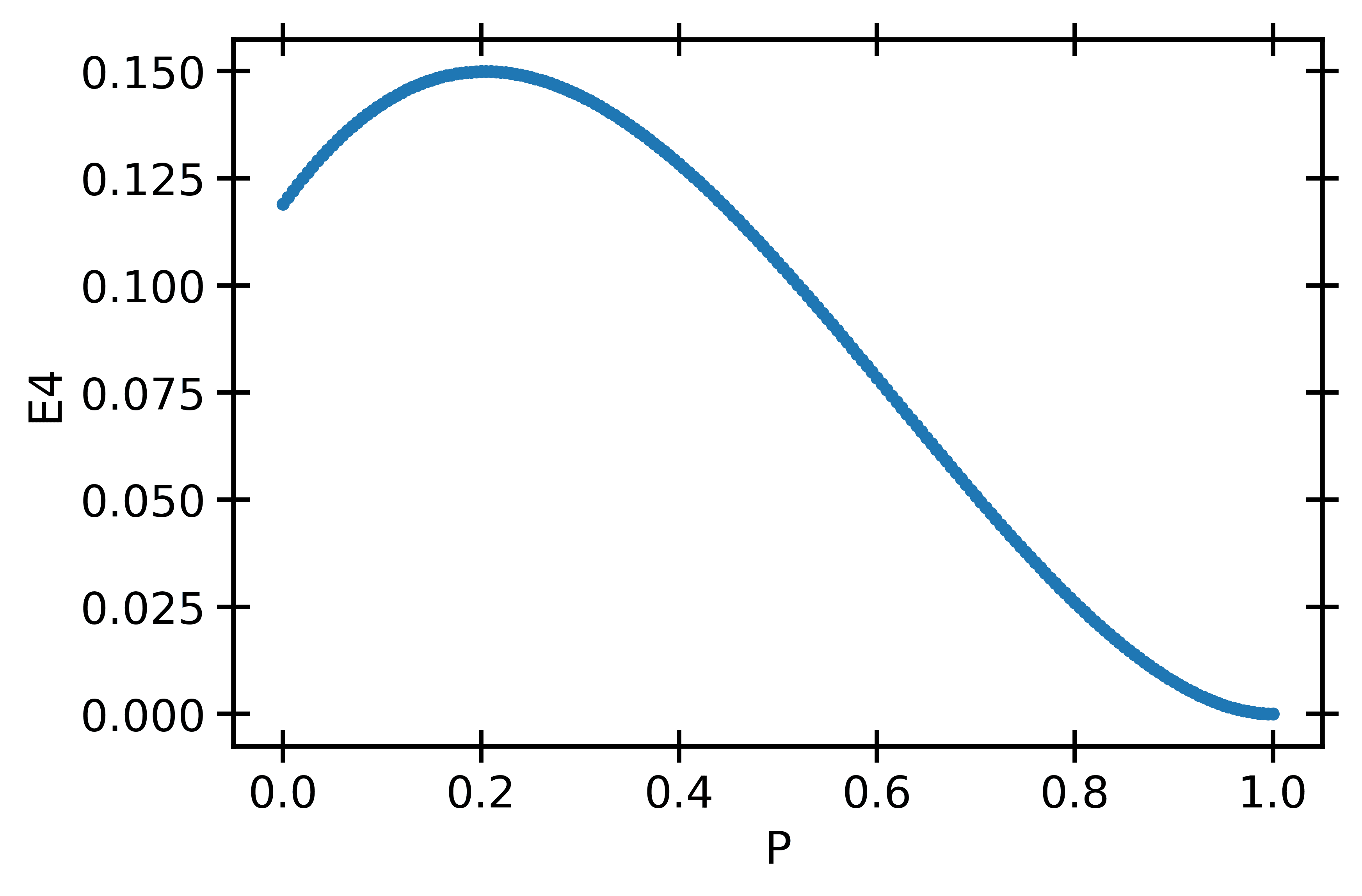}
    \end{subfigure}
    \\
    \begin{subfigure}[b]{0.49\textwidth}
        \centering
        \includegraphics[width=\textwidth]{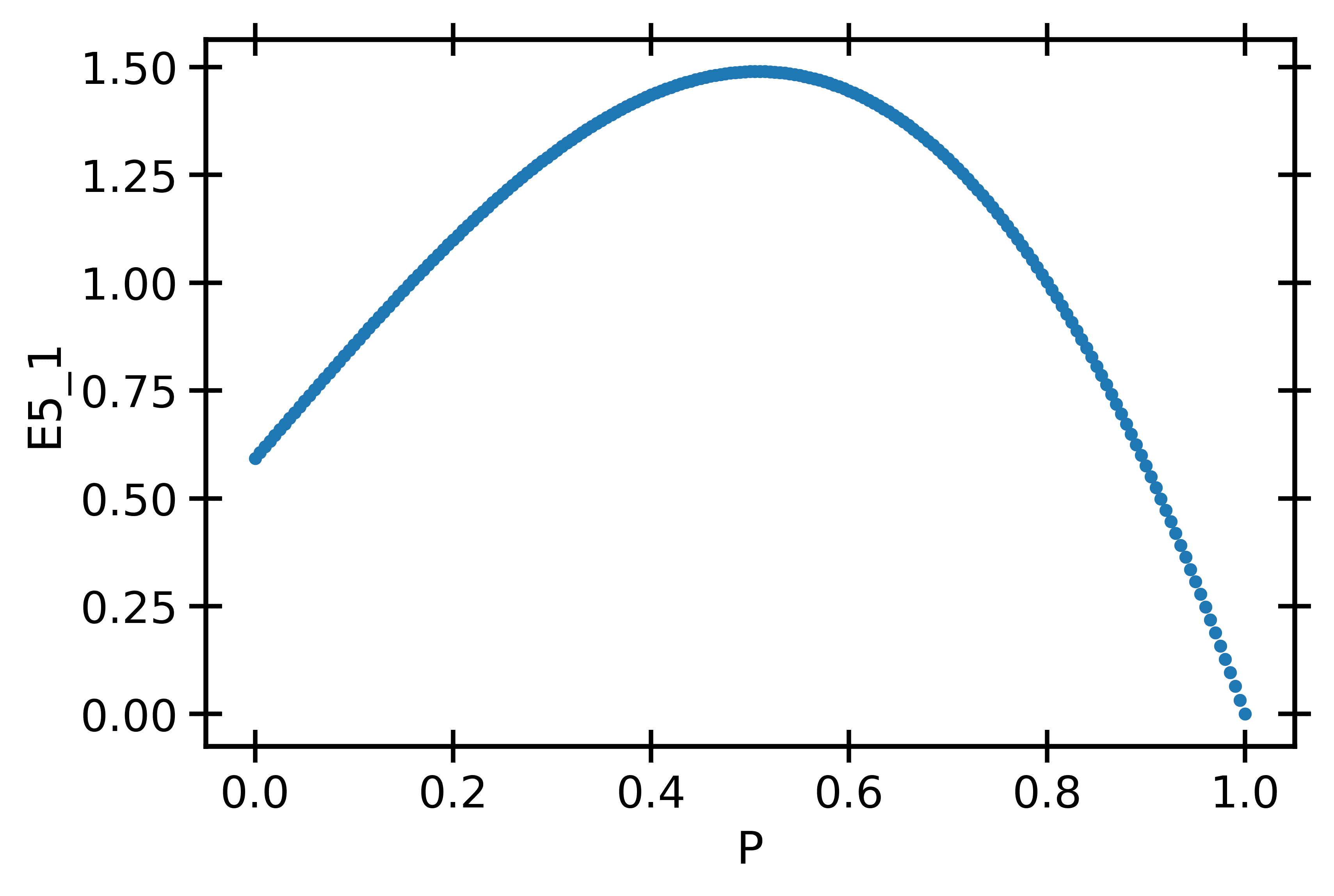}
    \end{subfigure}
    \begin{subfigure}[b]{0.49\textwidth}
        \centering
        \includegraphics[width=\textwidth]{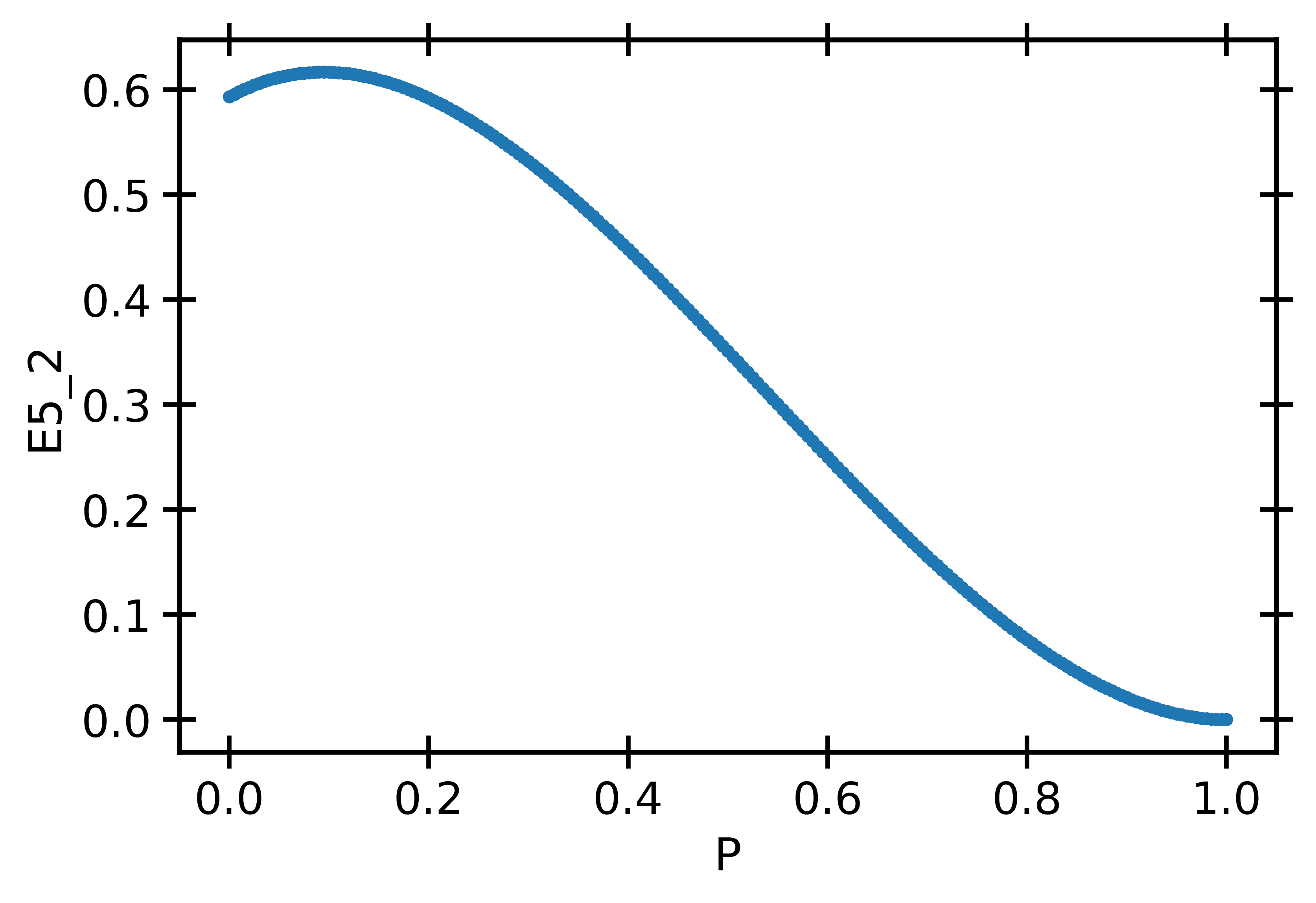}
    \end{subfigure}
    \\
    \begin{subfigure}[b]{0.49\textwidth}
        \centering
        \includegraphics[width=\textwidth]{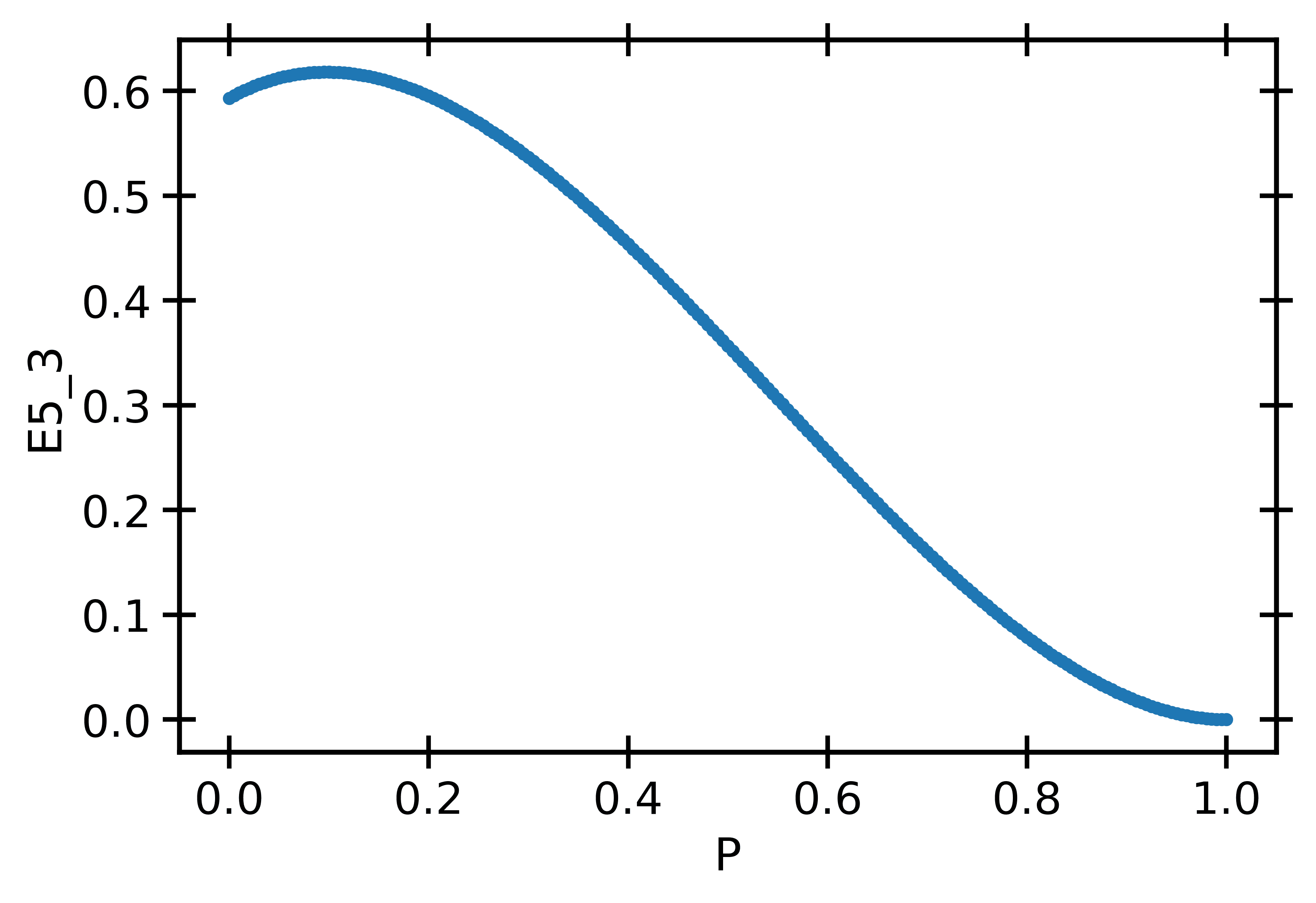}
    \end{subfigure}
    \caption{E3, E4, E5\_1, E5\_2 and E5\_3 in 
terms of the polarization $P$ for $S=3/2$. The error bars are smaller than 
the size of the symbols. }
 \label{fig.int_v4}
\end{figure}

\subsection{Spin $5/2$}
\begin{figure}[H]
    \centering
    \begin{subfigure}[b]{0.49\textwidth}
        \centering
        \includegraphics[width=\textwidth]{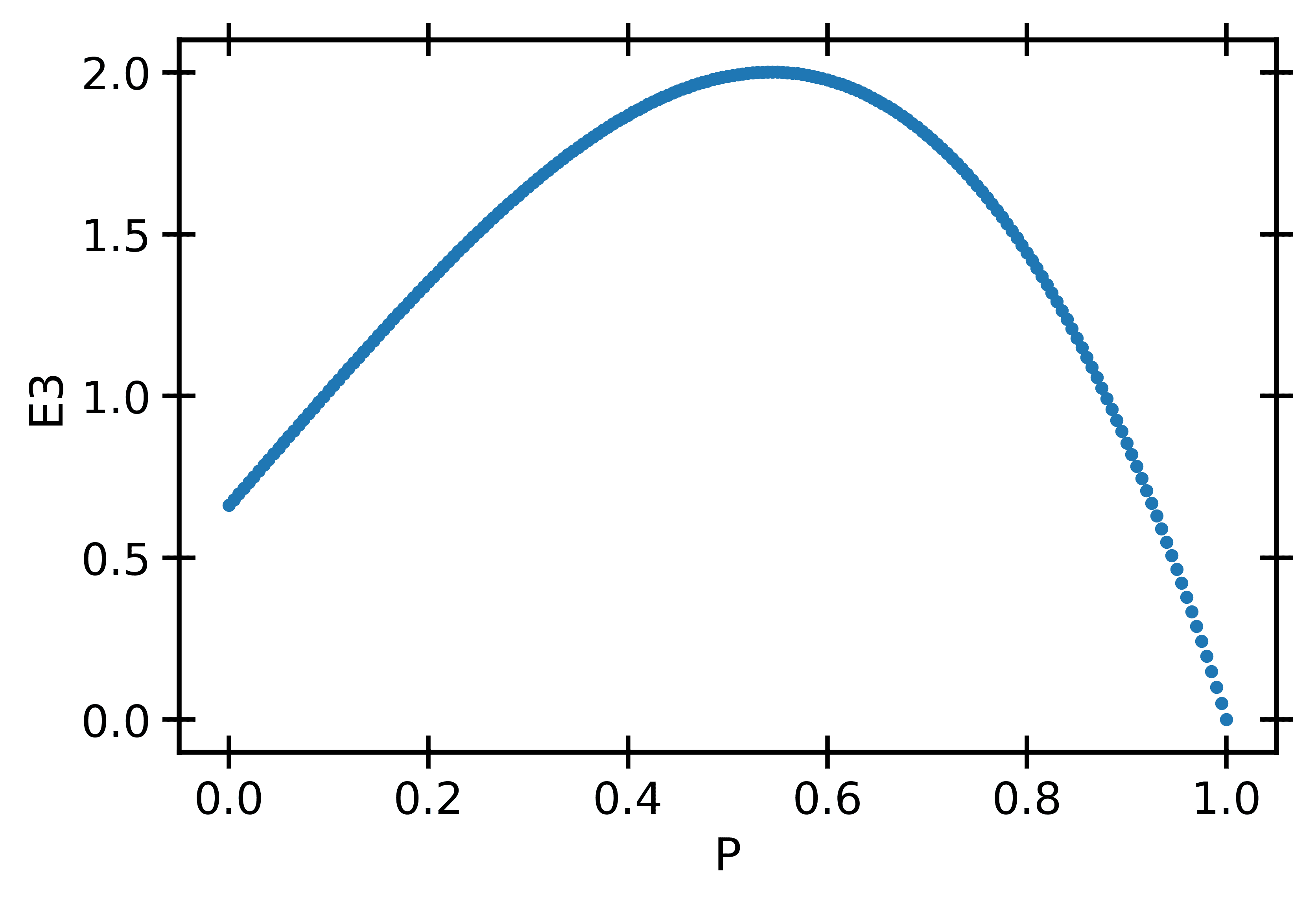}
    \end{subfigure}
    \begin{subfigure}[b]{0.49\textwidth}
        \centering
        \includegraphics[width=\textwidth]{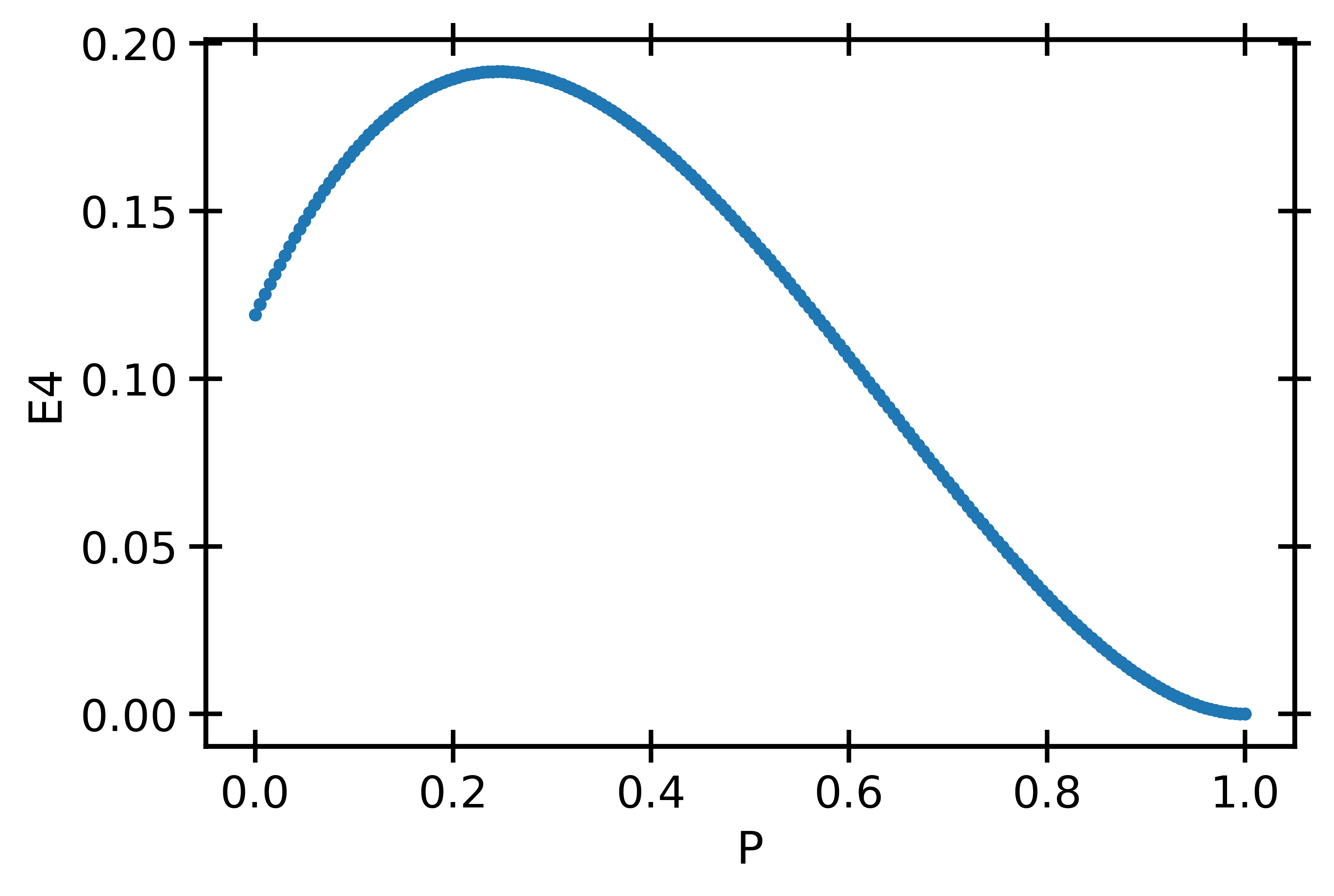}
    \end{subfigure}
    \\
    \begin{subfigure}[b]{0.49\textwidth}
        \centering
        \includegraphics[width=\textwidth]{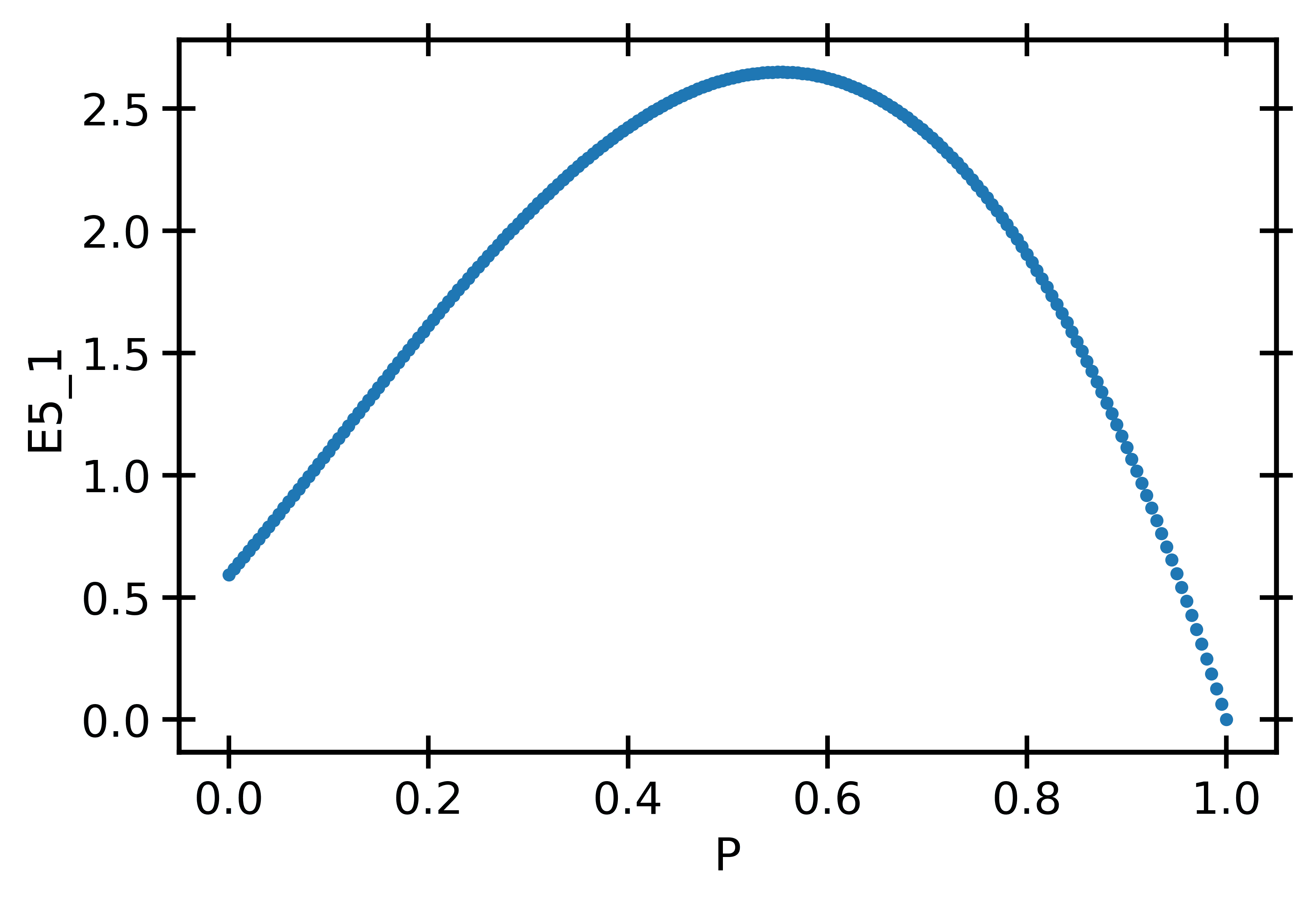}
    \end{subfigure}
    \begin{subfigure}[b]{0.49\textwidth}
        \centering
        \includegraphics[width=\textwidth]{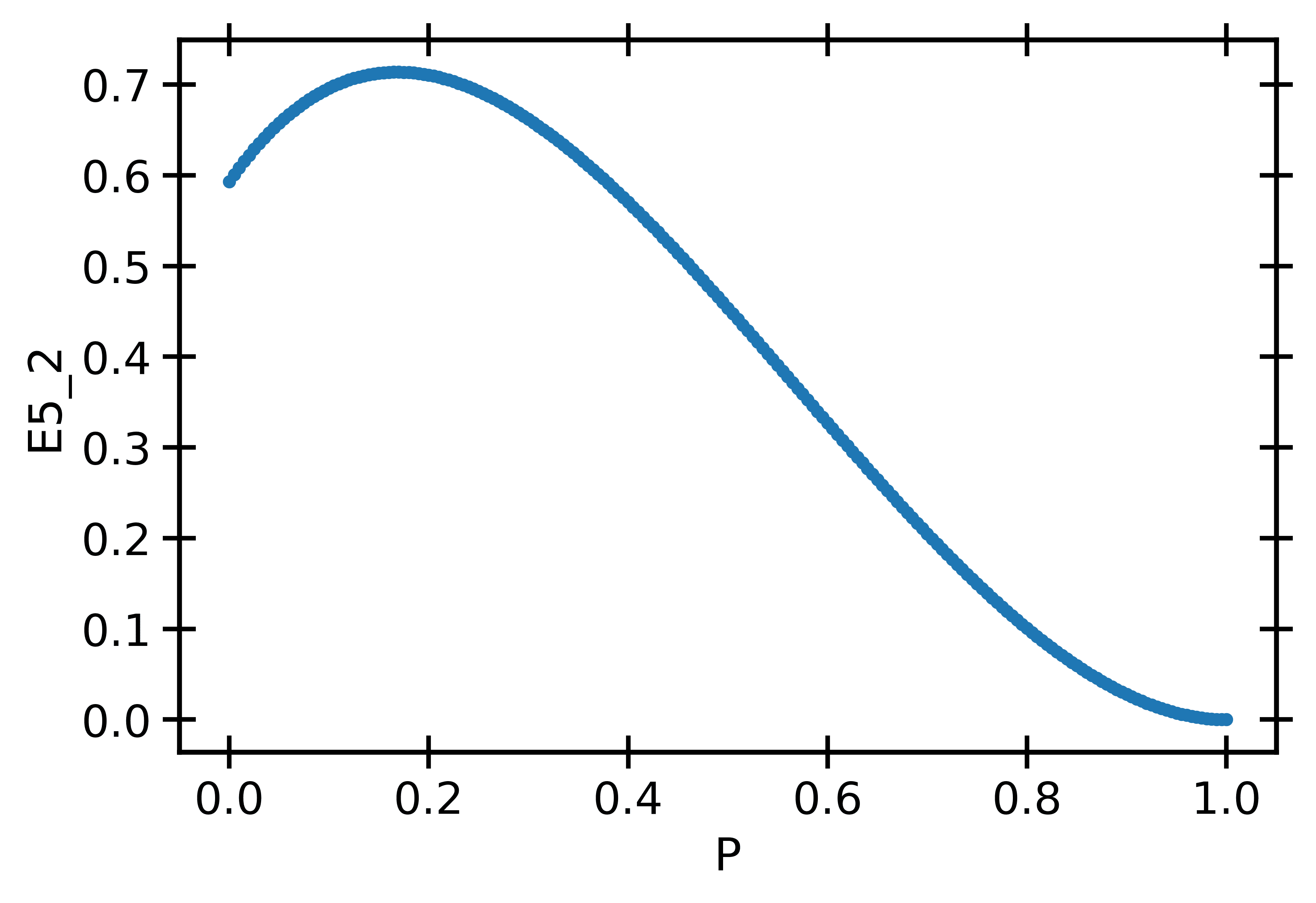}
    \end{subfigure}
    \\
    \begin{subfigure}[b]{0.49\textwidth}
        \centering
        \includegraphics[width=\textwidth]{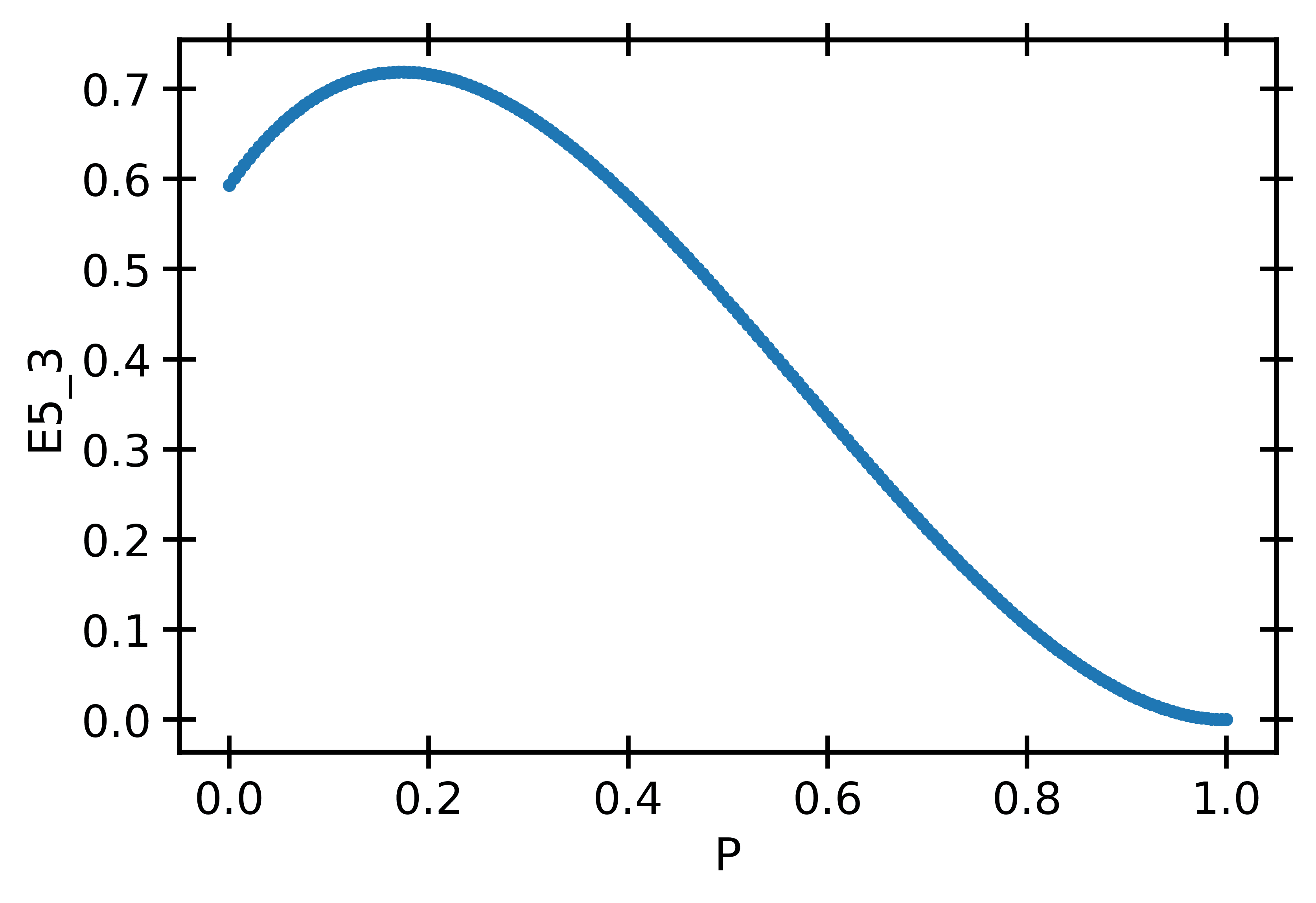}
    \end{subfigure}
    \caption{E3, E4, E5\_1, E5\_2 and E5\_3 in 
terms of the polarization $P$ for $S=5/2$. The error bars are smaller than 
the size of the symbols. }
\label{fig.int_v6}
\end{figure}

\subsection{Spin $7/2$}
\begin{figure}[H]
    \centering
    \begin{subfigure}[b]{0.49\textwidth}
        \centering
        \includegraphics[width=\textwidth]{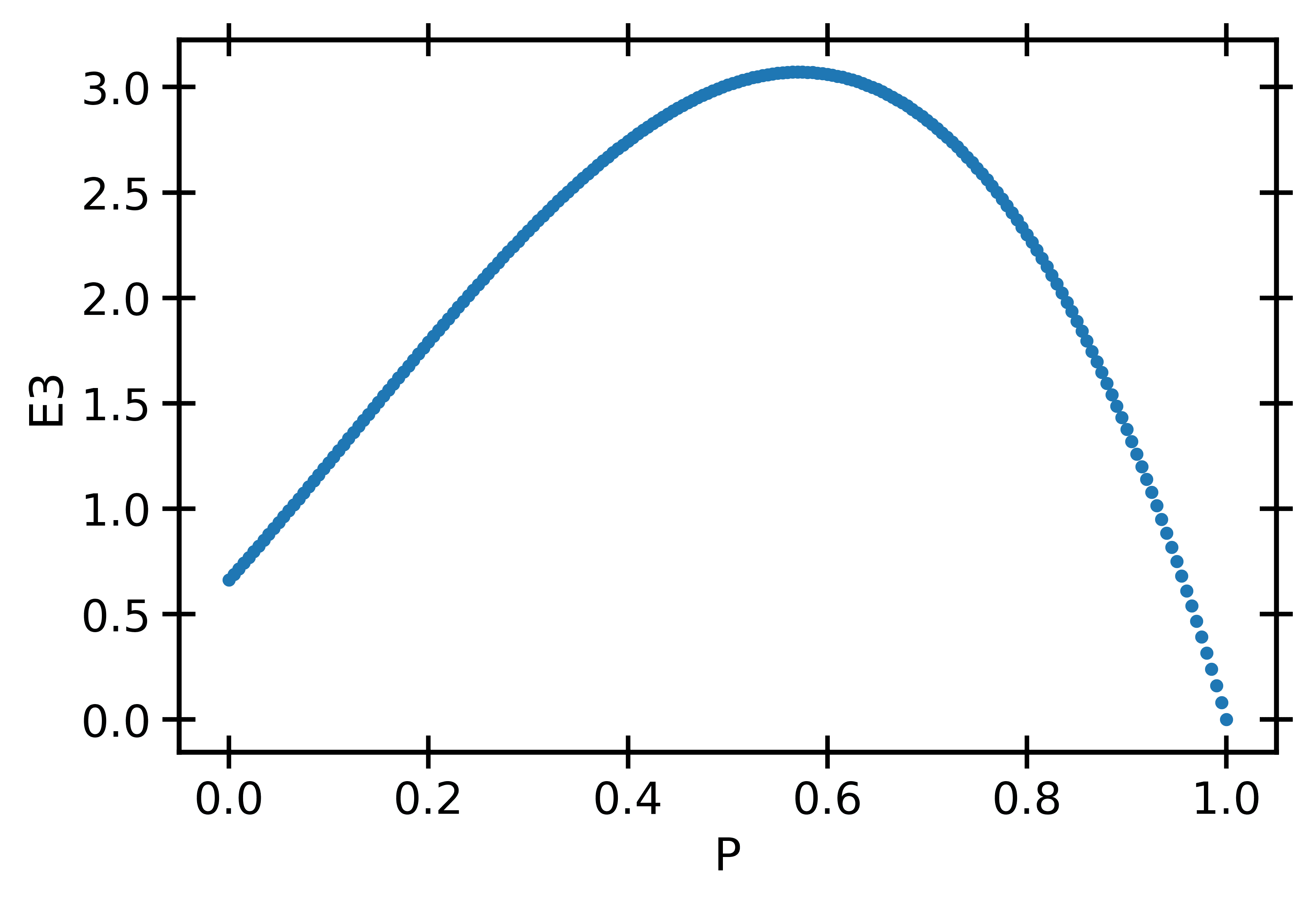}
    \end{subfigure}
    \begin{subfigure}[b]{0.49\textwidth}
        \centering
        \includegraphics[width=\textwidth]{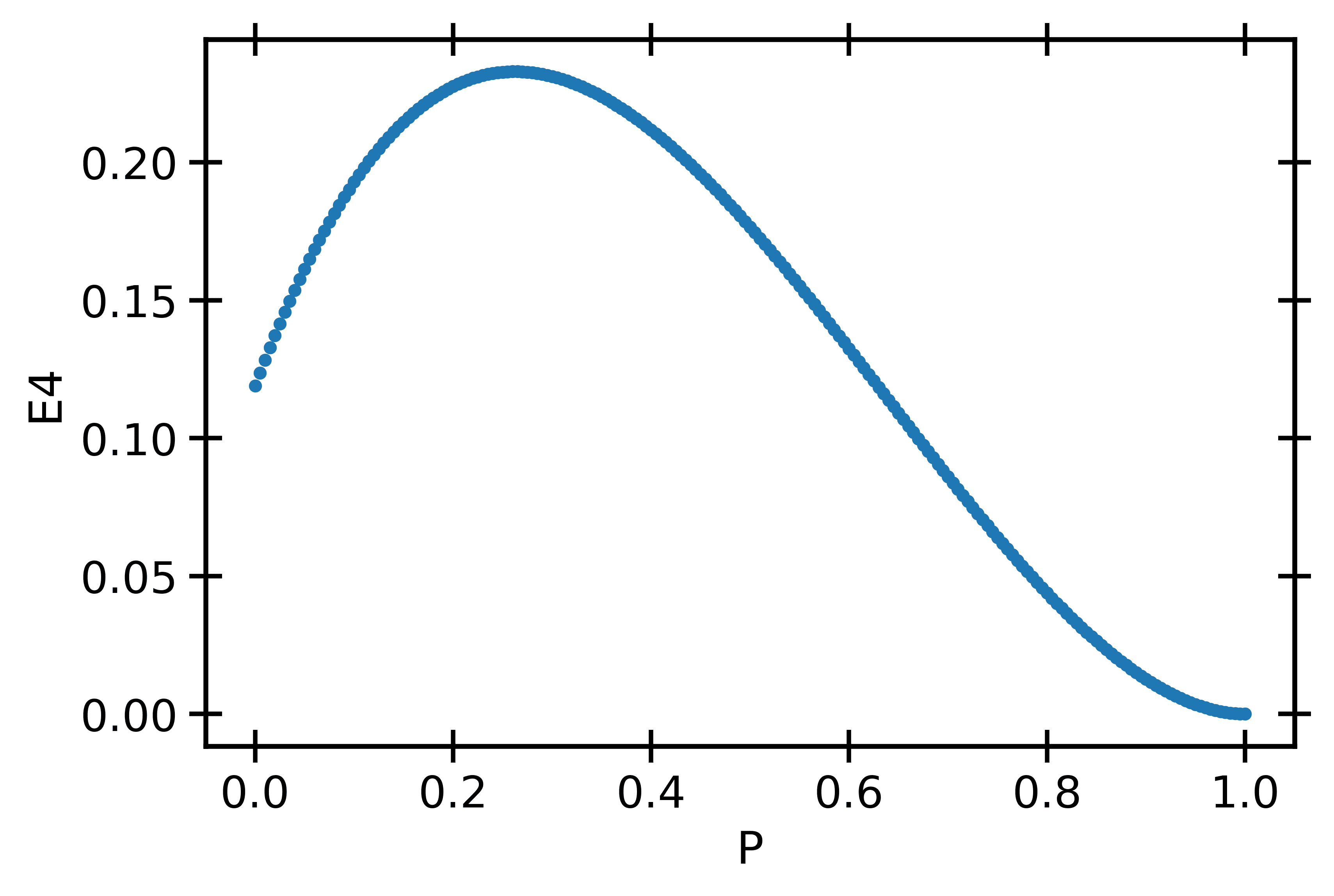}
    \end{subfigure}
    \\
    \begin{subfigure}[b]{0.49\textwidth}
        \centering
        \includegraphics[width=\textwidth]{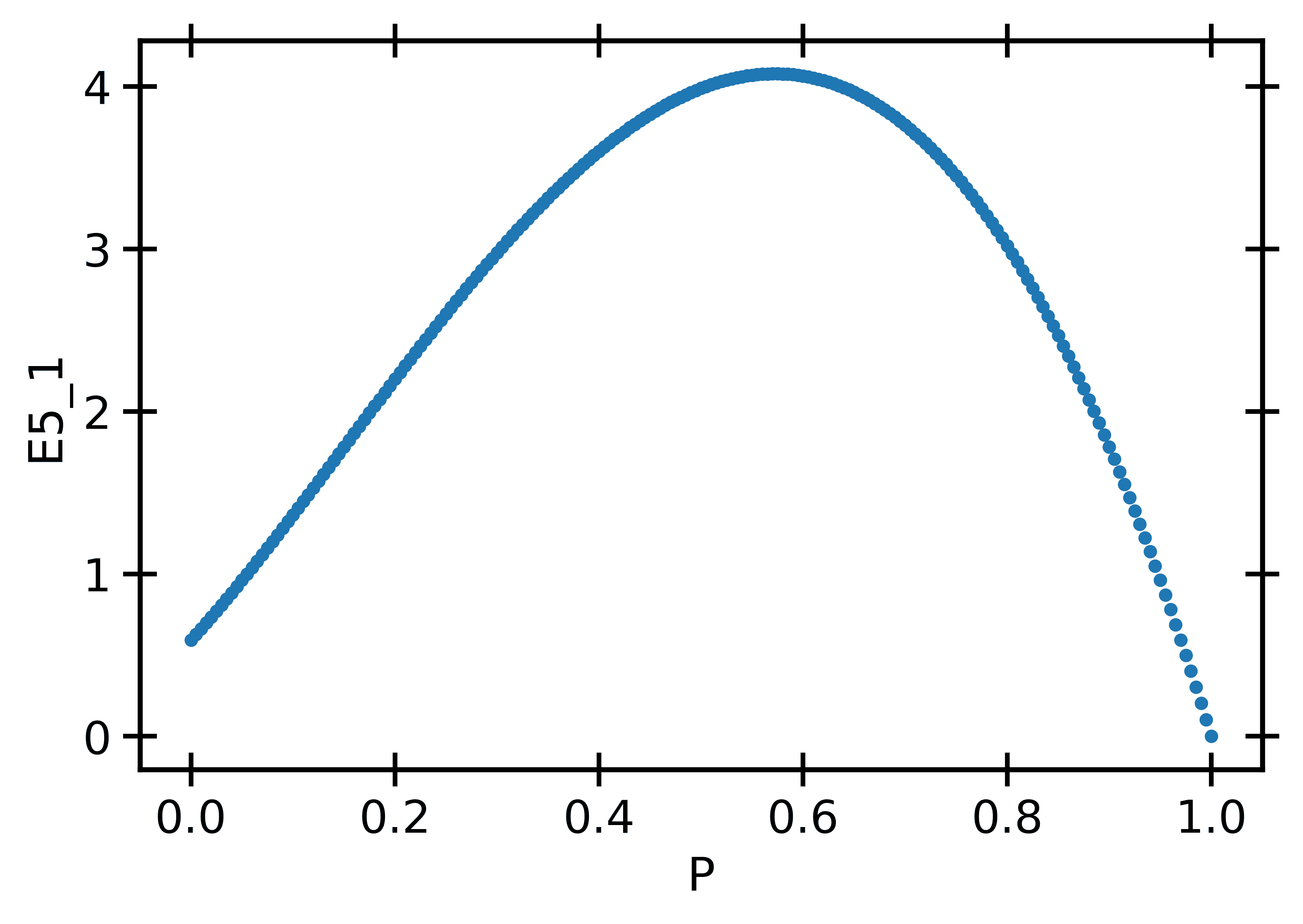}
    \end{subfigure}
    \begin{subfigure}[b]{0.49\textwidth}
        \centering
        \includegraphics[width=\textwidth]{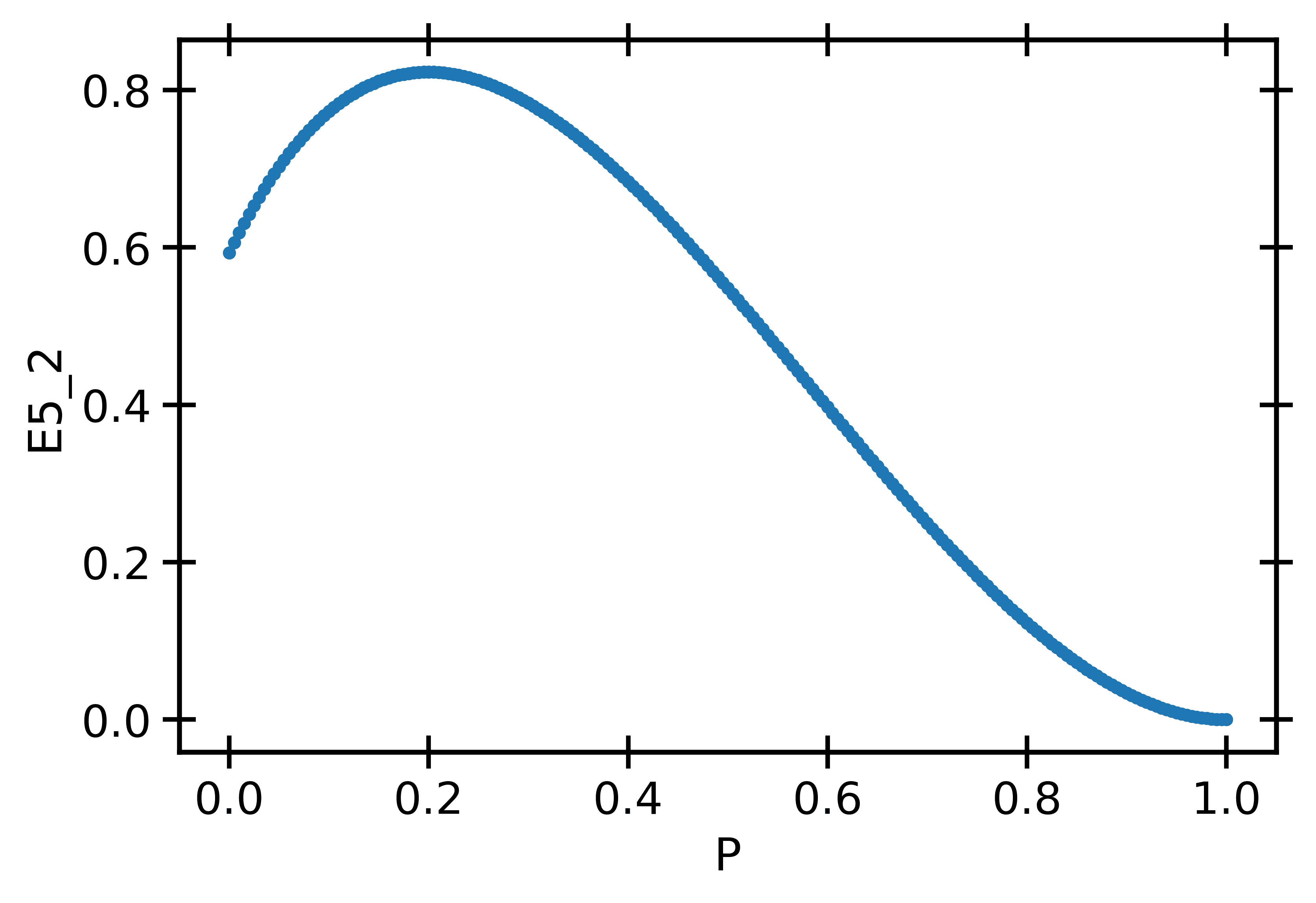}
    \end{subfigure}
    \\
    \begin{subfigure}[b]{0.49\textwidth}
        \centering
        \includegraphics[width=\textwidth]{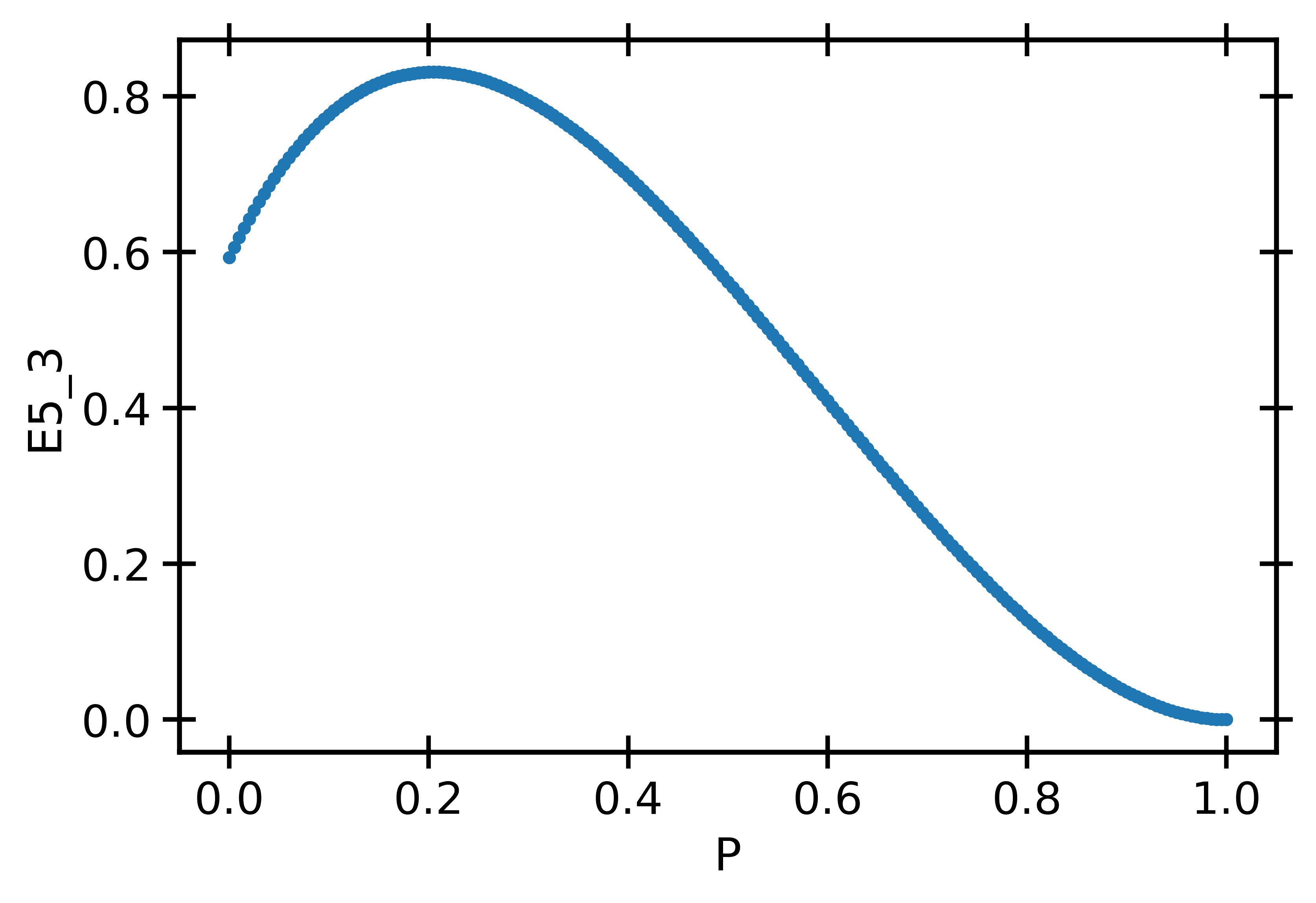}
    \end{subfigure}
    \caption{E3, E4, E5\_1, E5\_2 and E5\_3 in 
terms of the polarization $P$ for $S=7/2$. The error bars are smaller than 
the size of the symbols. }
   \label{fig.int_v8}
\end{figure}

\subsection{Spin $9/2$}
\begin{figure}[H]
    \centering
    \begin{subfigure}[b]{0.49\textwidth}
        \centering
        \includegraphics[width=\textwidth]{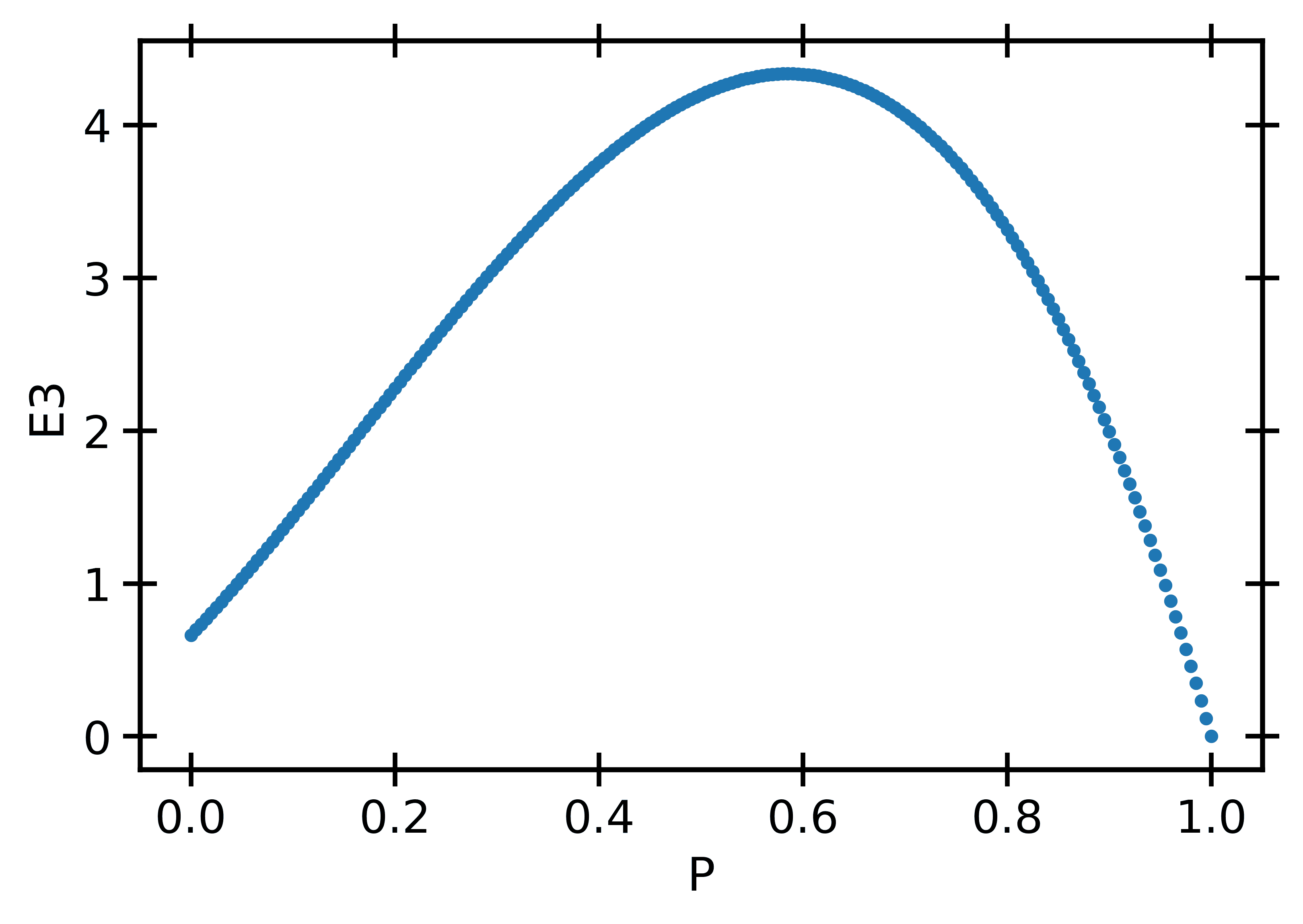}
    \end{subfigure}
    \begin{subfigure}[b]{0.49\textwidth}
        \centering
        \includegraphics[width=\textwidth]{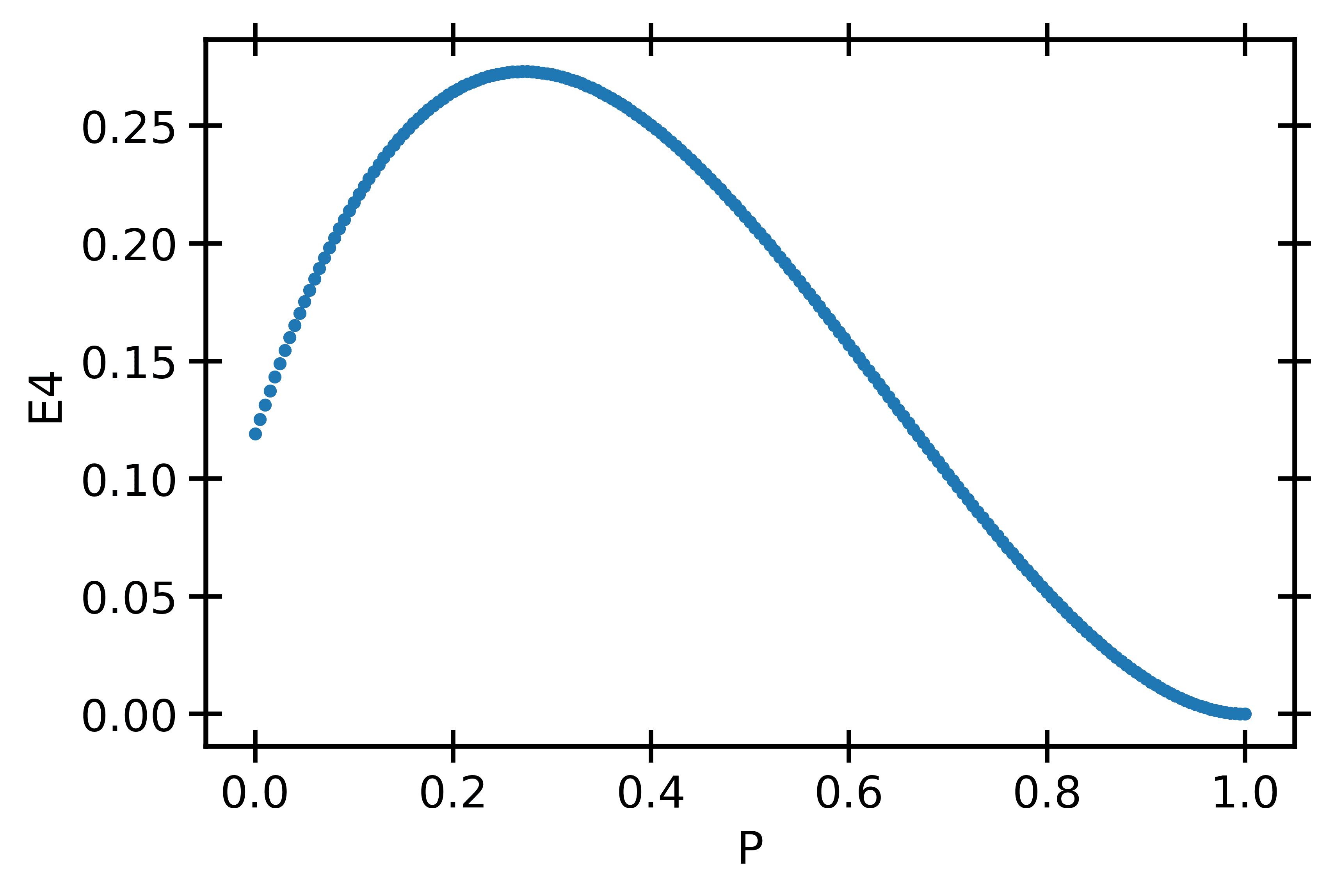}
    \end{subfigure}
    \\
    \begin{subfigure}[b]{0.49\textwidth}
        \centering
        \includegraphics[width=\textwidth]{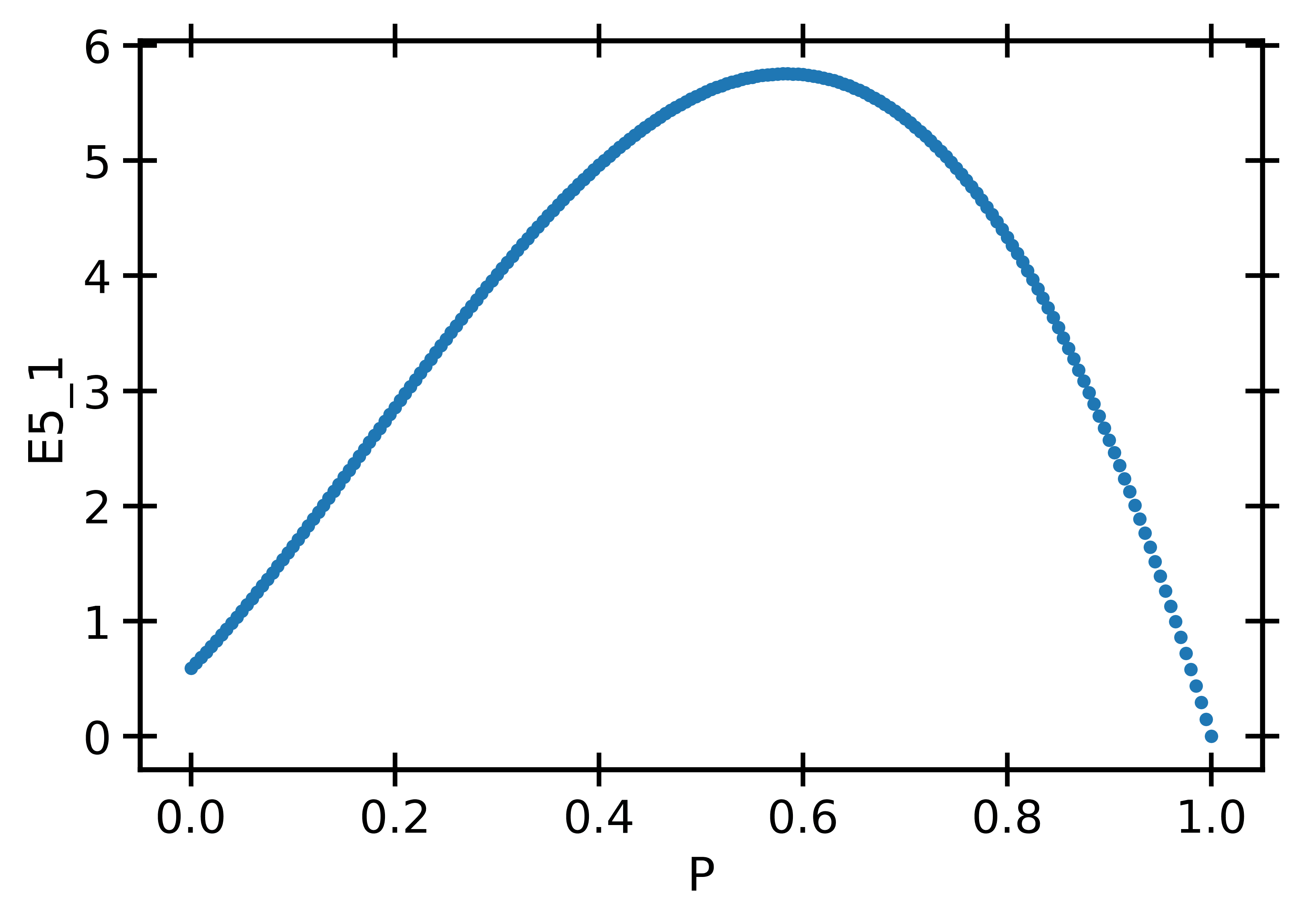}
    \end{subfigure}
    \begin{subfigure}[b]{0.49\textwidth}
        \centering
        \includegraphics[width=\textwidth]{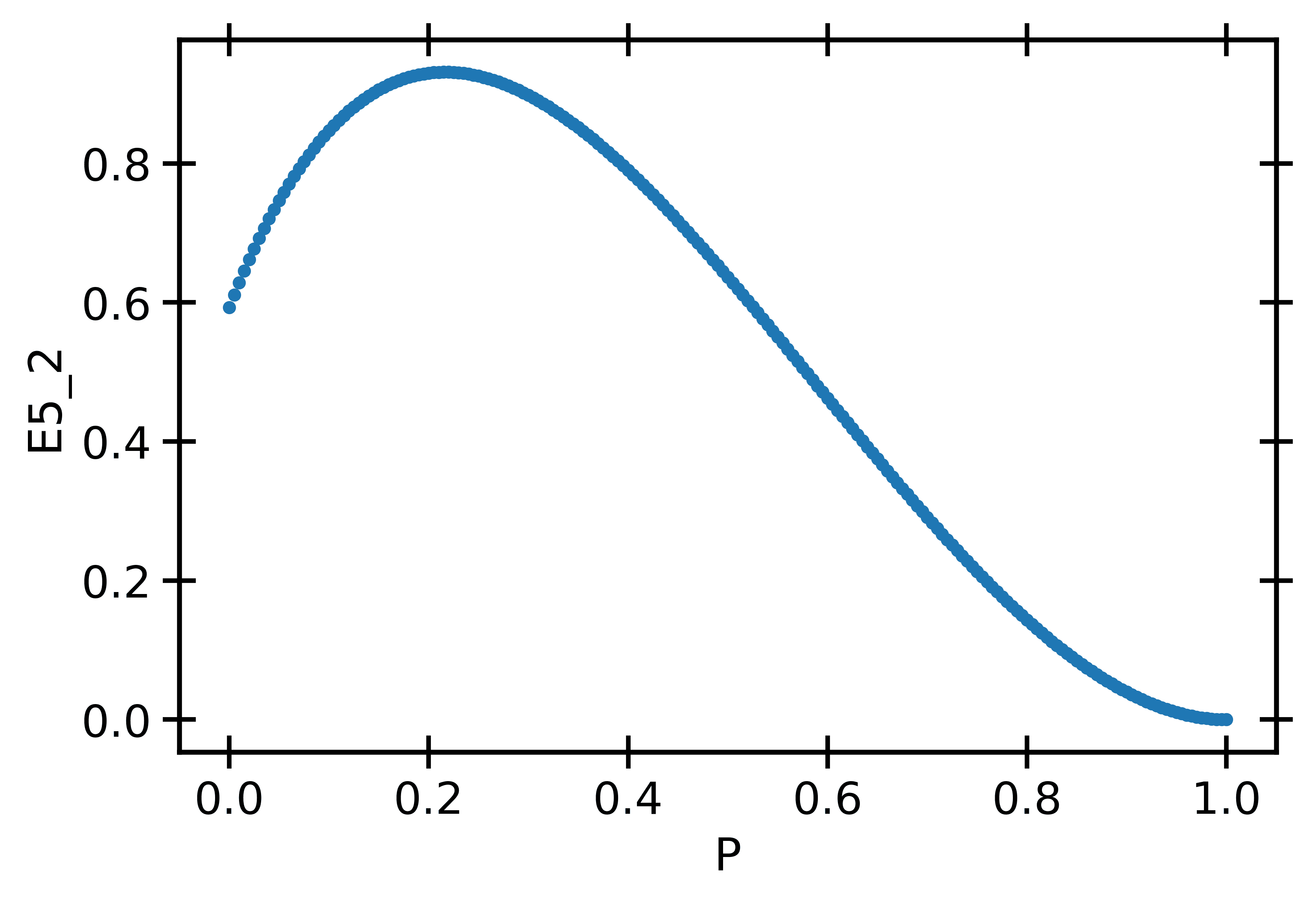}
    \end{subfigure}
    \\
    \begin{subfigure}[b]{0.49\textwidth}
        \centering
        \includegraphics[width=\textwidth]{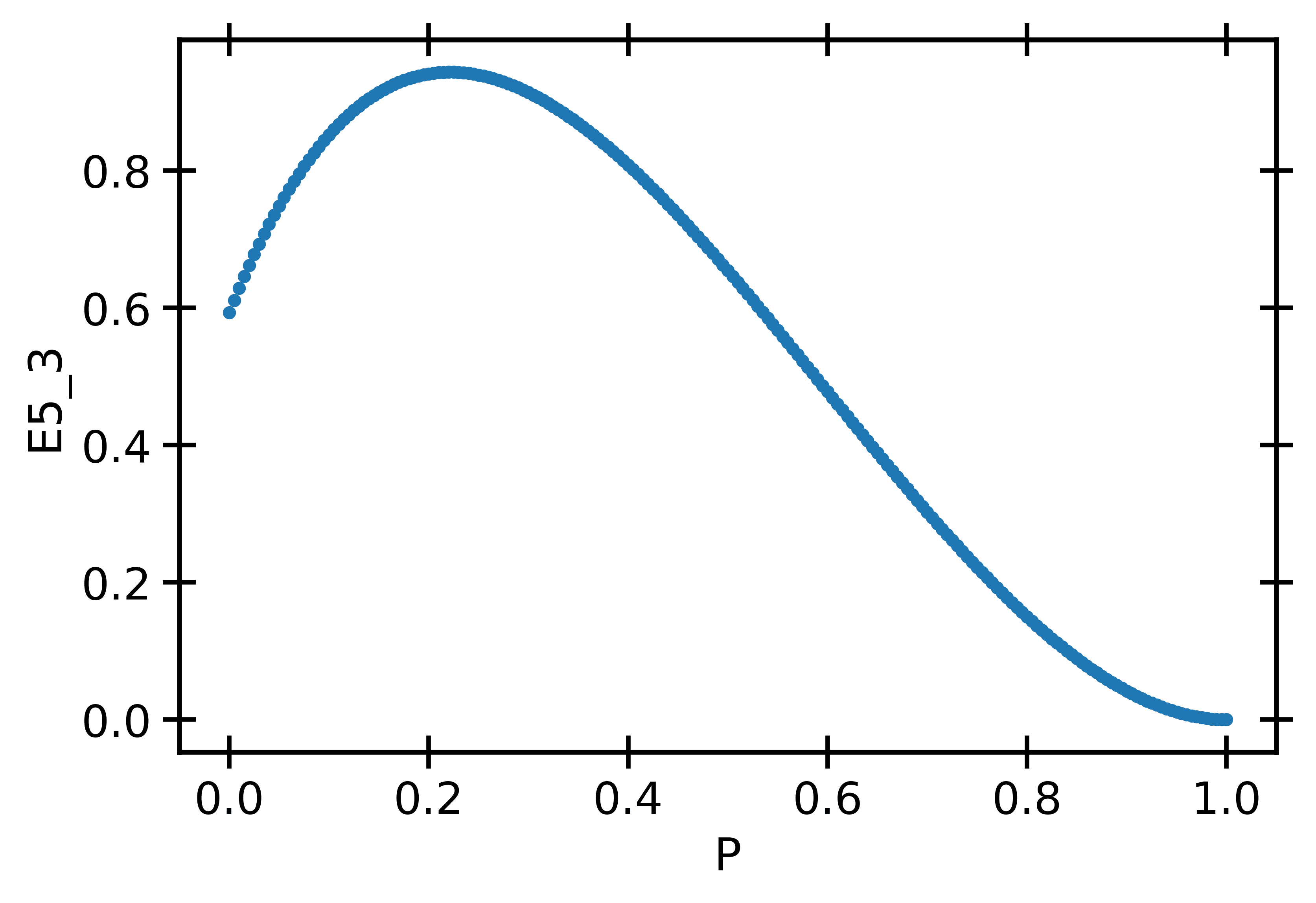}
    \end{subfigure}
    \caption{E3, E4, E5\_1, E5\_2 and E5\_3 in 
terms of the polarization $P$ for $S=9/2$. The error bars are smaller than 
the size of the symbols. }
\label{fig.int_v10}
\end{figure}

%\subsection{Spin $19/2$}
%\begin{figure}[H]
%    \centering
%    \begin{subfigure}[b]{0.49\textwidth}
%        \centering
%        \includegraphics[width=\textwidth]{E3_v20.png}
%    \end{subfigure}
%    \begin{subfigure}[b]{0.49\textwidth}
%        \centering
%        \includegraphics[width=\textwidth]{E4_v20.png}
%    \end{subfigure}
%    \\
%    \begin{subfigure}[b]{0.49\textwidth}
%        \centering
%        \includegraphics[width=\textwidth]{E5_1_v20.png}
%    \end{subfigure}
%    \begin{subfigure}[b]{0.49\textwidth}
%        \centering
%        \includegraphics[width=\textwidth]{E5_2_v20.png}
%    \end{subfigure}
%    \\
%    \begin{subfigure}[b]{0.49\textwidth}
%        \centering
%        \includegraphics[width=\textwidth]{E5_3_v20.png}
%    \end{subfigure}
%    \caption{\footnotesize{Representation of E3, E4, E5\_1, E5\_2 and E5\_3 in 
%terms of the polarization P for spin $19/2$. The error bars are so little that 
%cannot be seen.}}
%\label{fig.int_v20}
%\end{figure}

\end{widetext}

%\nocite{*}

\bibliography{aipsamp}% Produces the bibliography via BibTeX.

%\hrule

\end{document}